\documentclass[10pt,journal,compsoc,onecolumn]{IEEEtran}

\ifCLASSOPTIONcompsoc
  \usepackage[nocompress]{cite}
\else
  \usepackage{cite}
\fi

\usepackage[justification = centering,labelsep = period]{caption}
\usepackage{ragged2e}
\usepackage{amsmath,graphicx}
\usepackage{array}
\usepackage{textcomp}
\usepackage{stfloats}
\usepackage{verbatim}
\usepackage{graphicx}

\usepackage{psfrag,epsfig,subfigure,amsmath}
\usepackage{enumerate}
\usepackage{amsfonts,amssymb,amsmath,bm,paralist,theorem,cite,ifthen,color}
\usepackage{listings}

\usepackage{algorithm}
\usepackage{algpseudocode}
\usepackage{makecell}
\usepackage{dsfont}
\usepackage{url}

\usepackage{xcolor}
\usepackage[subject={Top1},author={YanmengWang},version=1]{pdfcomment}
\hypersetup{hidelinks}

\usepackage{threeparttable}
\usepackage{multirow}
\usepackage{booktabs}

\newtheorem{theorem}{Theorem}
\newtheorem{lemma}{Lemma}
\newtheorem{proposition}{Proposition}
\newtheorem{corollary}{Corollary}
\newtheorem{remark}{Remark}
\newtheorem{assumption}{Assumption}

\newtheorem{observation}{Observation}

\makeatletter
\renewcommand{\maketag@@@}[1]{\hbox{\m@th\normalsize\normalfont#1}}%
\makeatother

\allowdisplaybreaks[4]

\begin{document}

\title{Robust Federated Learning in Unreliable Wireless Networks: A Client Selection Approach}


\author{
Yanmeng~Wang,~\IEEEmembership{Member,~IEEE,}
Wenkai~Ji,
Jian~Zhou,~\IEEEmembership{Member,~IEEE,}
\\
Fu~Xiao,~\IEEEmembership{Senior Member,~IEEE,}
and~Tsung-Hui~Chang,~\IEEEmembership{Fellow,~IEEE}
\IEEEcompsocitemizethanks{
\IEEEcompsocthanksitem
Y. Wang, W. Ji, J. Zhou, and F. Xiao are with the School of Computer Science, Nanjing University of Posts and Telecommunications, Nanjing 210023, China.
(E-mail: hiwangym@gmail.com, jiwenkai540@gmail.com, zhoujian@njupt.edu.cn, xiaof@njupt.edu.cn)
\IEEEcompsocthanksitem
T.-H. Chang is with the School of Artificial Intelligence, The Chinese University of Hong Kong, Shenzhen 518172, China, and also with the Shenzhen Research Institute of Big Data, Shenzhen 518172, China.
(E-mail: tsunghui.chang@ieee.org)
\IEEEcompsocthanksitem
Corresponding author: Fu Xiao
\IEEEcompsocthanksitem
Published by IEEE Transactions on Mobile Computing (TMC)
}
}


\IEEEtitleabstractindextext{%
\begin{abstract}
\justifying
Federated learning (FL) has emerged as a promising distributed learning paradigm for training deep neural networks (DNNs) at the wireless edge, but its performance can be severely hindered by unreliable wireless transmission and inherent data heterogeneity among clients.
Existing solutions primarily address these challenges by incorporating wireless resource optimization strategies, often focusing on uplink resource allocation across clients under the assumption of homogeneous client-server network standards.
However, these approaches overlooked the fact that mobile clients may connect to the server via diverse network standards (e.g., 4G, 5G, Wi-Fi) with customized configurations, limiting the flexibility of server-side modifications and restricting applicability in real-world commercial networks.
This paper presents a novel theoretical analysis about how transmission failures in unreliable networks distort the effective label distributions of local samples, causing deviations from the global data distribution and introducing convergence bias in FL.
Our analysis reveals that a carefully designed client selection strategy can mitigate biases induced by network unreliability and data heterogeneity.
Motivated by this insight, we propose FedCote, a client selection approach that optimizes client selection probabilities without relying on wireless resource scheduling.
Experimental results demonstrate the robustness of FedCote in DNN-based classification tasks under unreliable networks with frequent transmission failures.
\end{abstract}

}

\maketitle

\IEEEdisplaynontitleabstractindextext
\IEEEpeerreviewmaketitle

\IEEEraisesectionheading{\section{Introduction}\label{sec:introduction}}

With rapid advancements in mobile communications and artificial intelligence (AI), edge AI, which leverages locally generated data to train deep neural networks (DNNs) at the wireless edge, has gained significant attention from both academia and industry \cite{meuser2024revisiting,tuli2022splitplace,zuo2024fluid,cui2024towards,fan2025ten}.
A prominent approach in this domain is federated learning (FL), where an edge server coordinates mobile clients in collaboratively training a shared DNN model while ensuring client privacy \cite{zhou2022privacy,shen2024fedconv,zhou2022pflf}.

However, FL faces a critical challenge due to ubiquitous data heterogeneity across clients, where training data are distributed in a non-i.i.d. and unbalanced manner.
If not addressed, data heterogeneity can severely degrade FL performance \cite{wang2023batch, wang2023beyond, li2024filling, chen2024learning,wang2022federated}.
Numerous FL algorithms have been proposed to mitigate this issue.
For example,
\texttt{FedProx} \cite{li2020federated} introduced a regularization term in the local objective function to control model divergence,
while \texttt{SCAFFOLD} \cite{karimireddy2020scaffold} employed control variates to correct local model drift.
\texttt{HFMDS} \cite{li2024feature} learned essential class-relevant features of real samples to generate an auxiliary synthetic dataset, which was shared among clients for local training, helping to alleviate data heterogeneity.
Additionally, \texttt{Aorta} \cite{xu2024overcoming} utilized the mixup data augmentation method in clients to balance class distributions and assigned aggregation weights based on local model quality, ensuring better models had greater influence during global aggregation.
Despite these advancements, these studies primarily focused on improving local training and global aggregation algorithms, often overlooking the influence of client selection on FL convergence.

In practice, client selection is a simple yet effective approach to mitigating the negative effects of data heterogeneity by ensuring that chosen clients collectively offer a more balanced and representative sample of the global data distribution \cite{cho2022towards,fu2023client,wang2023toward}.
However, current client selection approaches typically assumed reliable network conditions or relied on joint optimization with centralized configurations of uplink wireless resources for all clients.
The former neglected the inherent unreliability of network conditions at the wireless edge, while the latter overlooked the diverse, customized configurations and network standards (e.g., 4G, 5G, Wi-Fi) of mobile clients,
thereby limiting their practical implementation in commercial wireless networks
\cite{zheng2023federated, ye2023heterogeneous}.

\subsection{Related work}

In recent years, various client selection strategies have been proposed to optimize the FL process.
For example,
\texttt{Newt} \cite{zhao2022participant} combined local dataset size with the discrepancy between global and local models to assess client utility, selecting those with higher utility values.
In contrast, \texttt{POWER-OF-CHOICE} \cite{cho2022towards} prioritized clients with higher local loss values.
Additionally, \texttt{GS} \cite{ma2021client} utilized privacy-insensitive local label distribution to ensure that the aggregated label distribution from selected clients aligned with the global label distribution.
Furthermore, \texttt{FedCor} \cite{tang2022fedcor} leveraged client correlations to mitigate the effects of non-i.i.d. and unbalanced data in FL.
However, these approaches assumed ideal, lossless communication conditions between the server and clients,
limiting their practical implementation in commercial networks, particularly in resource-constrained wireless edge environments \cite{xu2025distributed, liu2024survey}.

\begin{figure}[!t]
\centering
\includegraphics[width= 3.5 in ]{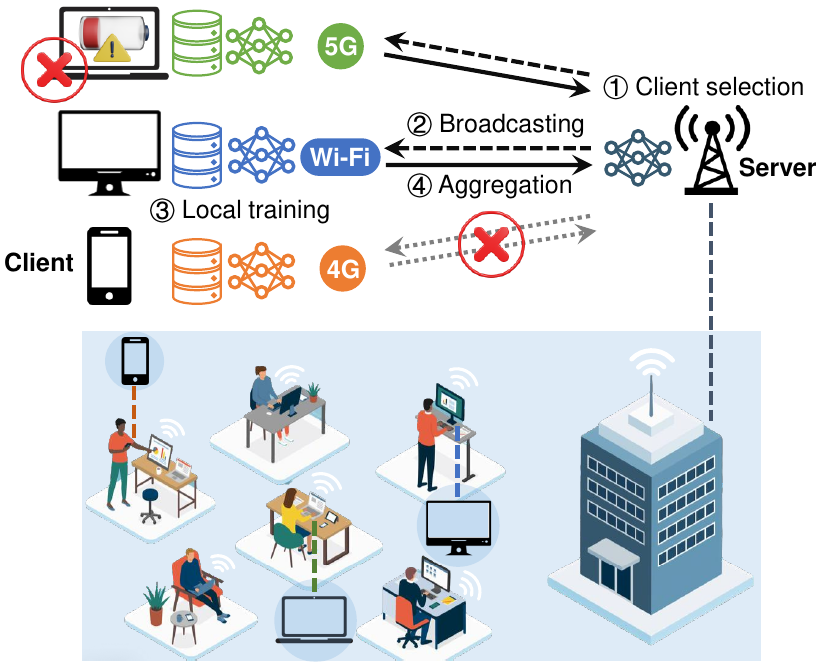}
\caption{FL in unreliable wireless networks with transmission failures.}
\label{fig:FL wireless networks}
\end{figure}

Recently, some studies have incorporated wireless communication conditions into client selection for FL.
These works typically assumed reliable downlink communication, with the server possessing sufficient power and bandwidth to broadcast the global model to clients, and primarily focused on optimizing uplink resource allocation.
For instance,
\texttt{CACS} \cite{qiao2020content} integrated channel capacity and local model updates into the client selection process, while \cite{luo2024adaptive} jointly optimized client selection probabilities and allocated bandwidth to address data heterogeneity and minimize FL convergence time.
Additionally, \cite{wu2024client} considered both transmission and energy consumption in a non-orthogonal multiple access (NOMA) system, optimizing client selection and resource allocation to minimize the overall time and energy cost of FL.
However, these studies still assumed reliable wireless conditions with lossless transmissions, which do not reflect the realities of practical implementations.
As illustrated in Fig. \ref{fig:FL wireless networks}, communication in real wireless networks between the server and clients is often unreliable, with frequent transmission failures caused by unstable channels or device-related issues.
These failures can intermittently disrupt the transfer of model parameters, leading to biased FL convergence \cite{salehi2021federated,wang2021robust}.

To mitigate the negative impacts of transmission failures on FL, some studies have focused on optimizing wireless resource allocation.
For instance,
\cite{chen2021joint} optimized uplink bandwidth and transmit power allocation for selected clients in frequency division multiple access (FDMA) systems, while \texttt{FedToe} \cite{wang2022quantized} adaptively adjusted uplink bandwidth, transmit power, and quantization bit allocation among clients.
Additionally, \cite{mahmoud2023federated} proposed an energy-efficient FL scheme by jointly optimizing uplink transmit power, bandwidth, and communication latency.
Some studies have extended this by jointly optimizing client selection and uplink resource allocation to alleviate the effects of unreliable networks.
For example,
\cite{zheng2023federated} optimized both client selection and uplink transmit power,
while \cite{chen2024robust} jointly optimized client selection, bandwidth allocation, and uplink transmit power.
Although these approaches offer significant improvements, they require centralized configuration of uplink communication resources across all mobile clients, which may pose deployment challenges in current commercial networks.
As shown in Fig. \ref{fig:FL wireless networks}, different mobile clients may connect to the server through diverse network standards, and devices may have user- or manufacturer-customized configurations, limiting the server's ability to modify them.
Different from above, \cite{salehi2021federated} optimized client selection and introduced a global aggregation scheme based on transmission failure probabilities to address both data heterogeneity and transmission failures.
However, this approach suffers from instability in high transmission failure conditions, as the transmission failure probability is incorporated into the denominator of the aggregation scheme.

It is worth noting that, beyond the aforementioned studies on digital communication systems, a parallel line of research investigates FL over over-the-air communication systems, where aggregation is performed in the analog domain \cite{sedaghat2023novel}.
In this setting, client selection also plays a critical role in improving performance. For instance, \cite{sun2021dynamic} jointly optimized client selection and transmit power under computation and communication energy constraints, while \cite{bereyhi2023device} considered joint optimization of client selection and linear receiver design to maximize the number of participating clients subject to estimation error constraints.
In this work, however, we focus on digital communication systems and regard over-the-air FL as an interesting direction for future extension.

\subsection{Contributions}

In this paper, we highlight the need for an innovative client selection strategy that effectively mitigates the negative impacts of data heterogeneity and {unreliable communication conditions}, while being practical and easily implementable in commercial networks without reconfiguring existing wireless resource allocation.
We begin by presenting a novel theoretical analysis of the impact of transmission failures on FL convergence.
Based on this analysis, we propose a new client selection approach, termed \texttt{FedCote} (\underline{Fed}erated learning algorithm with \underline{c}lient selection \underline{o}ptimization under \underline{t}ransmission failur\underline{e}),
which only optimizes client selection probabilities without reconfiguring existing network resource allocation, thereby enabling straightforward implementation in current commercial networks that support a diverse range of mobile devices and network standards.
Our main contributions are as follows:
\begin{enumerate}[1)]
\item
\textbf{Differentiating the causes of data heterogeneity}:
In contrast to previous studies on FL convergence, we theoretically distinguish between two sources of data heterogeneity: data feature distribution and label distribution. This differentiation facilitates a more comprehensive analysis of the impact of transmission failures on FL convergence.
To the best of our knowledge, this paper is the first to theoretically analyze FL convergence by separating data heterogeneity into these two components and jointly examining their effects, alongside transmission failures.

\item
\textbf{FL convergence analysis under data heterogeneity and transmission failure}:
We analyze a non-convex FL problem by considering both the common non-i.i.d. scenario with label distribution skew (where local label distributions vary across clients) and unreliable wireless networks.
The theoretical results show that transmission failures distort the effective appearance probabilities of local samples for each class (i.e., the effective label distribution), leading to a deviation from the true label distribution of the global dataset.
This distortion amplifies the negative effects of data heterogeneity and biases FL convergence.
Importantly, these findings are not network-specific and are applicable to various wireless networks affected by transmission failures.

\item
\textbf{\texttt{FedCote}}:
Based on the convergence analysis, we propose \texttt{FedCote}, a client selection approach designed to mitigate convergence bias by aligning the effective and actual label distributions.
Notably, \texttt{FedCote} optimizes only client selection probabilities and does not require reconfiguration of wireless resource allocation, making it simple to implement in commercial networks.

\item
\textbf{Experiments}:
We implement \texttt{FedCote} across various deep learning-based tasks, including handwritten-digit recognition and color image classification.
Experimental results demonstrate that \texttt{FedCote} outperforms benchmark schemes in unreliable wireless edge networks, highlighting its effectiveness.
\end{enumerate}

\textbf{Synopsis:}
Section \ref{sec:system model} presents the FL algorithm in wireless networks under transmission failures.
Section \ref{sec:convergence analysis} provides a theoretical analysis of the impact of transmission failures on FL convergence.
Building on the theoretical findings, Section \ref{Sec:Client Selection} proposes the client selection approach, termed \texttt{FedCote}.
The experimental evaluation of the proposed approach is detailed in Section \ref{sec:experimental results}.
Finally, Section \ref{section:Conclusion} concludes the paper with key findings and future research directions.
The key notations used in this paper are summarized in Table \ref{Summary of Notations}.

\begin{table}[!t]
\centering
\caption{Summary of Notations}
\begin{tabular}{ll}
\toprule
\textbf{Notation} & \textbf{Description} \\
\midrule
$N$ & Total number of clients \\
$K$ & Number of clients selected per iteration \\
$\mathcal{K}_r$ & Set of clients selected at iteration $r$ \\
$E$ & Number of local update steps per iteration \\
$R$ & Total number of iterations \\
$\gamma$ & Learning rate \\
$\mathcal{D}$, $\mathcal{D}_i$ & Global dataset; client $i$'s local dataset \\
$\nabla F$, $\nabla F_i$ & Gradient of the global and local cost functions \\
$p_i$, $s_i$ & Weight and selection probability of client $i$ \\
$\alpha_{g,c}$, $\alpha_{i,c}$ & Label distribution of class $c$ in global dataset and in client $i$'s local dataset \\
${\bar \alpha}_c$ & Effective label distribution of class $c$ in the global aggregation \\
$\mathbf{w}^{r,E}_{i}$, ${\bar {\mathbf{w}}}_{r}$ & Local model of client $i$ after $E$ local updates and the aggregated global model at iteration $r$ \\
$\epsilon^r_i$ & Transmission failure probability of client $i$ \\
${\bar \beta}^r_i$ & Effective appearance probability of client $i$ \\
$\mathds{1}^{r}_{i}$ & Indicator of successful reception of client $i$'s model at iteration $r$ \\
\bottomrule
\end{tabular}
\label{Summary of Notations}
\end{table}

\section{FL under Transmission Failure}\label{sec:system model}

\subsection{FL Algorithm}\label{sec:FL procedure}

Considering a wireless FL network as shown in Fig. \ref{fig:FL wireless networks}, a central server coordinates $N$ mobile clients to solve the following distributed learning problem:
\begin{equation}\label{objective function}
\min_{{\mathbf{w}} } \; F({\mathbf{w}})
=
\sum_{i = 1}^N p_i F_i({\mathbf{w}})
,
\end{equation}
where
$p_i = {|\mathcal{D}_i|}/{|\mathcal{D}|}$ is the weight of client $i$, with $\mathcal{D}_i$ denoting the local dataset of client $i$ and $\mathcal{D} = \bigcup_{i = 1}^N \mathcal{D}_i$ as the global dataset.
The (possibly) non-convex local cost function is defined as $F_i({\mathbf{w}}) = \mathbb{E}_{\xi_i \in \mathcal{D}_i}[\mathcal{L}({\mathbf{w}};\xi_i)]$, where $\mathbf{w}$ denotes the model parameters to be learned, $\xi_i$ refers to each sample in client $i$'s dataset, and $\mathcal{L}$ is the loss function.
The global cost function is similarly expressed as $F({\mathbf{w}}) = \mathbb{E}_{\xi \in \mathcal{D}}[\mathcal{L}({\mathbf{w}};\xi)]$.

We generally refer to the well-known \texttt{FedAvg} algorithm \cite{mcmahan2017communication} to outline the overall FL procedure, which consists of the following four steps during each ${r}$-th iteration:

\begin{enumerate}[1)]
\item
\textbf{Client selection}:
Due to the limited communication bandwidth at the wireless edge, the server typically selects a subset of $K$ mobile clients, denoted as ${\mathcal{K}}_{r}$, where $|{\mathcal{K}}_{r}| = K$, to participate in updating the global model in each iteration.
The clients in ${\mathcal{K}}_{r}$ are selected with replacement, according to the selection probability distribution $\{ s_1,\cdots,s_N \}$, where $s_i$ is the probability for selecting client $i$.

\item
\textbf{Broadcasting}:\label{sec:111}
The server broadcasts the latest global model ${\bar {\mathbf{w}}}_{r-1}$ to each selected client $i \in {\mathcal{K}}_{r}$.

\item
\textbf{Local model updating}:
Each selected client $i \in {\mathcal{K}}_{r}$ updates its local gradient parameters over $E$ successive steps of gradient descent\footnote{
When mini-batch stochastic gradient descent (SGD) is employed, the model gradient $\nabla F_{i}( {\mathbf{w}}^{r,t-1}_{i} )$ in (\ref{eq:local SGD_b}) is modified to $\nabla F_{i}( {\mathbf{w}}^{r,t-1}_{i}; {\bm \xi}^{r,t}_{i})$, where ${\bm \xi}^{r,t}_{i}$ represents the mini-batch samples and the local cost function is $F_{i}( {\mathbf{w}}^{r,t-1}_{i}; {\bm \xi}^{r,t}_{i}) = \mathbb{E}_{\xi_i \in {\bm \xi}^{r,t}_{i}}[\mathcal{L}({\mathbf{w}}^{r,t-1}_{i};\xi_i)]$.
For ease of illustration, we consider full gradient descent for theoretical development, while mini-batch SGD is used in experiments.
}.
Specifically, the local model is initialized as in \eqref{eq:local SGD_a} and subsequently updated according to \eqref{eq:local SGD_b}:
\begin{subequations}\label{eq:local updating}
\begin{align}
& \mathbf{w}^{r,0}_{i} = {\bar {\mathbf{w}}}_{r-1} ,
\label{eq:local SGD_a}\\
&
\mathbf{w}^{r,t}_{i} = {\mathbf{w}}^{r,t-1}_{i}  -  \gamma \nabla F_{i}( {\mathbf{w}}^{r,t-1}_{i} ),
\;
t = 1,\ldots, E,
\label{eq:local SGD_b}
\end{align}
\end{subequations}
where $\gamma>0$ is the learning rate and $E$ is the number of local updating steps.

\item
\textbf{Aggregation}:
The server collects the local model $\mathbf{w}_i^{r,E}$ from each selected client $i \in {\mathcal{K}}_{r}$, and subsequently aggregates them to generate a new global model by
\begin{align}\label{eq:global aggregation}
{\bar {\mathbf{w}}}_{r} = \frac{1}{K} \sum_{i \in {\mathcal{K}}_{r}} \mathbf{w}^{r,E}_{i}
.
\end{align}

\end{enumerate}

Under ideal and lossless wireless conditions, the \texttt{FedAvg} algorithm, by simply setting each client's selection probability $s_i = p_i$ and employing global aggregation as defined in \eqref{eq:global aggregation}, can yield an unbiased estimate of the aggregated model when all $N$ clients participate, resulting in
\begin{align}\label{eq:FedAvg unbiased estimate}
\mathbb{E}_{\mathcal{K}_{r}} [{\bar {\mathbf{w}}}_{r}] = \sum_{i = 1}^N p_i \mathbf{w}^{r,E}_{i}.
\end{align}
This unbiased estimate ensures that FL converges to an appropriate solution, even in the presence of non-i.i.d. data and with partial participation of $K$ clients per iteration \cite{liconvergence}.

However, the aforementioned FL scheme remains still far from practical implementation.
In real wireless networks, as depicted in Fig. \ref{fig:FL wireless networks}, transmission failures intermittently disrupt the delivery of model parameters between the server and clients.
These failures primarily stem from two factors:
a)
\emph{transmission channel conditions}, such as large-scale fading \cite{chen2021joint}, channel state information (CSI) errors \cite{wang2014outage}, and finite blocklength transmission \cite{xu2020transmission};
b)
\emph{device-related issues}, such as equipment malfunctions and battery drain\cite{yang2020power}.
Such unreliable wireless environments with transmission failures severely compromise the unbiased estimate in \eqref{eq:FedAvg unbiased estimate}, thereby adversely affecting the convergence properties and overall performance of FL \cite{wang2021robust,salehi2021federated}.
Thus, it is imperative to account for transmission failures in the design of wireless FL systems.

\subsection{Transmission Failure}

Previous research has extensively explored approaches for estimating transmission failure probability across various  network scenarios,
including quasi-static fading channels \cite{xi2011general}, 
multiple-input multiple-output (MIMO) channels with imperfect CSI \cite{park2012outage}, 
finite blocklength transmission with non-orthogonal multiple access (NOMA) \cite{xu2020transmission}, 
and systems beyond 5G \cite{yang2020power}. 
In this paper, we do not limit our analysis to specific network scenarios;
instead, we utilize $\epsilon^{r}_{i}$ to represent the transmission failure probability for client $i$ at iteration $r$ in our theoretical framework.
In practical wireless communication systems, this probability can be estimated on-the-fly by the server using methods such as leveraging instantaneous CSI feedback to empirically track failures over a sliding window \cite{goldsmith2005wireless},
inferring probabilities from Acknowledgment (ACK) / Negative ACK (NACK) reports via Bayesian or reinforcement learning techniques when CSI is limited \cite{alouini2000adaptive},
or monitoring pilot-signal quality and retransmission statistics under Automatic Repeat Request (ARQ) / Hybrid ARQ (HARQ) protocols \cite{zeng2019energy}.

We assume that downlink communication from the server to selected clients is reliable, given the server's sufficient power and bandwidth to broadcast the global model.
Our focus is on uplink transmission failures, which may arise from resource-constrained or unstable uplink channels, as well as device-related issues.
Such failures prevent the server from correctly receiving local models, resulting in the global model being aggregated as
\begin{align}\label{global_model_failure}
{\bar {\mathbf{w}}}_{r}
=
\frac{\sum_{{i} \in {\mathcal{K}}_{r}} \mathds{1}^{r}_{i} \mathbf{w}^{r,E}_{i} }{\sum_{{i} \in {\mathcal{K}}_{r}} \mathds{1}^{r}_{i}}
,
\end{align}
where ${\mathds{1}^{r}_{i}}=1$ indicates successful receipt of the local model from client ${i}$, and ${\mathds{1}^{r}_{i}}=0$ otherwise.
Given the transmission failure probability $\epsilon^{r}_{i}$ for client $i$, we have
\begin{align}
{\mathds{1}^{r}_{i}}
=
\begin{cases}
1, \; \text{with probability } 1 - \epsilon^{r}_{i},
\\
0, \; \text{with probability } \epsilon^{r}_{i}.
\end{cases}
\end{align}
A stable and resource-rich wireless network yields a lower $\epsilon^{r}_{i}$, whereas unstable or resource-limited uplink channels lead to a higher $\epsilon^{r}_{i}$.
Additionally, if no clients successfully transmit their local updates (i.e., $\mathds{1}^{r}_{i} = 0$ $\forall i \in \mathcal{K}_{r}$), retransmissions are initiated until at least one client's local model is correctly received by the server, thus  avoiding the denominator $\sum_{{i} \in {\mathcal{K}}_{r}} \mathds{1}^{r}_{i} = 0$.
The described FL algorithm under transmission failure is summarized in Algorithm \ref{algorithm:FL under transmission failure}.

\begin{algorithm}[!t]
\caption{FL under transmission failure}
\begin{algorithmic}[1]
\State Initialize global model ${\bar {\mathbf{w}}}_{0}$ by the server;
\For{each iteration $r=1,2,\cdots,R$}
\Statex $\quad\,$ \textbf{\texttt{// \textcircled{1} Client selection:}}
\State Server selects $K$ clients with replacement according to the selection probabilities $\{s_1,\cdots,s_N\}$;
\Statex $\quad\,$ \textbf{\texttt{// \textcircled{2} Broadcasting:}}
\State Server sends global model ${\bar {\mathbf{w}}}_{r-1}$ to selected clients;
\Statex $\quad\,$ \textbf{\texttt{// \textcircled{3} Local model updating:}}
\For {each client ${i} \in \mathcal{K}_r$} (\textbf{in parallel})
\State Update local model by \eqref{eq:local updating};
\State Upload updated local model $\mathbf{w}^{r,E}_{i}$ to the server;
\EndFor
\Statex $\quad\,$ \textbf{\texttt{// \textcircled{4} Aggregation:}}
\If {$\sum_{{i} \in {\mathcal{K}}_{r}} \mathds{1}^{r}_{i} = 0$}
\State {Repeat Step 7 for all selected clients in $\mathcal{K}_r$;}
\Else
\State {Server updates global model by \eqref{global_model_failure}.}
\EndIf
\EndFor
\end{algorithmic}
\label{algorithm:FL under transmission failure}
\end{algorithm}

\begin{figure}[!t]
\begin{minipage}[h]{1\linewidth}
\centering
\includegraphics[width= 2.55 in ]{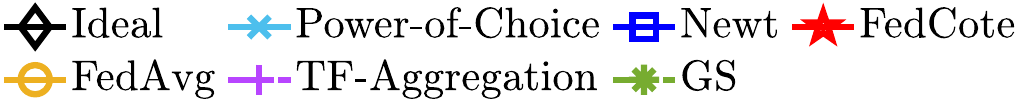}
\end{minipage}
\begin{minipage}[h]{1\linewidth}
\centering
\subfigure[i.i.d. data.]{
\includegraphics[width= 1.69 in ]{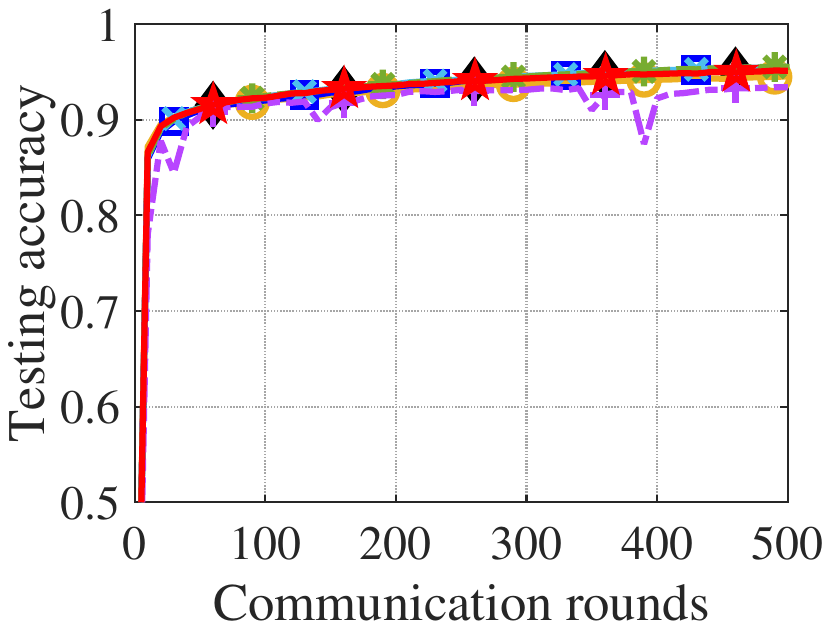}}
\subfigure[Non-i.i.d. data.]{
\includegraphics[width= 1.69 in ]{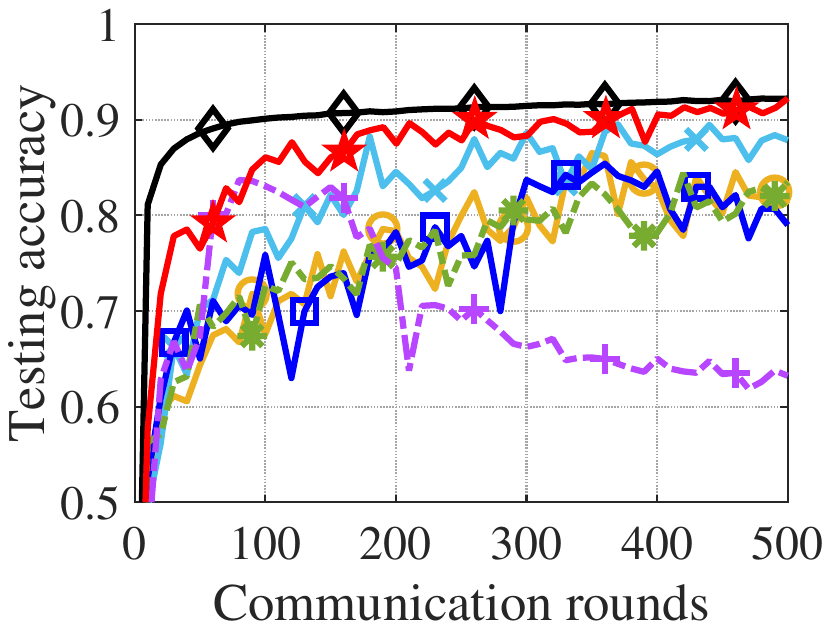}}
\caption{FL performance on unbalanced MNIST datasets under transmission failures (\emph{N} = 20, \emph{K} = 10, unbalanced ratio = 0.9).}
\label{fig:performance unbalanced MNIST}
\end{minipage}
\end{figure}

Fig. \ref{fig:performance unbalanced MNIST} illustrates the impact of transmission failures on FL performance by comparing the testing accuracy across various FL schemes under both i.i.d. and non-i.i.d. data distributions.
In this experiment, a DNN is trained on the MNIST dataset \cite{lecun1998gradient}, with clients experiencing varying levels of transmission failures and unbalanced local datasets.
Further details on the experimental setup are provided in Section \ref{sec:Parameter Setting}.
As shown in Fig. \ref{fig:performance unbalanced MNIST}, \texttt{FedAvg}, which employs a random selection strategy with $s_i = p_i$, exhibits significant performance degradation under non-i.i.d. data conditions, compared to the ideal case without transmission failures.
Additionally, advanced client selection schemes also perform poorly under non-i.i.d. conditions.
For example, as depicted in Fig. \ref{fig:performance unbalanced MNIST}(b), \texttt{Newt} \cite{zhao2022participant} and \texttt{GS} \cite{ma2021client} converge to biased solutions, while \texttt{POWER-OF-CHOICE} \cite{cho2022towards} experiences significant fluctuations and fails to reach a stable solution.
Furthermore, the client selection scheme proposed in \cite{salehi2021federated} (referred to as \texttt{TF-Aggregation} in this paper for ease of reference), designed to address transmission failures by incorporating failure probabilities into the global aggregation scheme and optimizing client selection, also exhibits instability under the non-i.i.d. data case.
These results highlight the critical need to understand the adverse impact of transmission failures on FL performance, particularly in non-i.i.d. scenarios, and to develop effective strategies to mitigate these effects.

In light of this, this paper proposes a robust FL scheme termed \texttt{FedCote}, which exhibits strong adaptability in unreliable wireless networks affected by transmission failures, as illustrated in Fig. \ref{fig:performance unbalanced MNIST}.
We first present a novel convergence analysis of Algorithm \ref{algorithm:FL under transmission failure} in Section \ref{sec:convergence analysis}.
Following this, Section \ref{Sec:Client Selection} introduces a new client selection scheme designed to enhance FL performance under transmission failures, without necessitating modifications to existing wireless resource configurations.

\section{How Transmission Failure affects FL?}\label{sec:convergence analysis}

\subsection{Causes of Data Heterogeneity}

In supervised learning, each sample consists of an input data and its corresponding label.
Let $\alpha_{i,c}$ denote the proportion of class-$c$ samples in client $i$'s local dataset,
and $\alpha_{g,c}$ $(= \sum_{i = 1}^N p_i \alpha_{i,c})$ represent the corresponding proportion in the global dataset.
Similarly, $\nabla F_{i,c}$ and $\nabla F_{g,c}$ denote the gradients of the cost function for class-$c$ samples in the local and global datasets, respectively.
For a set of samples with $C$ classes, the local and global gradients can be expressed as
\begin{subequations}\label{eq:nabla Fi F alpha c}
\begin{align}
& \nabla F_i({\mathbf{w}}) = \sum_{c=1}^C \alpha_{i,c} \nabla F_{i,c} ({\mathbf{w}}),
\label{eq:nabla Fi alpha c}
\\
&
\nabla F ({\mathbf{w}}) = \sum_{c=1}^C \alpha_{g,c} \nabla F_{g,c} ({\mathbf{w}}).
\label{eq:nabla F alpha c}
\end{align}
\end{subequations}

In most existing studies on the convergence analysis of FL for general non-convex optimization problems (e.g., \cite{wang2022quantized,wang2023batch,lian2017can,karimireddy2020scaffold,das2022faster}), data heterogeneity is commonly quantified using a constant that measures the difference between local and global gradients:
\begin{equation}\label{bound:Fi F Vi2}
\| \nabla F_i({\mathbf{w}}) - \nabla F ({\mathbf{w}}) \|^2 \leq V_i^2
,
\end{equation}
where a larger $V_i$ indicates greater data heterogeneity.
However, this approach provides a general characterization but offers limited insight into the underlying causes of data heterogeneity, thereby hindering a deeper analysis of how transmission failure affects FL.
To provide a more comprehensive understanding, we analyze data heterogeneity in detail and present Proposition \ref{proposition:data heterogeneity}.

\begin{proposition}\label{proposition:data heterogeneity}
The difference between local and global gradients can be bounded by two separate terms:
\begin{align}\label{ineq:nabla Fi - F}
\| \nabla F_i({\mathbf{w}})  -  \nabla F ({\mathbf{w}}) \|^2
\leq 
2
\bigg(
\underbrace{
\sum_{c=1}^C
\alpha_{i,c}
\| \nabla F_{i,c} ({\mathbf{w}})  -  \nabla F_{g,c} ({\mathbf{w}}) \|^2
}_{\text{\small (\ref{ineq:nabla Fi - F}a) related to data feature}}
+
\underbrace{
\chi^2_{\bm{\alpha}_i \| \bm{\alpha}_g}
\sum_{c=1}^C \alpha_{g,c} \| \nabla F_{g,c} ({\mathbf{w}}) \|^2
}_{\text{\small (\ref{ineq:nabla Fi - F}b) related to sample label}}
\bigg)
.
\end{align}
Here, term (\ref{ineq:nabla Fi - F}a) captures \textbf{data feature-related heterogeneity}, where the deviation between local and global gradients within a class, $\| \nabla F_{i,c} ({\mathbf{w}}) - \nabla F_{g,c} ({\mathbf{w}}) \|^2$, is influenced by the sample data characteristics such as local data insufficiency and feature shift across different data sources \cite{wang2022hetvis,li2021fedbn}.
Term (\ref{ineq:nabla Fi - F}b) reflects \textbf{label-related heterogeneity}, with the chi-square divergence $\chi^2_{\bm{\alpha}_i \| \bm{\alpha}_g} \triangleq \sum_{c=1}^C \frac{( \alpha_{i,c} - \alpha_{g,c} )^2}{\alpha_{g,c}}$ quantifying the mismatch between local and global label distributions.
\end{proposition}

\emph{Proof:}
See Appendix \ref{appendix:proof ineq nabla Fi - F}.
\hfill $\blacksquare$

Based on Proposition \ref{proposition:data heterogeneity}, data heterogeneity in FL arises from two sources: \emph{data feature distribution} and \emph{label distribution}.
To distinguish their respective effects on FL convergence, we introduce Assumptions \ref{assumption:class heterogeneity} and \ref{assumption:garident norm}, which provide bounds for the data feature-related and label-related terms in the right-hand side (RHS) of \eqref{ineq:nabla Fi - F}.

Notably, the classical bound in \eqref{bound:Fi F Vi2} characterizes the discrepancy between the local gradient evaluated over the entire local dataset and the global gradient computed on the full global dataset.
In contrast, Assumption \ref{assumption:class heterogeneity} refines this characterization by restricting the comparison to class-specific samples within both local and global datasets.
This refinement enables a more precise lens for the impact of analyzing data heterogeneity in the subsequent convergence analysis, and further motivates the novel optimization method introduced in \ref{Sec:Client Selection}, which alleviates the adverse effects of transmission failures solely through client selection.

\begin{assumption}\label{assumption:class heterogeneity}
Bounded gradient deviation within a class: $\| \nabla F_{i,c}({\mathbf{w}}) - \nabla F_{g,c} ({\mathbf{w}})\|^2 \leq V_{i,c}^2$, $\forall i \in [N]$.
\end{assumption}
\begin{assumption}\label{assumption:garident norm}
Bounded gradient norm: $\| \nabla F ({\mathbf{w}}) \|^2 \leq G^2$\cite{cho2022towards,qiao2020content,salehi2021federated}.
\end{assumption}

By combining \eqref{ineq:nabla Fi - F} with the above two assumptions, we derive Corollary \ref{lemma:class heterogeneity}, which will be employed in the convergence analysis of FL in Section \ref{sec:theorem FL failure} to quantify data heterogeneity.
\begin{corollary}\label{lemma:class heterogeneity}
Under Assumptions \ref{assumption:class heterogeneity} and \ref{assumption:garident norm}, the difference between the local and global gradients is bounded by
\begin{align}\label{ineq:nabla Fi - F 2}
\| \nabla F_i({\mathbf{w}}) - \nabla F ({\mathbf{w}}) \|^2
\leq &
2 \sum_{c=1}^C
\alpha_{i,c} V_{i,c}^2
+ 2 \chi^2_{\bm{\alpha}_i \| \bm{\alpha}_g} G^2
,
\end{align}
where the two terms on the RHS of \eqref{ineq:nabla Fi - F 2} quantify the effects of data feature-related and label-related heterogeneity, respectively.
\end{corollary}

\subsection{Convergence Rate of FL with Transmission Failure}\label{sec:theorem FL failure}

\subsubsection{Assumptions}

In addition to Assumptions \ref{assumption:class heterogeneity} and \ref{assumption:garident norm}, we adopt the following standard assumptions on $F_i$\cite{wang2022quantized,cho2022towards,qiao2020content,chen2021joint,salehi2021federated}:
\begin{assumption}\label{assumption:Fi lower bound}
The value of each local cost function $F_i({\mathbf{w}})$ is lower bounded: $F_i({\mathbf{w}}) \geq \underline{F} > -\infty$.
\end{assumption}
\begin{assumption}\label{assumption:L continuous}
Each local functions $F_i$ is differentiable, and its gradient $\nabla F_i$ is Lipschitz continuous with constant $L$:
$\forall$${\mathbf{w}}$ and ${\mathbf{w}'}$, $F_i({\mathbf{w}'}) \leq F_i({\mathbf{w}}) + ({\mathbf{w}'} - {\mathbf{w}})^T \nabla  F_i({\mathbf{w}}) + \frac{L}{2} \| {\mathbf{w}'} - {\mathbf{w}} \|_2^2$.
\end{assumption}

\subsubsection{Theoretical Results}

For clarity of exposition, we first consider fixed transmission failure probabilities throughout the training process, i.e., $\epsilon^{r}_{i} = \epsilon_{i}$ for all $r = 1, \ldots, R$.
This simplification is sufficient to capture the key insights into how transmission failures influence FL convergence.
The extension to the more general case is straightforward and is provided in Appendix \ref{proof: theorem unfixed failure prob}.

To analyze the convergence of FL under transmission failures, we begin with Lemma \ref{lemma:beta}, which quantifies the expected value of the global model aggregated using  \eqref{global_model_failure}, taking into account both partial participation and transmission failures.

\begin{lemma}\label{lemma:beta}
In the FL procedure described in Algorithm \ref{algorithm:FL under transmission failure}, the global aggregation in (\ref{global_model_failure}) satisfies
\begin{align}\label{average_beta}
\mathbb{E}_{{\mathcal{K}}_{r}, \mathds{1}^{r}_{i}}
\bigg[
\frac{ \sum_{{i} \in {\mathcal{K}}_{r}} \mathds{1}^{r}_{i}
\mathbf{w}^{r,E}_{i}
 }{ \sum_{{i} \in {\mathcal{K}}_{r}} \mathds{1}^{r}_{i}} \bigg|
\sum_{{i} \in {\mathcal{K}}_{r}} \mathds{1}^{r}_{i} \neq 0
\bigg]
=
\sum_{i=1}^N {\bar \beta}_i
\mathbf{w}^{r,E}_{i}
,
\end{align}
where ${\bar \beta}_i \in [0,1]$ and $\sum_{i=1}^N {\bar \beta}_i = 1$.
Here, $\mathbb{E}[\cdot]$ denotes the expectation taken over $\mathcal{K}_{r}$ and $\{ \mathds{1}^{r}_{i} \}$.
If the transmission failure probability $\epsilon_i = 0$ $\forall i \in [N]$, then $\mathds{1}^{r}_{i}=1$ and ${\bar \beta}_i = s_i$.
\end{lemma}

\emph{Proof:}
The result in Lemma \ref{lemma:beta} follows from discussions in \cite[Lemma 2]{wang2022quantized}.
For ease of reference, the proof is presented in Appendix \ref{sec:Proof_lemma_beta}.
\hfill $\blacksquare$

From \eqref{average_beta}, ${\bar \beta}_i$ can be interpreted as the effective appearance probability of client $i$ in the global aggregation.
When employing a random selection scheme with $s_i = p_i$, as in the \texttt{FedAvg} algorithm, ${\bar \beta}_i$ deviates from the weight $p_i$ defined in \eqref{objective function} due to transmission failures, rendering the unbiased estimate in \eqref{eq:FedAvg unbiased estimate} invalid.
Thus, instead of using \eqref{eq:FedAvg unbiased estimate}, which holds under ideal conditions, we employ \eqref{average_beta} in our analysis to study FL convergence under transmission failures.
The main convergence result is as follows.
\begin{theorem}\label{theorem:convergence FL failure}
Suppose Assumptions \ref{assumption:class heterogeneity} to \ref{assumption:L continuous} hold.
Given the number of selected clients $K$, assume the total number of gradient descent updates $T = RE$ is sufficiently large with $T \geq K^{3}$, while the number of local update steps $E$ remains relatively small with $E \leq T^{\frac{1}{4}}/K^{\frac{3}{4}}$, where $R$ denotes the number of iterations in Algorithm 1.
If the learning rate is set to $\gamma = K^{\frac{1}{2}} / (6L{T}^{\frac{1}{2}})$, then the convergence of FL under transmission failures, as described in Algorithm 1, is upper bounded by
\begin{align}\label{eq:convergence bound of FL failure}
&
\frac{1}{R} \sum_{r=1}^R
\mathbb{E}[ \left\| \nabla  F({\bar {\mathbf{w}}}_{r-1}) \right\|^2 ]
\notag \\
\leq &
\frac{2484L}{55(TK)^{\frac{1}{2}}}
\left( \mathbb{E}[F({\bar {\mathbf{w}}}_{0})]
- \underline{F} \right)
+ \bigg( \frac{276}{55(TK)^{\frac{1}{4}}} + \frac{24}{55(TK)^{\frac{1}{2}}} + \frac{4}{55(TK)^{\frac{3}{4}}} \bigg)
\underbrace{
\sum_{i = 1}^{N} {\bar \beta}_i \sum_{c=1}^C
\left( \alpha_{i,c} V_{i,c}^2
+ \chi^2_{\bm{\alpha}_i \| \bm{\alpha}_g} G^2 \right)
}_{\text{\small (\ref{eq:convergence bound of FL failure}a) caused by non-i.i.d. data}}
\notag \\
&
+
\underbrace{
\frac{828}{55} \chi^2_{\bm{\bar \beta}\|\mathbf{p}} \sum_{c=1}^C \sum_{i = 1}^N p_i \alpha_{i,c} V_{i,c}^2
+
\frac{828}{55}
\chi^2_{\bm{\bar \alpha} \| \bm{\alpha}_g} G^2
}_{\text{\small (\ref{eq:convergence bound of FL failure}b) caused by transmission failure and non-i.i.d. data}}
,
\end{align}
where in term (\ref{eq:convergence bound of FL failure}b), $\chi^2_{\bm{\bar \beta}\|\mathbf{p}} \triangleq \sum_{i = 1}^N \frac{({\bar \beta}_i - p_i)^2}{p_i}$ represents the chi-square divergence \cite{wang2020tackling} between the effective appearance probabilities $\{ \bar{\beta}_i \}$ and the weights $\{ p_i \}$.
Meanwhile, $\chi^2_{\bm{\bar \alpha} \| \bm{\alpha}_g} \triangleq \sum_{c=1}^C \frac{( \sum_{i = 1}^N (p_i - {\bar \beta}_i) \alpha_{i,c})^2}{\alpha_{g,c}} = \sum_{c=1}^C \frac{( \alpha_{g,c} - \sum_{i = 1}^N {\bar \beta}_i \alpha_{i,c})^2}{\alpha_{g,c}}$ quantifies the divergence between the actual global label distribution $\{ \alpha_{g,c} \}$ and the effective global label distribution $\{ {\bar \alpha}_c \}$, where ${\bar \alpha}_c \triangleq \sum_{i = 1}^N {\bar \beta}_i \alpha_{i,c}$.
\end{theorem}

\emph{Proof:}
Unlike existing studies on client selection for FL \cite{cho2022towards,ma2021client,qiao2020content,chen2021joint,salehi2021federated}, we explore more practical FL scenarios in wireless edge networks, where transmission failures, non-i.i.d. data, and non-convex learning problems coexist.
Furthermore, in contrast to \cite{wang2022quantized} and \cite{wang2020tackling}, which also address FL with non-i.i.d. data and non-convex learning problems, we refine data heterogeneity into data feature-related and label-related components and employ non-fixed selection probabilities (whereas \cite{wang2022quantized} and \cite{wang2020tackling} set $s_i = p_i$).
Our derivation builds on the analytical frameworks in \cite{wang2022quantized} and \cite{wang2020tackling}, integrating these  considerations to enhance the comprehensiveness of convergence analysis.
The detailed derivation process is presented in Appendix \ref{appendix:proof theorem convergence FL failure}.
\hfill $\blacksquare$

From the RHS of \eqref{eq:convergence bound of FL failure}, it is evident that the convergence of Algorithm \ref{algorithm:FL under transmission failure} is influenced by various parameters, including the number of selected clients $K$, data heterogeneity factors, $\{ V_{i,c}^2 \}$ and $\{ \chi^2_{\bm{\alpha}_i \| \bm{\alpha}_g} \}$, as well as the divergences $\chi^2_{\bm{\bar \beta}\|\mathbf{p}}$ and $\chi^2_{\bm{\bar \alpha} \| \bm{\alpha}_g}$.
The divergences $\chi^2_{\bm{\bar \beta}\|\mathbf{p}}$ and $\chi^2_{\bm{\bar \alpha} \| \bm{\alpha}_g}$ arise from transmission failures.
As shown in Lemma \ref{lemma:beta}, the transmission failure probability $\epsilon_{i}$ directly affects the effective appearance probability ${\bar \beta}_i$ of client $i$ in global aggregation.
A lower $\epsilon_{i}$ increases ${\bar \beta}_i$, while a higher $\epsilon_{i}$ decreases it.
Consequently, transmission failures cause ${\bar \beta}_i$ to deviate from the original weights $p_i$, leading to some clients being overrepresented and others underrepresented.
Terms (\ref{eq:convergence bound of FL failure}a) and (\ref{eq:convergence bound of FL failure}b) therefore indicate that both non-i.i.d. data and transmission failures degrade FL convergence.
Moreover, we discover several important insights as follows.
\begin{itemize}
\item
First, in an ideal wireless scenario without transmission failures, where ${\bar \beta}_i = p_i$ as stated in Lemma \ref{lemma:beta}, the divergence factors in term (\ref{eq:convergence bound of FL failure}b), $\chi^2_{\bm{\bar \beta}\|\mathbf{p}}$ and $\chi^2_{\bm{\bar \alpha} \| \bm{\alpha}_g}$, reduce to zero.
Consequently, term (\ref{eq:convergence bound of FL failure}b) is eliminated, but term (\ref{eq:convergence bound of FL failure}a) still impedes FL convergence due to non-i.i.d. data.

\item
Second, when local datasets are i.i.d., the local and global label distributions are identical (i.e., $\alpha_{i,c} = \alpha_{g,c}$), and the within-class deviation between local and global gradients, $V_{i,c}^2$ (as defined in Assumption \ref{assumption:class heterogeneity}), approaches zero.
This causes both terms (\ref{eq:convergence bound of FL failure}a) and (\ref{eq:convergence bound of FL failure}b) to vanish.
As a result, i.i.d. data mitigates the negative impact of transmission failures, enabling proper FL convergence despite their presence, as illustrated in Fig. \ref{fig:performance unbalanced MNIST}(a).

\item
Third, the term (\ref{eq:convergence bound of FL failure}b), arising from transmission failures and non-i.i.d. data, does not decrease with increasing $T$.
This indicates that transmission failures exacerbate the adverse effects of non-i.i.d. data, leading FL to converge to a biased solution, as observed in Fig. \ref{fig:performance unbalanced MNIST}(b).

\item
Lastly, combining term (\ref{eq:convergence bound of FL failure}b) with Corollary \ref{lemma:class heterogeneity} reveals that $\chi^2_{\bm{\bar \beta}\|\mathbf{p}}$ and $\chi^2_{\bm{\bar \alpha} \| \bm{\alpha}_g}$, arising from transmission failures, amplify the negative effects of data feature-related and label-related heterogeneity, respectively, thereby biasing FL convergence.
Furthermore, the impacts of data feature-related and label-related heterogeneity depends on the values of the average gradient deviation within a class, ${\sum}_{c=1}^C {\sum}_{i = 1}^N p_i \alpha_{i,c} V_{i,c}^2$, and the squared norm of model gradient, $G^2$.
To compare their relative impacts on FL convergence bias, a detailed comparison of the empirical value of ${\sum}_{c=1}^C {\sum}_{i = 1}^N p_i \alpha_{i,c} V_{i,c}^2$ and $G^2$ is provided in the subsequent Section \ref{sec:Dominant Influence Label Hetero}.
\end{itemize}

\subsection{Dominant Influence of Label-Related Heterogeneity}\label{sec:Dominant Influence Label Hetero}

To quantify the specific influence of data feature-related (first) and label-related (second) components in term (\ref{eq:convergence bound of FL failure}b) on biasing FL convergence, we conduct experimental tests to estimate the empirical ratio between ${\sum}_{c=1}^C {\sum}_{i = 1}^N p_i \alpha_{i,c} V_{i,c}^2$ and $G^2$.
Based on Proposition \ref{proposition:data heterogeneity} and Corollary \ref{lemma:class heterogeneity}, we estimate the within-class deviation between local and global gradients, $V_{i,c}^2$, using $\| \nabla F_{i,c}({\bar {\mathbf{w}}}_{r-1}) - \nabla F_{g,c} ({\bar {\mathbf{w}}}_{r-1})\|^2$, and the squared norm of model gradient, $G^2$, using $\sum_{c=1}^C \alpha_{g,c} \| \nabla F_{g,c} ({\bar {\mathbf{w}}}_{r-1}) \|^2$, where ${\bar {\mathbf{w}}}_{r-1}$ represents the global model aggregated at the $(r-1)$-th iteration.

Experiments are conducted on two distinct tasks: training a 3-layer fully connected DNN on the MNIST dataset \cite{lecun1998gradient} and training a ResNet-20 model with Group Normalization (GN) \cite{hsieh2020non} on the CIFAR-10 dataset \cite{krizhevsky2009learning}.
To mitigate the impact of transmission failures, we assume an ideal and lossless wireless environment.
Additionally, we consider an extreme non-i.i.d scenario with label distribution skew, where each client holds samples from only one class.
We examine both the balanced condition, where all clients have an equal number of samples, and the unbalanced condition, with half of the clients holding 90\% of the total training samples.
For further details on local label and weight distributions for each client, refer to Appendix \ref{sec:label_distribution_experiments_Vic_G_value}.
All results are averaged over five independent experiments, with $N=K=20$ and $E=1$.

\begin{figure}[!t]
\begin{minipage}[h]{1\linewidth}
\centering
\includegraphics[width= 3.3 in ]{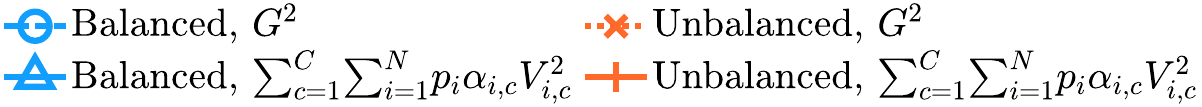}
\end{minipage}
\begin{minipage}[h]{1\linewidth}
\centering
\subfigure[MNIST.]{
\includegraphics[width= 1.69 in ]{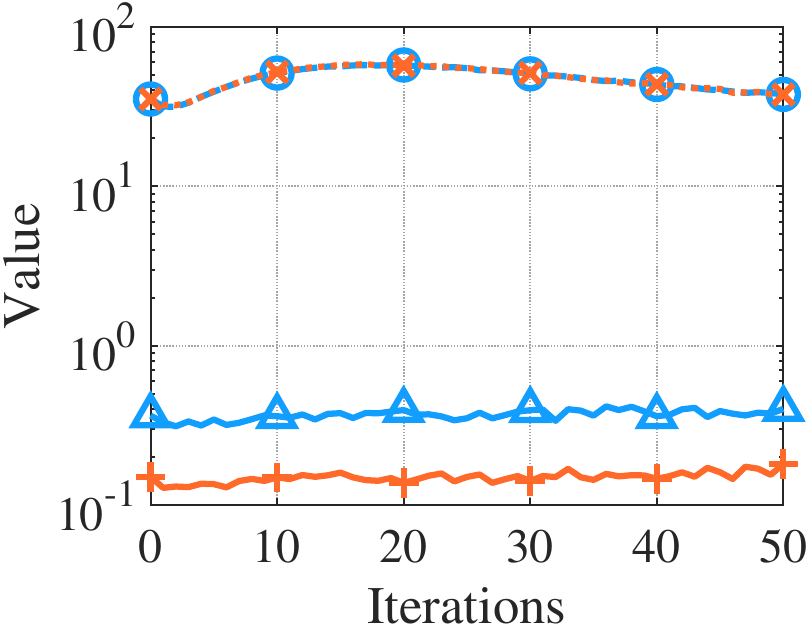}}
\hspace{1cm}
\subfigure[CIFAR-10.]{
\includegraphics[width= 1.69 in ]{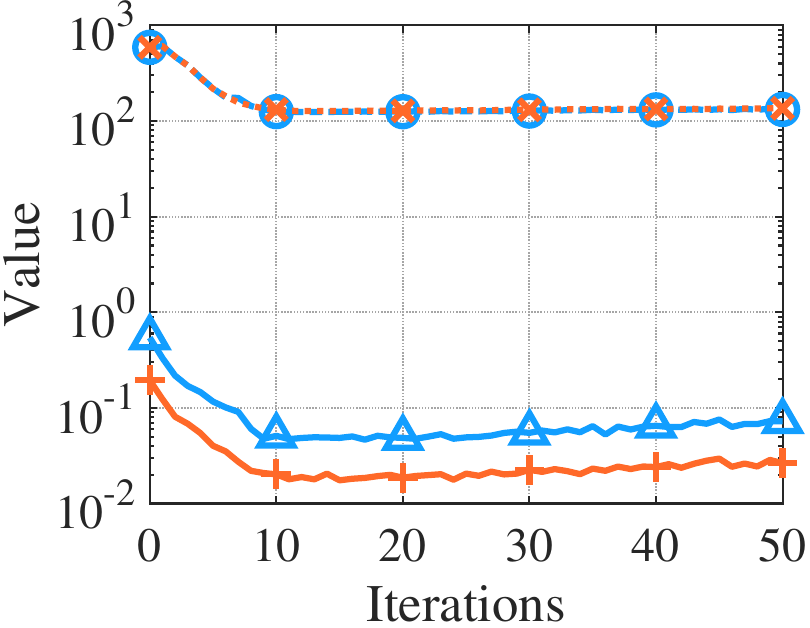}}
\caption{Comparison of the values of $G^2$ and ${\sum}_{c=1}^C {\sum}_{i = 1}^N p_i \alpha_{i,c} V_{i,c}^2$.}
\label{fig:value Vic G MNIST CIFAR-10}
\end{minipage}
\end{figure}

\begin{figure}
\centering
\begin{minipage}[h]{1\linewidth}
\centering
\includegraphics[width= 2.5 in ]{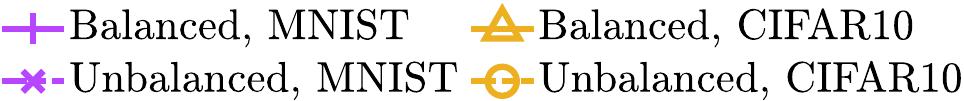}
\vspace{0.1cm}
\end{minipage}
\begin{minipage}[!t]{0.3\linewidth}
\centering
\includegraphics[width= 1.69 in ]{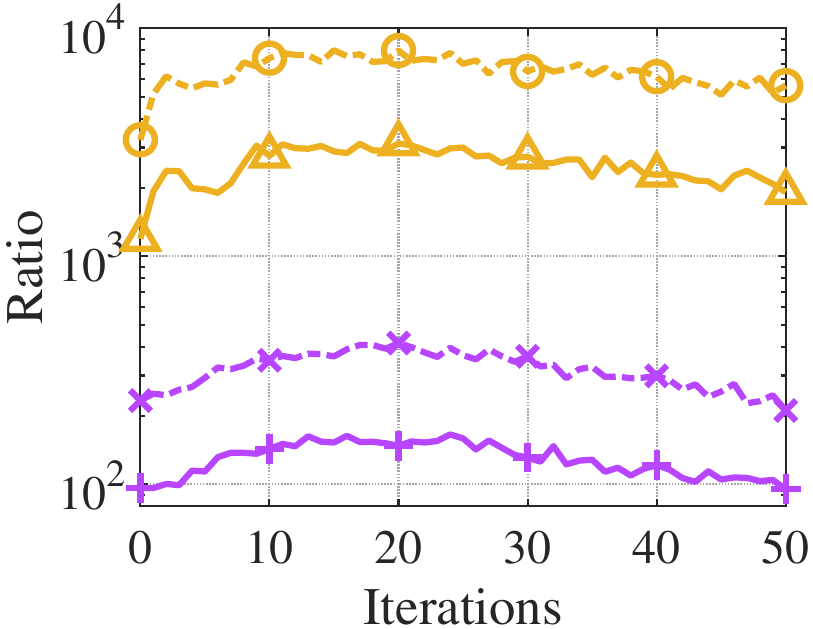}
\caption{Ratio of ${G^2}$ to $\qquad$ ${{\sum}_{c=1}^C {\sum}_{i = 1}^N p_i \alpha_{i,c} V_{i,c}^2}$.}
\label{fig:G Vic ratio MNIST CIFAR-10}
\end{minipage}
\hspace{-0.5cm}
\begin{minipage}[!t]{0.34\linewidth}
\centering
\includegraphics[width= 1.69 in ]{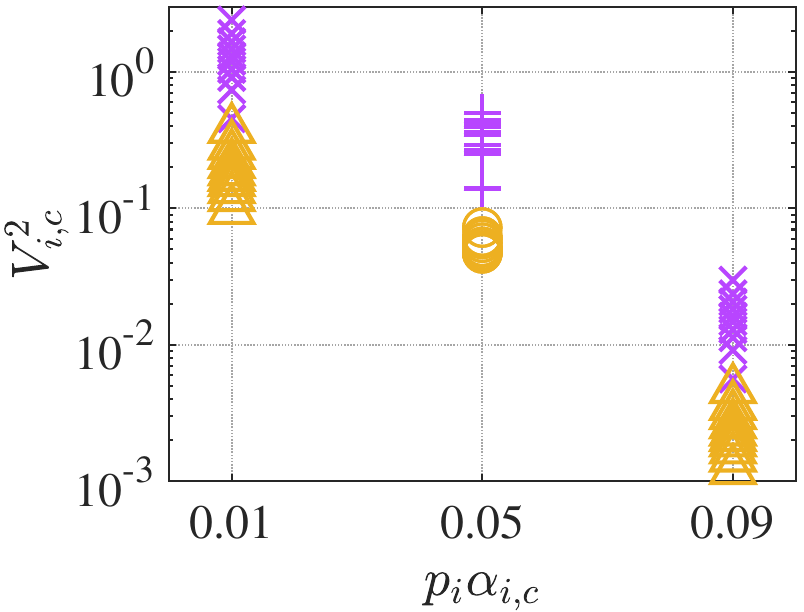}
\caption{Value of $V_{i,c}^2$ per client with respect to $p_i \alpha_{i,c}$ (markers consistent with Fig. \ref{fig:G Vic ratio MNIST CIFAR-10}).}
\label{fig:Vic value versus pi alpha}
\end{minipage}
\end{figure}

Fig. \ref{fig:value Vic G MNIST CIFAR-10}(a) and Fig. \ref{fig:value Vic G MNIST CIFAR-10}(b) depict the trends of ${\sum}_{c=1}^C {\sum}_{i = 1}^N p_i \alpha_{i,c} V_{i,c}^2$ and $G^2$ over the initial 50 iterations for the balanced and unbalanced MNIST and CIFAR-10 datasets, respectively.
The corresponding ratio of $G^2$ to ${\sum}_{c=1}^C {\sum}_{i = 1}^N p_i \alpha_{i,c} V_{i,c}^2$ is shown in Fig. \ref{fig:G Vic ratio MNIST CIFAR-10}.

\begin{observation}\label{observation:G larger than Vic}
As shown in Fig. \ref{fig:value Vic G MNIST CIFAR-10} and Fig. \ref{fig:G Vic ratio MNIST CIFAR-10}, the squared norm of model gradient, $G^2$, is orders of magnitude greater (hundreds to thousands of times larger) than the average gradient deviation within a class, ${\sum}_{c=1}^C {\sum}_{i = 1}^N p_i \alpha_{i,c} V_{i,c}^2$.
This indicates that in non-i.i.d scenario with label distribution skew, the label-related heterogeneity component in (\ref{eq:convergence bound of FL failure}b), $\chi^2_{\bm{\bar \alpha} \| \bm{\alpha}_g} G^2$, is the dominant factor in biasing FL convergence.
\end{observation}

To further analyze the cause why ${\sum}_{c=1}^C {\sum}_{i = 1}^N p_i \alpha_{i,c} V_{i,c}^2$ is small, we plot in Fig. \ref{fig:Vic value versus pi alpha} the median of each client's $V_{i,c}^2$ taken over all iterations alongside with its corresponding $p_i \alpha_{i,c}$.
As can be observed from Fig. \ref{fig:Vic value versus pi alpha}, in balanced scenario, $V_{i,c}^2$, is almost same for all clients at the value of low level.
In unbalanced scenario, although $V_{i,c}^2$ is higher for the clients with insufficient training samples (i.e., with smaller $p_i \alpha_{i,c}$), the multiplication $p_i \alpha_{i,c} V_{i,c}^2$ still has minor contribution to the summation ${\sum}_{c=1}^C {\sum}_{i = 1}^N p_i \alpha_{i,c} V_{i,c}^2$.

According to Theorem \ref{theorem:convergence FL failure}, when the divergence between the effective and actual label distributions $\chi^2_{\bm{\bar \alpha} \| \bm{\alpha}_g} = 0$, the label-related (second) component in term (\ref{eq:convergence bound of FL failure}b) vanishes, indicating that the influence of label-related heterogeneity on biasing FL convergence is eliminated.
Consequently, based on Observation \ref{observation:G larger than Vic}, term (\ref{eq:convergence bound of FL failure}b) becomes negligible, enabling FL to converge to an appropriate stationary solution.
From this, we derive the following result:

\begin{corollary}\label{corollary:convergence FL failure}
Under the same conditions as Theorem \ref{theorem:convergence FL failure} and incorporating Observation \ref{observation:G larger than Vic}, if $\chi^2_{\bm{\bar \alpha} \| \bm{\alpha}_g}=0$, we approximately have
\begin{align}\label{eq:convergence bound of FL failure corollary}
&
\frac{1}{R} \sum_{r=1}^R
\mathbb{E}[ \left\| \nabla  F({\bar {\mathbf{w}}}_{r-1}) \right\|^2 ]
\notag \\
\leq &
\frac{2484L}{55(TK)^{\frac{1}{2}}}
\left( \mathbb{E}[F({\bar {\mathbf{w}}}_{0})]
- \underline{F} \right)
+ \bigg( \frac{276}{55(TK)^{\frac{1}{4}}} + \frac{24}{55(TK)^{\frac{1}{2}}} + \frac{4}{55(TK)^{\frac{3}{4}}} \bigg)
\cdot
\sum_{i = 1}^{N} {\bar \beta}_i \sum_{c=1}^C
\left( \alpha_{i,c} V_{i,c}^2
+ \chi^2_{\bm{\alpha}_i \| \bm{\alpha}_g} G^2 \right)
.
\end{align}
\end{corollary}

From the RHS of \eqref{eq:convergence bound of FL failure corollary}, we observe that with $\chi^2_{\bm{\bar \alpha} \| \bm{\alpha}_g}=0$, the FL algorithm can achieve a \emph{linear speed-up} with respect to the number of selected clients $K$, even in the presence of transmission failures.
This highlights the importance of balancing the effective appearance probabilities of each class's local samples.
Aligning these probabilities with the label distribution of global dataset enhances the robustness and convergence speed of FL in unreliable wireless networks.
\begin{remark}
To the best of our knowledge, the influence of label-related heterogeneity on FL convergence under transmission failures, as presented in Theorem \ref{theorem:convergence FL failure} and Corollary \ref{corollary:convergence FL failure}, has not been discovered in previous literature.
Furthermore, these results are not limited to any specific type of network and can be applied to FL deployments in various unreliable networks experiencing transmission failures.
\end{remark}

\begin{remark}\label{remark:online}
Theorem \ref{theorem:convergence FL failure} and Corollary \ref{corollary:convergence FL failure} readily extend to the general case with dynamically changed transmission failure probabilities.
For example, the upper bound for Corollary \ref{corollary:convergence FL failure} can be adapted by replacing $\chi^2_{\bm{\bar \alpha} \| \bm{\alpha}_g}=0$ with $\chi^2_{\bm{\bar \alpha}_r \| \bm{\alpha}_g}=0$, and ${\sum}_{i = 1}^{N} {\bar \beta}_i$ with $\frac{1}{R}{\sum}_{i = 1}^{N} {\bar \beta}^{r}_{i}$,
where ${\bar \beta}^{r}_{i}$ denotes the effective appearance probability of client $i$ in the global aggregation at iteration $r$, and $\chi^2_{\bm{\bar \alpha}_r \| \bm{\alpha}_g}= \sum_{c=1}^C \frac{( \alpha_{g,c} - \sum_{i = 1}^N {\bar \beta}^{r}_{i} \alpha_{i,c})^2}{\alpha_{g,c}}$ quantifies the divergence between the actual global label distribution and the effective one at iteration $r$.
Further details are given in Appendix \ref{proof: theorem unfixed failure prob}.
\end{remark}

\section{Client Selection Optimization}\label{Sec:Client Selection}

Since transmission failures are inevitably occur at the wireless edge, we formulate an optimization problem in this section to minimize their impact through client selection, thereby accelerating FL convergence.

\subsection{Different Optimization Strategies}\label{Sec:Different Optimization Strategies}

As discussed in previous Observation \ref{observation:G larger than Vic} and Corollary \ref{corollary:convergence FL failure}, minimizing the divergence $\chi^2_{\bm{\bar \alpha} \| \bm{\alpha}_g}$ can significantly enhance FL convergence under transmission failure.
According to Theorem \ref{theorem:convergence FL failure}, the divergence between the effective and actual global label distributions, $\chi^2_{\bm{\bar \alpha} \| \bm{\alpha}_g}$, is expressed as
\begin{subequations}\label{eq:definition chi square alpha}
\begin{align}
\chi^2_{\bm{\bar \alpha} \| \bm{\alpha}_g}
= &
\sum_{c=1}^C \frac{\Big( \sum_{i = 1}^N (p_i - {\bar \beta}_i) \alpha_{i,c} \Big)^2}{\alpha_{g,c}}
\label{eq:definition chi square alpha a}
\\
= &
\sum_{c=1}^C \frac{\Big( \alpha_{g,c} - \sum_{i = 1}^N {\bar \beta}_i \alpha_{i,c} \Big)^2}{\alpha_{g,c}}
\label{eq:definition chi square alpha b}
,
\end{align}
\end{subequations}
where \eqref{eq:definition chi square alpha a} and \eqref{eq:definition chi square alpha b} provide two equivalent representations of $\chi^2_{\bm{\bar \alpha} \| \bm{\alpha}_g}$, with the global label distribution $\alpha_{g,c} = \sum_{i = 1}^N p_i \alpha_{i,c}$ as specified in \eqref{eq:nabla Fi F alpha c}.

From \eqref{eq:definition chi square alpha}, the minimum value of $\chi^2_{\bm{\bar \alpha} \| \bm{\alpha}_g}$ is zero, and the effective appearance probability of each client, ${\bar \beta}_i$, plays a crucial role in achieving this minimum.
Accordingly, the detailed formulation of ${\bar \beta}_i$ is derived in Proposition \ref{proposition:bar beta i formulation}.

\begin{proposition}\label{proposition:bar beta i formulation}
Based on the selection scheme described in Section \ref{sec:FL procedure}, the selection set $\mathcal{K}_r$ has $C^{N+K-1}_K$ ($=\frac{(N+K-1)!}{(N-1)!K!}$) possible combinations, denoted by $\mathcal{K}_r^z,  z = 1,\ldots,C^{N+K-1}_K$ \cite{baker2013discrete}.
Thus, from \eqref{average_beta}, the effective appearance probability ${\bar \beta}_i$ for each client $i \in [N]$ is given by
\begin{align}\label{eq:bar beta i formulation}
{\bar \beta}_i
=
&
\sum_{z=1}^{C^{N+K-1}_K}
\underbrace{
\frac{K!}{\prod_{i \in [N]} n_{z,i}!} \prod_{i \in \mathcal{K}_r^z} s_i
}_{\triangleq \Pr(\mathcal{K}_r^z)}
\cdot
\underbrace{
\frac{n_{z,i}(1 - \epsilon_i)}{1 - \prod_{j \in \mathcal{K}_r^z} \epsilon_j}
\sum_{k = 0}^{K-1}
\bigg(
\frac{1}{K C^{K-1}_{k}}
\sum_{\mathcal{K}' \subseteq \mathcal{K}_r^z \backslash i, \atop |\mathcal{K}'| = k}
\prod_{i' \in \mathcal{K}'} \epsilon_{i'}
\bigg)
}_{\triangleq \beta_{z,i}}
,
\end{align}
where
$\Pr(\mathcal{K}_r^z)$ represents the selection probability of each set $\mathcal{K}_r^z$,
and $\beta_{z,i}$ is the effective appearance probability of client $i$'s local model when $\mathcal{K}_r^z$ is selected.
Here, $\beta_{z,i} = 0$ if $i \notin \mathcal{K}_r^z$ and $\sum_{i=1}^N \beta_{z,i}=1$.
The variable $n_{z,i}$ is the number of times client $i$ appears in $\mathcal{K}_r^z$, where $n_{z,i} = 0$ if $i \notin \mathcal{K}_r^z$, and $\frac{K!}{\prod_{i \in [N]} n_{z,i}!}$ gives the number of distinct permutations of selected clients in $\mathcal{K}_r^z$.
The notation $\mathcal{K}_r^z \backslash i$ refers to the set obtained by removing one occurrence of $i$ from $\mathcal{K}_r^z$.
\end{proposition}

\emph{Proof:}
See Appendix \ref{sec:Appearance Probability}.
\hfill $\blacksquare$

According to \eqref{eq:bar beta i formulation}, ${\bar \beta}_i$ depends on both the selection probabilities $\{ s_i \}$ and the transmission failure probabilities $\{ \epsilon_i \}$.
Moreover, ${\bar \beta}_i$ is influenced by the differences in clients' transmission failure probabilities, as described in Proposition \ref{proposition:beta_zi_failure_diff}.

\begin{proposition}\label{proposition:beta_zi_failure_diff}
The effective appearance probability ${\bar \beta}_i$ for each client $i \in [N]$ can also be expressed as
\begin{align}\label{eq:beta_zi_failure_prob_diff}
{\bar \beta}_i
= &
s_i
\bigg(
1 +
\sum_{i'=1, i' \neq i}^N s_{i'} (\epsilon_{i'} - \epsilon_i)
\cdot 
\sum_{j_1, \, j_2, \, \cdots, \, j_{K-2} = 1}^{N}
\prod_{k=1}^{K-2} s_{j_k}
\frac{1 + \sum_{k=1}^{K-2} \prod_{k'=1}^{k} \epsilon_{j_{k'}}}{1 - \epsilon_{i}\epsilon_{i'} \prod_{k=1}^{K-2} \epsilon_{j_k}}
\bigg)
,
\end{align}
where $(\epsilon_{i'} - \epsilon_i)$ captures the difference in transmission failure probabilities between two distinct clients $i$ and $i' \in [N]$.
\end{proposition}

\emph{Proof:}
See Appendix \ref{sec:proposition beta_zi_failure_diff}.
\hfill $\blacksquare$

Based on Propositions \ref{proposition:bar beta i formulation} and \ref{proposition:beta_zi_failure_diff}, two strategies can be employed to minimize $\chi^2_{\bm{\bar \alpha} \| \bm{\alpha}_g}$:
\begin{enumerate}[(i)]
\item
\textbf{Optimizing wireless resource allocation:}
This strategy aims to equalize transmission failure probabilities across all clients, such that $\epsilon_i = \epsilon_{i'}$, $\forall i,i' \in [N]$ \cite{wang2022quantized}.
As a result, the effective appearance probability ${\bar \beta}_i = s_i$ according to Proposition \ref{proposition:beta_zi_failure_diff}.
When the classical \texttt{FedAvg} algorithm is applied, using a simple random selection strategy with $s_i = p_i$, we have ${\bar \beta}_i = p_i$, which leads to $\chi^2_{\bm{\bar \alpha} \| \bm{\alpha}_g} = 0$ based on \eqref{eq:definition chi square alpha a}.
However, this strategy requires network reconfiguration and may be infeasible if some clients have $\epsilon_i = 1$, limiting its practicality.

\item
\textbf{Optimizing client selection:}
In this strategy, the selection probabilities $\{s_i\}$ in \eqref{eq:bar beta i formulation} are optimized to adjust each client's effective appearance probability ${\bar \beta}_i$, ensuring that the effective appearance probability of each class's samples, $\sum_{i = 1}^N {\bar \beta}_i \alpha_{i,c}$, matches its proportion in the global dataset, $\alpha_{g,c}$.
This leads to $\chi^2_{\bm{\bar \alpha} \| \bm{\alpha}_g} = 0$ based on \eqref{eq:definition chi square alpha b}.
This strategy mitigates the impact of transmission failures without altering the existing network configuration.
Predefined wireless resource allocation schemes, such as resource blocks assigned to each selected client in FDMA systems or time slots allocated in time division multiple access (TDMA) systems, can be directly applied, facilitating straightforward implementation.
Thus, this paper focuses on optimizing client selection.
\end{enumerate}

\subsection{Proposed FedCote}

We first consider a static scenario, where client locations remains fixed and each client¡¯s transmission failure probability is constant throughout the training process.
The dynamic case, with time-varying transmission failure probabilities, will be addressed in Section \ref{section:Online Scheduling}.

\subsubsection{Optimization Problem Formulation}

As discussed in previous Section \ref{Sec:Different Optimization Strategies}, optimizing client selection probabilities to minimize $\chi^2_{\bm{\bar \alpha} \| \bm{\alpha}_g} = 0$ is a straightforward approach to mitigate the impact of transmission failure on FL convergence.
Accordingly, based on \eqref{eq:definition chi square alpha b}, we formulate the following optimization problem for determining the client selection probabilities:
\begin{subequations}\label{eq:optimization problem}
\begin{align}
\min_{s_i, i \in [N]} \quad
&
\sum_{c=1}^C \frac{( \alpha_{g,c} - \sum_{i = 1}^N {\bar \beta}_i \alpha_{i,c})^2}{\alpha_{g,c}},
\label{eq:objective function}
\\
\text{s.t.} \quad
&
\sum_{i=1}^N s_i = 1,
\label{eq:sum constraint}
\\
&
s_i \geq 0, \;\; i \in [N].
\label{eq:si0 ei0}
\end{align}
\end{subequations}

\begin{remark}
In the optimization problem (\ref{eq:optimization problem}), clients are required to share only their local label distributions $\{ \alpha_{i,c} \}$, which are minimally privacy-sensitive.
If the label distributions are still considered privacy-sensitive, protection mechanisms like secure multiparty computation can be employed \cite{ma2021client}.
However, since this paper primarily investigates the effects of transmission failure and label distribution skew on FL convergence, privacy protection will not be discussed extensively.
\end{remark}

\subsubsection{Optimal Solution}

\emph{1) The i.i.d. data case:}
In this scenario, the global and local datasets share identical label distributions (i.e., $\alpha_{i,c} = \alpha_{g,c}$), resulting in a zero numerator in \eqref{eq:objective function}, and the objective function \eqref{eq:objective function} directly attains its minimum value of zero.
Moreover, if a client's transmission failure probability $\epsilon_i$ is close to one, it is highly unlikely to successfully upload its local model, making its selection ineffective and reducing selection efficiency.
Based on these observations, we establish Proposition \ref{proposition:iid solution}.

\begin{proposition}\label{proposition:iid solution}
In the i.i.d. data case, any set of selection probabilities satisfying (\ref{eq:sum constraint}) and (\ref{eq:si0 ei0}) constitutes an optimal solution to \eqref{eq:optimization problem}.
Consequently, a random selection strategy is optimal, with selection probabilities given by
\begin{align}\label{eq:selection probabilities}
s_i =
\begin{cases}
\frac{p_i}{\sum_{j \in [N], \epsilon_{j} < 1} p_{j}}, & \text{if } \, \epsilon_i \leq \epsilon_{th}, \\
0, & \text{if } \, \epsilon_i > \epsilon_{th},
\end{cases}
\end{align}
where $\epsilon_{th}$ denotes the transmission failure probability threshold for excluding clients with excessively high failure probabilities.
The solution in (\ref{eq:selection probabilities}) can also serve as an initialization for optimizing $\{ s_i \}$ under non-i.i.d. data.
\end{proposition}

\noindent
\emph{2) The non-i.i.d. data case:}
We first initialize the selection probabilities $\{ s_i \}$ using \eqref{eq:selection probabilities}.
Given the smooth and continuous gradient of \eqref{eq:objective function} due to the squared terms, we employ gradient descent as in \eqref{eq:gradient descent} to update the selection probabilities for clients with $\epsilon_i \leq \epsilon_{th}$.
This is followed by gradient projection \cite{boyd2004convex} as in \eqref{eq:gradient projection} to enforce the equality constraint in \eqref{eq:sum constraint}:
\begin{subequations}
\begin{align}
s_i
= &
s_i - \eta \frac{\partial {\rm(\ref{eq:objective function})}}{\partial s_i}
,
\label{eq:gradient descent}
\\
s_i
= &
s_i + \frac{1- \sum_{j \in [N], \epsilon_{j} < \epsilon_{th}} s_{j}}{N - \sum_{j \in [N], \epsilon_{j} < \epsilon_{th}} 1}
,
\label{eq:gradient projection}
\end{align}
\end{subequations}
where $\eta>0$ denotes the step size, and the detailed expression for ${\partial {\rm(\ref{eq:objective function})}}/{\partial s_i}$ is provided in Appendix \ref{sec:Objective Function Gradient}.
By iterating between \eqref{eq:gradient descent} and \eqref{eq:gradient projection}, we obtain the optimal selection probabilities.

The proposed client selection optimization approach for FL under transmission failures, named as \texttt{FedCote}, is summarized in Algorithm \ref{algorithm:client selection optimization}.
It can be seamlessly integrated into the overall FL procedure in Algorithm 1,
by directly applying the optimized selection probabilities $\{ s_i \}$ for client selection in Step 1.
This integration provides a plug-and-play mechanism to enhance FL robustness without modifying the underlying network configuration.

\begin{algorithm}[!t]
\caption{\texttt{FedCote}: Client selection optimization}
\begin{algorithmic}[1]
\Statex \textbf{\texttt{// Client-side:}}
\For {each client ${i} \in [N]$}
\State Upload local label distributions $\{ \alpha_{i,c} \}$ to the server;
\EndFor
\Statex \textbf{\texttt{// Server-side:}}
\State Initialize client selection probabilities $\{ s_i \}$ using \eqref{eq:selection probabilities};
\If {$\alpha_{i,c} = \alpha_{g,c}$, $\forall i \in [N]$ \textbf{(i.i.d. data case)}}
\State {\textbf{break}};
\Else { \textbf{(non-i.i.d. data case)}}
\State {Estimate transmission failure probabilities $\{ \epsilon_i \}$;}
\State $j \gets 0$;
\While {$j <$ maximum number of update steps}
\State {Compute effective appearance probabilities $\{ {\bar \beta}_i \}$ and $\{ \beta_{z,i} \}$ using \eqref{eq:bar beta i formulation};}
\State Update $\{s_i\}$ using gradient descent in \eqref{eq:gradient descent};
\State Project $\{s_i\}$ using \eqref{eq:gradient projection};
\State $j \gets j+1$;
\EndWhile
\EndIf
\renewcommand{\algorithmicensure}{\textbf{Output:}}
\Ensure Optimized client selection probabilities $\{s_i\}$.
\end{algorithmic}
\label{algorithm:client selection optimization}
\end{algorithm}

\subsection{Online Scheduling for Dynamic Scenario}\label{section:Online Scheduling}

In this subsection, we investigate the dynamic scenario where clients' positions vary over time and the transmission failure probabilities $\epsilon^{r}_{i}$ change across iterations during training.
To adapt to this setting, client selection probabilities $\{ s_i \}$ are optimized at each iteration $r$.
This online scheduling framework provides better adaptability to general dynamic environments.
Following Remark \ref{remark:online}, the client selection probabilities are obtained by solving the following optimization problem at each iteration $r$:
\begin{align}
\min_{s_i, i \in [N]} \quad
&
\sum_{c=1}^C \frac{( \alpha_{g,c} - \sum_{i = 1}^N {\bar \beta}^{r}_{i} \alpha_{i,c})^2}{\alpha_{g,c}},
\label{eq:optimization problem online}
\\
\text{s.t.} \quad
&
\eqref{eq:sum constraint}, \;
\eqref{eq:si0 ei0},
\notag
\end{align}
where ${\bar \beta}^{r}_{i}$ denotes the effective appearance probability of client $i$ in the global aggregation at iteration $r$, obtained from \eqref{eq:bar beta i formulation} by substituting $\epsilon_{i}$ with $\epsilon^{r}_{i}$.
The optimization problem in \eqref{eq:optimization problem online} can be solved analogously to Algorithm \ref{algorithm:client selection optimization}, with the optimized client selection probabilities determining the selection set $\mathcal{K}_r$ at each iteration.

\subsection{Enhanced FedCote for Reducing Computation Complexity}\label{section:Enhanced FedCote}

According to Proposition \ref{proposition:bar beta i formulation}, for a total of $N$ clients, the number of distinct combinations of $\mathcal{K}_r^z$ (i.e., $C^{N+K-1}_K$) grows rapidly as the number of selected clients, $K$, increases.
This leads to higher computational complexity for calculating the effective appearance probability ${\bar \beta}_i$ in \eqref{eq:bar beta i formulation}, thereby increasing the computational cost of client selection optimization in \eqref{eq:optimization problem}.
Fortunately, the optimization problem in \eqref{eq:optimization problem} possesses an advantageous property, as stated in Proposition \ref{proposition:difference increasing K}.

\begin{proposition}\label{proposition:difference increasing K}
Let $[{\bar \beta}_i]_{K}$ denote the effective appearance probability ${\bar \beta}_i$, as defined in (\ref{eq:bar beta i formulation}), when selecting $K$ clients from the total $N$ clients per iteration.
As $K$ increases, the difference between $[{\bar \beta}_i]_{K}$ and $[{\bar \beta}_i]_{K+1}$ gradually decreases.
Consequently, the gap between the optimal solutions of (\ref{eq:optimization problem}) for selecting $K$ and $(K+1)$ clients per iteration also diminishes as $K$ grows.
\end{proposition}

\emph{Proof:}
See Appendix \ref{sec:proposition difference increasing K}.
\hfill $\blacksquare$

Building on Proposition \ref{proposition:difference increasing K}, the computational complexity of solving the client selection optimization problem in \eqref{eq:optimization problem} for larger values of $K$ can be significantly reduced by approximating its optimal solution using a smaller value, $K_{\rm apx}$ ($K_{\rm apx} < K$).
This computation-efficient variant of \texttt{FedCote} is referred to as \texttt{FedCote-II}.

\section{Experimental Results}\label{sec:experimental results}

\subsection{Parameter Settings}\label{sec:Parameter Setting}

\subsubsection{FL Scenario}

In the simulations, a central server coordinates $N=20$ mobile clients to collaboratively train DNN models. The clients are connected through different network standards, including 4G, 5G, Wi-Fi (2.4\,GHz), and Wi-Fi (5\,GHz), as summarized in Table~\ref{table:network standard}.
The 4G and 5G base stations are deployed outdoors at the center of a 200\,m-radius cell, mounted at a height of 20\,m. The Wi-Fi access points (APs) are placed indoors at the center of a $20 \times 20$\,m area, with an installation height of 3\,m. The cellular base stations are positioned 10\,m from the indoor area, resulting in a 30\,m separation between the base stations and the Wi-Fi APs.

\begin{table}[!t]
\centering
\caption{Network standard assigned to each client.}\label{table:network standard}
\begin{tabular}{c|ccccc}
\hline
\multirow{2}{*}{\textbf{Network Standard}} & \multicolumn{2}{c}{\textbf{Wi-Fi}} & \multirow{2}{*}{\textbf{4G}} & \multirow{2}{*}{\textbf{5G}}  \\
& (2.4\,GHz) & (5\,GHz) & &  \\
\hline
\textbf{Client Index} & 1, 5, 9, 13, 17 & 2, 6, 10, 14, 18 & 3, 7, 11, 15, 19 & 4, 8, 12, 16, 20 \\
\hline
\end{tabular}
\end{table}

Two training scenarios are considered based on client mobility:
\begin{itemize}
\item
\textbf{Static Scenario}:
Clients with indices $1 - 8$ are uniformly distributed indoors, while the remaining 12 clients are uniformly distributed outdoors.

\item
\textbf{Dynamic Scenario}:
Clients start from the positions defined in the static scenario.
Half of them move randomly either from indoors to the outdoor cell edge or vice versa, with a walking speed of 1.5\,m/s, while the others remain stationary throughout training.
\end{itemize}

\subsubsection{Datasets and DNN Models}

We consider training DNN models using the following two widely adopted datasets.
\begin{itemize}
\item
\textbf{MNIST dataset} \cite{lecun1998gradient}:
This dataset comprises 60000 training samples and 10000 testing samples, categorized into ten classes.
A 3-layer fully connected network with dimensions $784 \times 30 \times 10$ is trained for digit classification, where the hidden layer contains 30 neurons and the total number of parameters is 23860.
Based on empirical values, training is conducted with a batch size of 128, a fixed learning rate of $\gamma = 0.05$, and $R=500$ iterations.

\item
\textbf{CIFAR-10 dataset} \cite{krizhevsky2009learning}:
This dataset includes 50000 training samples and 10000 testing samples, also categorized into ten classes.
For this dataset, the ResNet-20 model with Group Normalization (GN) \cite{hsieh2020non} is employed for image classification, comprising 269722 parameters.
Training is performed with a batch size of 128, an initial learning rate of $\gamma=0.1$ that decays to 0.01 after 6000 iterations, and a total of $R=10000$ iterations.
\end{itemize}

During local model updates, the number of local updating steps per iteration is set to $E=5$.

\subsubsection{Data Distributions}\label{section:Data distributions}

The simulations consider two types of dataset distributions: i.i.d. and non-i.i.d.
\begin{itemize}
\item
\textbf{i.i.d.}:
Training samples are shuffled and randomly distributed among the clients.

\item
\textbf{Non-i.i.d.}:
Each client is assigned data samples from only two specific classes.
Higher-indexed clients assigned samples from classes with higher labels.
For instance, clients 1$\sim$4 hold samples from classes 1 and 2, clients 5$\sim$8 from classes 3 and 4, and so forth.
\end{itemize}

Both i.i.d. and non-i.i.d. types are evaluated two settings: balanced and unbalanced.
\begin{itemize}
\item
\textbf{Balanced}:
Each client is allocated an equal number of samples, i.e., $p_i=\frac{1}{N}$, $\forall i \in [N]$.

\item
\textbf{Unbalanced}:
The number of samples allocated to clients varies, leading to differing values of $p_i$ among clients.
Specifically, the unbalanced ratio $u$ represents the proportion of training samples assigned to clients with even indices, while the remaining $(1-u)$ proportion is allocated to clients with odd indices.
A larger $u$ indicates a higher degree of imbalance, with $u=0.5$ corresponding to the balanced case.

\end{itemize}

Despite the varying data distribution settings in the FL procedure, the performance of the aggregated global model is assessed by evaluating the training loss and testing accuracy using the entire training and testing datasets, respectively, to evaluate its convergence capability and generalizability.

\subsubsection{Transmission Failure Simulation}

In the simulation, frequency division multiple access (FDMA) is assumed for uplink transmission.
The channel capacity of client $i \in {\mathcal{K}}_{r}$ is given by
\begin{align}\label{channel_capacity}
{C}^{r}_{i} = W_{i} \log_2 \left( 1 + \frac{P_{i} | {h}^{r}_{i} |^2 }{{W_{i} N_0}} \right) \; \text{bps,}
\end{align}
where $W_{i}$ and $P_{i}$ denote the allocated bandwidth and transmit power of client $i$, respectively, $N_0$ is the noise power spectrum density, and ${h}^{r}_{i}$ is the uplink channel coefficient between the server and client $i$.
The latter is modeled using the classical path-loss model with shadowing\cite{goldsmith2005wireless}:
\begin{align}
[ | {h}^{r}_{i} |^2 ]_{\rm dB}
= - [PL_0(d_0)]_{\rm dB} - \lambda [d^{r}_{i}]_{\rm dB} - \psi_{\rm wall} + \psi_{\rm shadow} ,
\end{align}
where $[x]_{\rm dB}$ denotes $x$ in dB, $\lambda$ is the path loss exponent, $d^{r}_{i}$ (m) is the distance between client $i$ and the server, $\psi_{\rm shadow} \sim \mathcal{N}(0, \sigma^2_{\rm shadow})$ represents log-normal shadowing, and $\psi_{\rm wall}$ denotes wall penetration loss.
The free-space path loss is defined as
$[PL_0(d_0)]_{\rm dB} = 20 \log_{10}(d_0) + 20 \log_{10}(f) + 32.44$,
where $d_0$ (km) is the reference distance and $f$ (MHz) is the carrier frequency \cite{balanis2016antenna}.

Assuming an uplink delay constraint $\tau_i$, the transmission rate $R_{i}$ is defined as the ratio between the local model size and $\tau_i$.
Each model parameter is represented using 32-bit floating-point precision, and the model size is obtained by multiplying the total number of parameters by 32.
Then, according to the channel coding theorem \cite{goldsmith2005wireless}, if $C^{r}_{i} \leq R_{i}$, the server cannot successfully decode $\mathbf{w}_i^{r,E}$, resulting in transmission failure.
Thus, the transmission failure probability is
\begin{align}\label{outage_probability}
\epsilon^{r}_{i} \triangleq \Pr \left( C^{r}_{i} \leq R_{i} \right).
\end{align}

The simulation parameter value for different network standards are listed in Table \ref{Parameter_value} \cite{3gpp_tr38901,etsi_en301893,ieee80211_channelmodels}.

\begin{table}[!t]
\centering
\caption{Simulation parameters for different network standards}\label{Parameter_value}
\begin{tabular}{c|ccccc}
\hline
\multirow{2}{*}{\textbf{Parameter}} & \multicolumn{2}{c}{\textbf{Wi-Fi}} & \multirow{2}{*}{\textbf{4G}} & \multirow{2}{*}{\textbf{5G}} \\
& (2.4\,GHz) & (5\,GHz) & & \\
\hline
$W_{i}$ (MHz) & 10 & 10 & 1.8 & 2.88 \\
$P_{i}$ (dBm) & 23 & 23 & 20 & 23 \\
$f$ (GHz) & 2.4 & 5 & 2.6 & 3.5 \\
$N_0$ (dBm/Hz) & -174 & -174 & -174 & -174 \\
$d_0$ (m) & 1 & 1 & 1 & 1 \\
$\lambda$ & 3 & 3 & 3 & 3  \\
$\psi_{\rm wall}$ (dB) & 12 & 18 & 10 & 15 \\
$\sigma_{\rm shadow}$ & \multicolumn{4}{c}{4 for $d_{i} \leq 100$\,m; 8 for $d_{i} > 100$\,m} \\
$\tau_i$ & \multicolumn{4}{c}{0.1\,s (MNIST); 1\,s (CIFAR-10)} \\
$\epsilon_{th}$ in (\ref{eq:selection probabilities}) & 0.85 & 0.85 & 0.85 & 0.85 \\
\hline
\end{tabular}
\end{table}

\subsubsection{Baselines}\label{section:Baselines}

To comprehensively evaluate the effectiveness of the proposed method, we select baselines from multiple perspectives, including client selection, global aggregation, and local training, along with an ideal scheme, for comparison with the proposed \texttt{FedCote}.
The details of each baseline are presented below.
\begin{itemize}
\item
\textbf{\texttt{FedAvg}} \cite{mcmahan2017communication}:
This scheme employs Algorithm \ref{algorithm:FL under transmission failure} with a random selection strategy, where $s_i = p_i$.

\item
\textbf{\texttt{Power-of-Choice}} \cite{cho2022towards}:
This scheme modifies the client selection in Step 3 of Algorithm \ref{algorithm:FL under transmission failure} as follows:
1) It selects $\tilde{K}$ clients ($\tilde{K} \geq K$) without replacement, based on the selection probability $s_i = p_i$, to form a candidate set $\tilde{K}_r$;
2) From this candidate set, it selects the $K$ clients with the largest local training loss $F_i(\mathbf{w}^{r-1,E}_{i})$.
In the experiments, when $K=10$, we set $\tilde{K}=15$; when $K=N=20$, we use $\tilde{K}=K=20$.

\item
\textbf{\texttt{Newt}} \cite{zhao2022participant}:
At the $r$-th iteration, this scheme measures the difference between the local model of client $i$ and the global model as
$e^{-p_i} \| \mathbf{w}^{r-1,E}_{i} - {\bar {\mathbf{w}}}_{r-1} \|$, and selects $K$ clients with the largest differences for local training.
The global model is then aggregated using a weighted average as
$
{\bar {\mathbf{w}}}_{r}
=
\frac{\sum_{{i} \in {\mathcal{K}}_{r}} p_i \mathbf{w}^{r,E}_{i} }{\sum_{{i} \in {\mathcal{K}}_{r}} p_i}
$.
For fair comparison, we modify the aggregation scheme to account for transmission failures as
\begin{equation}\label{eq:weighted aggregation failure}
{\bar {\mathbf{w}}}_{r}
=
\frac{\sum_{{i} \in {\mathcal{K}}_{r}} \mathds{1}^{r}_{i} p_i \mathbf{w}^{r,E}_{i} }{\sum_{{i} \in {\mathcal{K}}_{r}} \mathds{1}^{r}_{i} p_i}
.
\end{equation}

\item
\textbf{\texttt{GS}} \cite{ma2021client}:
This scheme partitions clients into groups, each containing at most $K$ clients.
The grouping is optimized to ensure that the aggregated label distribution of the clients within a group $\mathcal{G}$, $\{ \sum_{{i} \in \mathcal{G}} \alpha_{i,c} \}_{c=1}^C$, closely matches the global label distribution, $\{ \alpha_{g,c} \}_{c=1}^C$.
At each iteration, one group is selected for local training, and the global model is aggregated via a weighted average.
Similar to \texttt{Newt}, this scheme does not consider transmission failures.
Therefore, we adopt the modified aggregation scheme in \eqref{eq:weighted aggregation failure}.

\item
\textbf{\texttt{TF-Aggregation}} \cite{salehi2021federated}:
This baseline incorporates transmission failures by modifying the global aggregation (Step 12) in Algorithm \ref{algorithm:FL under transmission failure} as
\begin{equation}\label{eq:optimization problem TO}
{\bar {\mathbf{w}}}_{r} = \frac{1}{K} \sum_{{i} \in {\mathcal{K}}_{r}}
\mathds{1}^{r}_{i} \frac{p_i}{s_i \left(1- \epsilon^{r}_{i} \right)} \mathbf{w}^{r,E}_{i}
,
\end{equation}
where $\epsilon^{r}_{i}$ is the transmission failure probability of client $i$ at iteration $r$.
This design yields an unbiased estimate of the global model under full participation, i.e., $\mathbb{E}_{\mathcal{K}_{r}, \mathds{1}^{r}_{i}} [{\bar {\mathbf{w}}}_{r}] = \sum_{i = 1}^N p_i \mathbf{w}^{r,E}_{i}$ \cite[Lemma 2]{salehi2021federated}.
Unlike our formulation in \eqref{eq:optimization problem TO}, this baseline optimizes client selection via $\min_{s_i, i \in [N]} \sum_{i=1}^N \frac{p_i}{s_i (1- \epsilon^{r}_{i})}$, subject to \eqref{eq:sum constraint} and \eqref{eq:si0 ei0}.
Since the denominator grows unbounded as $\epsilon^{r}_{i} \to 1$, we introduce a threshold $\epsilon_{th}$, consistent with \eqref{eq:selection probabilities} and revise the optimization problem for fair comparison as follows:
\begin{align}
\min_{s_i, i \in [N]} \quad
&
\sum_{i=1, \epsilon^{r}_{i} \leq \epsilon_{th}}^N \frac{p_i}{s_i \left(1- \epsilon^{r}_{i} \right)},
\label{eq:optimization problem FedAvg TO}
\\
\text{s.t.} \quad
&
\eqref{eq:sum constraint}, \eqref{eq:si0 ei0}.
\notag
\end{align}

\item
\textbf{\texttt{FedNova}} \cite{wang2020tackling}:
This scheme modifies global aggregation by normalizing local gradients.
To ensure fairness under transmission failures, we adapt aggregation rule as
\begin{equation}\label{eq:FedNova aggregation}
{\bar {\mathbf{w}}}_{r}
=
{\bar {\mathbf{w}}}_{r-1}
- \frac{\eta_{\rm eff}}{\sum_{{i} \in {\mathcal{K}}_{r}} \mathds{1}^{r}_{i}}
\sum_{{i} \in {\mathcal{K}}_{r}} \mathds{1}^{r}_{i} \frac{{\bar {\mathbf{w}}}_{r-1} - \mathbf{w}^{r,E}_{i}}{E_i}
,
\end{equation}
where $\eta_{\rm eff} = \frac{\sum_{{i} \in {\mathcal{K}}_{r}} \mathds{1}^{r}_{i} E_i}{\sum_{{i} \in {\mathcal{K}}_{r}} \mathds{1}^{r}_{i}} $ denotes the effective local update steps, $E_i$ is the number of local updates performed by client $i$.

\begin{table*}[!t]
\centering
\caption{
Performance of various FL schemes on the balanced MNIST dataset (\emph{K} = 10).
}\label{table:FL performance MNIST}
\begin{tabular}{c|cc|cc}
\hline
\multirow{2}{*}{\textbf{FL scheme}} & \multicolumn{2}{c|}{\textbf{i.i.d.}} & \multicolumn{2}{c}{\textbf{Non-i.i.d.}} \\
& \textbf{Training loss} & \textbf{Testing accuracy (\%)} & \textbf{Training loss} & \textbf{Testing accuracy (\%)} \\
\hline
\texttt{FedAvg}
&  \textbf{0.1271 \scriptsize{$\pm$ 0.0050}}&  \textbf{95.86 \scriptsize{$\pm$ 0.1027}} &  0.5843 \scriptsize{$\pm$ 0.1100} &  80.89 \scriptsize{$\pm$ 3.6126} \\
\texttt{Power-of-Choice}
& 0.1396 \scriptsize{$\pm$ 0.0031} & 95.46 \scriptsize{$\pm$ 0.1038} & 0.3487 \scriptsize{$\pm$ 0.1086} & 88.89 \scriptsize{$\pm$ 3.9024} \\
\texttt{Newt}
& 0.1390 \scriptsize{$\pm$ 0.0032} & 95.51 \scriptsize{$\pm$ 0.2055} & 0.6156 \scriptsize{$\pm$ 0.4047} & 82.99 \scriptsize{$\pm$ 8.5893} \\
\texttt{GS}
& 0.1383 \scriptsize{$\pm$ 0.0041} & 95.54 \scriptsize{$\pm$ 0.1264} & 0.4558 \scriptsize{$\pm$ 0.0816} & 85.54 \scriptsize{$\pm$ 2.9953} \\
\texttt{TF-Aggregation}
& 3.6379$\times$10$^{\text{34}}$ \scriptsize{$\pm$ 2.0891$\times$10$^{\text{34}}$} &  92.35 \scriptsize{$\pm$ 0.4967} & 8.6953e$\times$10$^{\text{34}}$\scriptsize{$\pm$ 6.8145e$\times$10$^{\text{34}}$} &  77.52 \scriptsize{$\pm$ 13.3359} \\
\texttt{FedNova}
&  0.1271 \scriptsize{$\pm$ 0.0050} & 95.85 \scriptsize{$\pm$ 0.1064} &  0.5843 \scriptsize{$\pm$ 0.1101} &  80.89 \scriptsize{$\pm$ 3.6126} \\
\texttt{FedYOGI}
& 0.1619 \scriptsize{$\pm$ 0.0049} & 94.99 \scriptsize{$\pm$ 0.0773} & 0.4711 \scriptsize{$\pm$ 0.1381} & 85.06 \scriptsize{$\pm$ 4.7914} \\
\texttt{FedProx}
& 0.1279 \scriptsize{$\pm$ 0.0050} & 95.81 \scriptsize{$\pm$ 0.1055} & 0.5840 \scriptsize{$\pm$ 0.1101} & 80.87 \scriptsize{$\pm$ 3.6306} \\
\texttt{SCAFFOLD}
& 8.8665$\times$10$^{\text{34}}$ \scriptsize{$\pm$ 6.3840$\times$10$^{\text{34}}$} &  92.56 \scriptsize{$\pm$ 0.7903} & 4.4726$\times$10$^{\text{34}}$ \scriptsize{$\pm$ 7.3000$\times$10$^{\text{34}}$}& 54.46 \scriptsize{$\pm$ 19.7468} \\
\textbf{\texttt{FedCote} (Ours)}
& 0.1289 \scriptsize{$\pm$ 0.0035} & 95.72 \scriptsize{$\pm$ 0.0738} & \textbf{0.2174 \scriptsize{$\pm$ 0.0138}} & \textbf{93.17 \scriptsize{$\pm$ 0.4425}} \\
\hline
\texttt{Ideal}
& 0.1190 \scriptsize{$\pm$ 0.0028} & 96.03 \scriptsize{$\pm$ 0.1337} & 0.2794 \scriptsize{$\pm$ 0.0363} & 91.27 \scriptsize{$\pm$ 1.3622} \\
\hline
\end{tabular}
\end{table*}

\begin{table*}[!t]
\centering
\caption{
Performance of various FL schemes on the balanced CIFAR-10 dataset (\emph{K} = 10).
}\label{table:FL performance CIFAR10}
\begin{tabular}{c|cc|cc}
\hline
\multirow{2}{*}{\textbf{FL scheme}} & \multicolumn{2}{c|}{\textbf{i.i.d.}} & \multicolumn{2}{c}{\textbf{Non-i.i.d.}} \\
& \textbf{Training loss} & \textbf{Testing accuracy (\%)} & \textbf{Training loss} & \textbf{Testing accuracy (\%)} \\
\hline
\texttt{FedAvg}
& \textbf{0.1518 \scriptsize{$\pm$ 0.0061}} & \textbf{84.90 \scriptsize{$\pm$ 0.4441}} & 0.6554 \scriptsize{$\pm$ 0.0923} & 71.27 \scriptsize{$\pm$ 3.0723} \\
\texttt{Power-of-Choice}
& 0.1769 \scriptsize{$\pm$ 0.0711} & 83.69 \scriptsize{$\pm$ 2.4187} & 0.5952 \scriptsize{$\pm$ 0.0827} & 72.47 \scriptsize{$\pm$ 2.5712} \\
\texttt{Newt}
& 0.3923 \scriptsize{$\pm$ 0.0248}          & 80.16 \scriptsize{$\pm$ 0.7204} & 0.9577 \scriptsize{$\pm$ 0.4828} & 67.81 \scriptsize{$\pm$ 5.8911} \\
\texttt{GS}
& 0.1657 \scriptsize{$\pm$ 0.0060}          & 84.51 \scriptsize{$\pm$ 0.4738} & 0.7325 \scriptsize{$\pm$ 0.1396} & 69.16 \scriptsize{$\pm$ 3.1439} \\
\texttt{TF-Aggregation}
& NaN                          & 10.00 \scriptsize{$\pm$ 0.00}   & NaN                 & 10.00 \scriptsize{$\pm$ 0.00}      \\
\texttt{FedNova}
& 0.1535 \scriptsize{$\pm$ 0.0045}          & 84.88 \scriptsize{$\pm$ 0.3470} & 0.6663 \scriptsize{$\pm$ 0.0996} & 70.87 \scriptsize{$\pm$ 3.2911} \\
\texttt{FedYOGI}
& 0.1888 \scriptsize{$\pm$ 0.0092}          & 84.00 \scriptsize{$\pm$ 0.2981} & 0.7704 \scriptsize{$\pm$ 0.0585} & 67.89 \scriptsize{$\pm$ 2.4015} \\
\texttt{FedProx}
& 0.1510 \scriptsize{$\pm$ 0.0064}          & 84.78 \scriptsize{$\pm$ 0.4652} & 0.6754 \scriptsize{$\pm$ 0.1000} & 70.79 \scriptsize{$\pm$ 3.2738} \\
\texttt{SCAFFOLD}
& 3.8897$\times$10$^{\text{10}}$ \scriptsize{$\pm$ 8.6957$\times$10$^{\text{10}}$} & 9.87 \scriptsize{$\pm$ 0.4210} & 7.6962$\times$10$^{\text{15}}$ \scriptsize{$\pm$ 1.6933$\times$10$^{\text{16}}$} & 10.52 \scriptsize{$\pm$ 1.0351}\\
\textbf{\texttt{FedCote} (Ours)}
& 0.1662 \scriptsize{$\pm$ 0.0003} & 84.15 \scriptsize{$\pm$ 0.8670} & \textbf{0.5289 \scriptsize{$\pm$ 0.0965}} & \textbf{75.26 \scriptsize{$\pm$ 3.4509}} \\
\hline
\texttt{Ideal}
& 0.0399 \scriptsize{$\pm$ 0.0095}          & 86.22 \scriptsize{$\pm$ 0.2428} & 0.4212 \scriptsize{$\pm$ 0.0630} & 77.04 \scriptsize{$\pm$ 2.0662} \\
\hline
\end{tabular}
\end{table*}

\item
\textbf{\texttt{FedYOGI}} \cite{reddi2020adaptive}:
This scheme integrates an adaptive optimization strategy into global aggregation.
After computing the global update as
\begin{equation}\label{eq:FedYOGI model update}
\Delta_{r}
=
\frac{\sum_{{i} \in {\mathcal{K}}_{r}} \mathds{1}^{r}_{i} ( \mathbf{w}^{r,E}_{i} - {\bar {\mathbf{w}}}_{r-1} )}{\sum_{{i} \in {\mathcal{K}}_{r}} \mathds{1}^{r}_{i}}
,
\end{equation}
the first- and second-order moment estimates are updated by $m_r = \beta_1 m_{r-1} + (1-\beta_1)\Delta_{r}$
and $v_r = v_{r-1} - (1-\beta_2)\Delta_{r}^2 {\rm sign}(v_{r-1} - \Delta_{r}^2)$, respectively.
Then, the global model is updated as
\begin{equation}\label{eq:FedYOGI aggregation}
{\bar {\mathbf{w}}}_{r}
=
{\bar {\mathbf{w}}}_{r-1}
+ \gamma \frac{m_r}{\sqrt{v_r} + \tau}
,
\end{equation}
where $\tau$ controls adaptivity, with smaller values corresponding to stronger adaptivity.

\item
\textbf{\texttt{FedProx}} \cite{li2020federated}:
This scheme extends \texttt{FedAvg} by incorporating a proximal term into each client's local objective, defined as $\min_{\mathbf{w}} F_i({\mathbf{w}}) + \frac{\mu}{2}\| \mathbf{w} - {\bar {\mathbf{w}}}_{r-1}\|^2$,
where $\mu$ is the proximal coefficient.
The server then aggregates the local models through weighted averaging as in \eqref{eq:weighted aggregation failure}.

\item
\textbf{\texttt{Scaffold}} \cite{karimireddy2020scaffold}:
This scheme introduces control variates, $\mathbf{c}$ for the server and $\mathbf{c}_i$ for each client $i$.
During local training, client $i$ updates its model as
$\mathbf{w}^{r,t}_{i} = {\mathbf{w}}^{r,t-1}_{i}  -  \gamma_l ( \nabla F_{i}( {\mathbf{w}}^{r,t-1}_{i} ) - \mathbf{c}_i + \mathbf{c} )$,
where $\gamma_l$ is the local learning rate.
After training, the client updates its control variate as
$\mathbf{c}_i^{+} = \mathbf{c}_i - \mathbf{c} + \tfrac{1}{K\gamma_l} ({\bar {\mathbf{w}}}_{r-1} - \mathbf{w}^{r,E}_{i})$.
Then, with global learning rate $\gamma_g$, the server then aggregates both models and control variates as
\begin{subequations}
\begin{align}\label{eq:scaffold_global}
&
{\bar {\mathbf{w}}}_{r}
=
{\bar {\mathbf{w}}}_{r-1}
+ \gamma_g \frac{\sum_{{i} \in {\mathcal{K}}_{r}} \mathds{1}^{r}_{i} ( \mathbf{w}^{r,E}_{i} - {\bar {\mathbf{w}}}_{r-1} )}{\sum_{{i} \in {\mathcal{K}}_{r}} \mathds{1}^{r}_{i}}
, \\
&
\mathbf{c}^{r}
= \mathbf{c}^{r-1}
+ \frac{1}{N}\sum_{{i} \in {\mathcal{K}}_{r}} \mathds{1}^{r}_{i} ( \mathbf{c}_i^{+} - \mathbf{c}_i )
.
\end{align}
\end{subequations}

\item
\textbf{\texttt{Ideal}}:
The ideal scheme performs FL using Algorithm \ref{algorithm:FL under transmission failure} with $s_i = p_i$, and does not suffer from transmission failures,  i.e., $\mathds{1}^{r}_{i} = 1$ in \eqref{global_model_failure}.
This scheme serves as the performance upper bound in the simulations.
\end{itemize}

For fair comparison, the performance of each FL scheme is averaged over five independent experiments.

\subsection{Performance of FedCote in Static Scenario}\label{section:FedCote Robustness}

In this subsection, we evaluate the robustness of the proposed \texttt{FedCote} in the static scenario across various data distributions.
We consider a partial participation setting with $K=10$ to evaluate its learning performance.

\subsubsection{Balanced Dataset}\label{section:Balanced dataset}

We first assess the robustness of \texttt{FedCote} in balanced scenarios.
Tables \ref{table:FL performance MNIST} and \ref{table:FL performance CIFAR10} compare the training loss and testing accuracy of various FL schemes on balanced MNIST and CIFAR-10 datasets under both i.i.d. and non-i.i.d. settings.
The reported results represent the mean and standard deviation across five independent experiments.

On the MNIST dataset, as shown in Table \ref{table:FL performance MNIST}, \texttt{FedCote} achieves the lowest training loss and the highest testing accuracy in both i.i.d. and non-i.i.d. settings.
Its performance approaches that of the ideal scheme without transmission failures and surpasses the vanilla \texttt{FedAvg} with random client selection by 12.28\% in testing accuracy under the non-i.i.d. setting.
In contrast, other FL baselines exhibit significant performance degradation under non-i.i.d. conditions.
Specifically, \texttt{FedAvg} performs poorly due to its reliance on the weight $p_i$ as the selection probability, which neglects the effects of transmission failures.
\texttt{Power-of-Choice}, while selecting clients with the largest local training losses, indirectly prioritizes clients with higher transmission failure probabilities.
This partially mitigates the impact of uneven client participation caused by varying transmission conditions.
However, its testing accuracy under non-i.i.d. conditions remains 4.28\% lower than that of \texttt{FedCote}, as it does not consider balancing the effective appearance probabilities of each class's samples under transmission failures.
\texttt{Newt} exhibits poor and unstable performance, characterized by lower testing accuracy and higher variance, due to its reliance on client selection based on local-global model discrepancies, which is ineffective under transmission failures.
\texttt{GS}, which matches the aggregated label distribution to the global distribution, also overlooks transmission failures, resulting in suboptimal performance.
Although \texttt{TF-Aggregation} accounts for clients' transmission failure probabilities $\{\epsilon_i\}$, it places $(1 - \epsilon_i)$ in the denominator of the global aggregation scheme \eqref{eq:optimization problem TO}.
This design becomes unstable when clients exhibit high transmission failure probabilities, impairing global model convergence and markedly increasing training loss in both i.i.d. and non-i.i.d. settings;
refer to Appendix \ref{sec:convergence TF-Aggregation} for detailed theoretical analysis.
Besides, advanced FL schemes such as \texttt{FedNova}, \texttt{FedYOGI}, \texttt{FedProx}, and \texttt{SCAFFOLD}, which refine aggregation or local updates to alleviate data heterogeneity, also suffer performance degradation in the non-i.i.d. case.
This is because they overlook transmission failures and thus cannot effectively mitigate their adverse impact.
Notably, \texttt{SCAFFOLD} performs particularly poorly, with training losses diverging to extremely large values and testing accuracy becoming unstable.
This instability indicates that its control variate mechanism is highly sensitive to transmission failures and may require further modification.

A similar pattern emerges on the CIFAR-10 dataset, as shown in Table \ref{table:FL performance CIFAR10}.
In both i.i.d. and non-i.i.d. data cases, \texttt{FedCote} consistently outperforms all evaluated FL schemes, achieving the highest testing accuracy.
Notably, under non-i.i.d. settings, \texttt{FedCote} improves testing accuracy by 3.99\% compared to vanilla \texttt{FedAvg} and by 2.79\% compared to the next-best baseline, \texttt{Power-of-Choice}.
Furthermore, the increased complexity of the CIFAR-10 dataset and the ResNet-20 model amplifies the instability of \texttt{TF-Aggregation}, preventing it from converging due to gradient explosions in its global aggregation scheme.

\begin{figure}[!t]
\centering
\includegraphics[width= 3.2 in ]{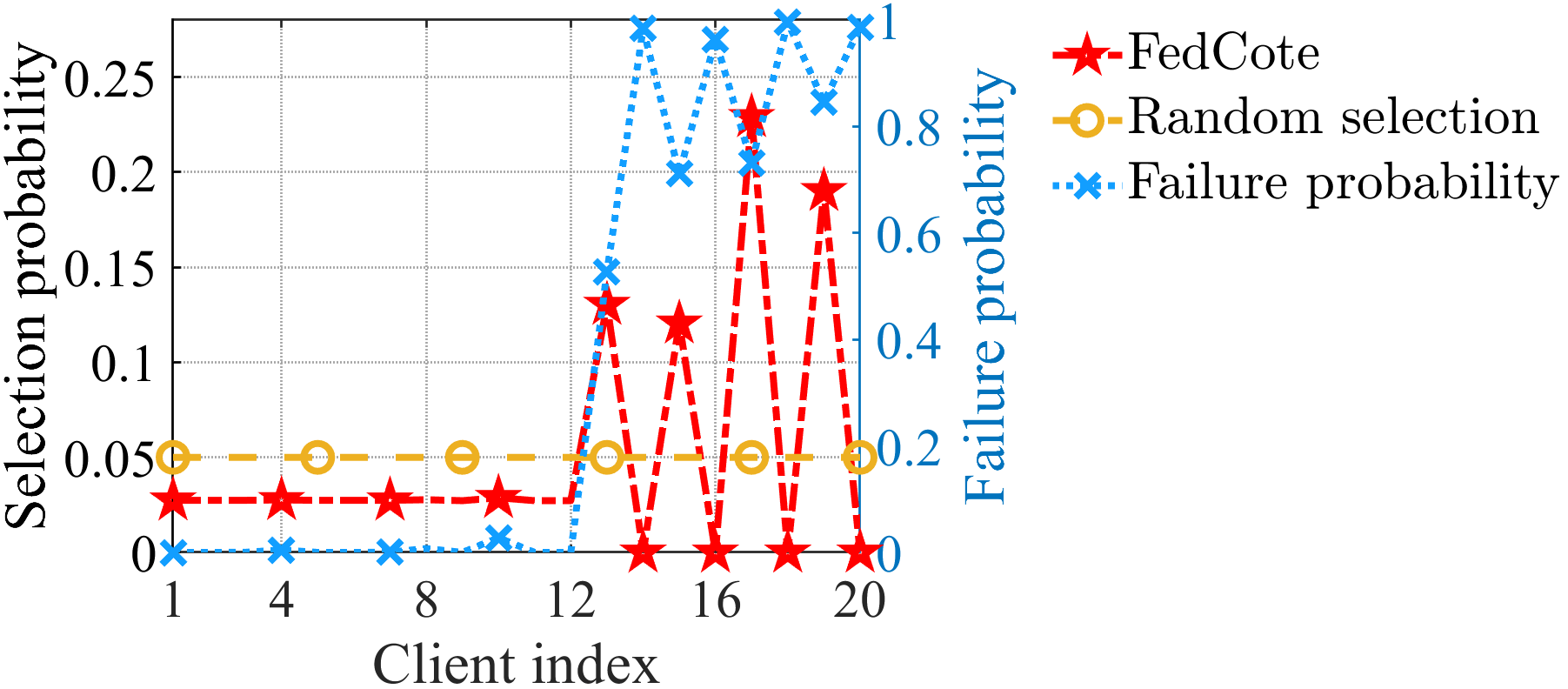}
\caption{Selection probabilities of FedCote for non-i.i.d. balanced datasets (MNIST, \emph{K} = 10).}
\label{fig:selection prob K10}
\end{figure}

To analyze why \texttt{FedCote} performs robustly under non-i.i.d. data cases, Fig. \ref{fig:selection prob K10} presents the selection probabilities for each client, $s_i$, optimized by \texttt{FedCote} using Algorithm \ref{algorithm:client selection optimization}.
Unlike the random selection strategy (orange line), where $s_i = p_i = \frac{1}{N}$ in the balanced case, \texttt{FedCote} (red line) optimizes selection probabilities by considering both transmission failure probabilities, $\{\epsilon_i\}$, and local label distributions.
By comparing the red and blue lines, it is evident that \texttt{FedCote} adjusts the selection probabilities for each clients in response to varying transmission failure probabilities, thereby balancing the effective appearance probabilities of each class's samples.
Furthermore, if $\epsilon_i$ exceeds the threshold $\epsilon_{th}$ (set to 0.85 in our experiments), \texttt{FedCote} assigns $s_i=0$ to exclude inefficient client selections.
Through this adaptive optimization, \texttt{FedCote} determines optimal selection probabilities for each client under diverse transmission conditions, effectively mitigating the negative impacts of transmission failures, particularly in challenging non-i.i.d. environments.

\subsubsection{Unbalanced Dataset}

Table \ref{table:FL performance MNIST unbalanced} and Table \ref{table:FL performance CIFAR10 unbalanced} further validate the effectiveness of \texttt{FedCote} on unbalanced datasets in challenging non-i.i.d. scenarios.
As described in Section \ref{section:Data distributions}, a higher imbalance ratio $u$ indicates a greater degree of dataset imbalance, with $u=0.5$ representing the balanced case.
As shown in Table \ref{table:FL performance MNIST unbalanced} and Table \ref{table:FL performance CIFAR10 unbalanced}, under non-i.i.d. settings, the testing accuracy of \texttt{FedAvg}, \texttt{Newt}, and \texttt{GS} significantly deteriorates on both MNIST and CIFAR-10 datasets as the imbalance ratio increases.
This performance decline is attributed to their inability to compensate for the loss of samples from clients with high transmission failure probabilities.
Consistent with the observations from balanced datasets discussed in Section \ref{section:Balanced dataset}, \texttt{Power-of-Choice} demonstrates slightly better performance due to its client selection strategy based on local training losses.
This strategy partially mitigates the impact of varying transmission failure probabilities across clients.
\texttt{TF-Aggregation} continues to perform poorly in unbalanced scenarios, due to the instability introduced by its global aggregation scheme, which impedes the convergence of the global model.
Moreover, advanced FL schemes, including \texttt{FedNova}, \texttt{FedYOGI}, \texttt{FedProx}, and \texttt{SCAFFOLD}, also experience performance degradation, as they neglect transmission failures and therefore cannot mitigate their negative effects.

In contrast, the proposed \texttt{FedCote} demonstrates robust performance under imbalanced conditions.
At extreme imbalance ratios ($u=0.8$ and $0.9$), \texttt{FedCote} achieves testing accuracy comparable to the ideal scheme on the MNIST dataset, with only a slight performance decline observed on CIFAR-10.
This performance degradation on CIFAR-10 can be attributed to its more complex image samples and the complete loss of samples from clients with $\epsilon_i > \epsilon_{th}$, which negatively impacts the generalization ability of trained model.
Nevertheless, \texttt{FedCote} consistently outperforms other FL baselines across all imbalance ratios, owing to its ability to effectively optimize client selection probabilities, thereby mitigating the adverse effects of both data imbalance and transmission failures.

As a brief summary, the proposed \texttt{FedCote} can effectively identify optimal client selection probabilities across diverse data distribution scenarios.
The experimental results underscore its robustness and adaptability, particularly in challenging non-i.i.d. environments, establishing \texttt{FedCote} as a superior client selection approach in FL compared to existing baselines.

\begin{table*}[!t]
\centering
\caption{
Testing accuracy (\%) of various FL schemes on the unbalanced MNIST datasets (Non-i.i.d., \emph{K} = 10).
}\label{table:FL performance MNIST unbalanced}
\begin{tabular}{c|c|cccc}
\hline
\multirow{2}{*}{\textbf{FL scheme}} & \textbf{Balanced} & \multicolumn{4}{c}{\textbf{Unbalanced}} \\
& $u = 0.5$ & $u = 0.6$ & $u = 0.7$ & $u = 0.8$ & $u = 0.9$ \\
\hline
\texttt{FedAvg}
& 80.90 \scriptsize{$\pm$ 3.6126} & 79.54 \scriptsize{$\pm$ 3.4957} & 80.60 \scriptsize{$\pm$ 3.2926} & 81.28 \scriptsize{$\pm$ 4.1780} & 81.10 \scriptsize{$\pm$ 2.4614} \\
\texttt{Power-of-Choice}
& 88.89 \scriptsize{$\pm$ 3.9024} & 88.20 \scriptsize{$\pm$ 1.6957} & 89.41 \scriptsize{$\pm$ 2.4485} & 89.44 \scriptsize{$\pm$ 1.8655} & 87.86 \scriptsize{$\pm$ 3.1834} \\
\texttt{Newt}
& 82.99 \scriptsize{$\pm$ 8.5893} & 83.17 \scriptsize{$\pm$ 10.1543} & 80.47 \scriptsize{$\pm$ 3.9466} & 83.15 \scriptsize{$\pm$ 9.9776} & 78.98 \scriptsize{$\pm$ 7.6076} \\
\texttt{GS}
& 85.54 \scriptsize{$\pm$ 2.9953} & 85.47 \scriptsize{$\pm$ 2.9188} & 84.99 \scriptsize{$\pm$ 3.3257} & 84.31 \scriptsize{$\pm$ 3.5246} & 83.46 \scriptsize{$\pm$ 3.9220} \\
\texttt{TF-Aggregation}
& 9.80 \scriptsize{$\pm$ 0.00} & 9.80 \scriptsize{$\pm$ 0.00} & 41.66 \scriptsize{$\pm$ 43.6243} & 63.95 \scriptsize{$\pm$ 34.9012} & 63.22 \scriptsize{$\pm$ 33.7940}
\\
\texttt{FedNova}
& 80.90 \scriptsize{$\pm$ 3.6126} & 79.53 \scriptsize{$\pm$ 3.5007} & 80.60 \scriptsize{$\pm$ 3.2888} & 81.28 \scriptsize{$\pm$ 4.1812} & 81.10 \scriptsize{$\pm$ 2.4679} \\
\texttt{FedYOGI}
& 85.06 \scriptsize{$\pm$ 4.7914} & 82.91 \scriptsize{$\pm$ 3.4434} & 81.45 \scriptsize{$\pm$ 4.7325} & 81.84 \scriptsize{$\pm$ 4.6295} & 82.42 \scriptsize{$\pm$ 4.5243} \\
\texttt{FedProx}
& 80.87 \scriptsize{$\pm$ 3.6306} & 79.54 \scriptsize{$\pm$ 3.5344} & 80.61 \scriptsize{$\pm$ 3.2781} & 81.33 \scriptsize{$\pm$ 4.1319} & 81.09 \scriptsize{$\pm$ 2.4654} \\
\texttt{SCAFFOLD}
& 59.23 \scriptsize{$\pm$ 16.6187} & 42.16 \scriptsize{$\pm$ 9.8092} & 64.86 \scriptsize{$\pm$ 10.8535} & 62.05 \scriptsize{$\pm$ 11.4378} & 59.71 \scriptsize{$\pm$ 8.7906} \\
\textbf{\texttt{FedCote} (Ours)}
& \textbf{93.17 \scriptsize{$\pm$ 0.4425}} & \textbf{93.00 \scriptsize{$\pm$ 0.3544}} & \textbf{92.99 \scriptsize{$\pm$ 0.3877}} & \textbf{92.85 \scriptsize{$\pm$ 0.5035}} & \textbf{92.31 \scriptsize{$\pm$ 0.6076}} \\
\hline
\texttt{Ideal}
& 91.27 \scriptsize{$\pm$ 1.3622} & 92.63 \scriptsize{$\pm$ 0.3125} & 92.63 \scriptsize{$\pm$ 0.3178} & 92.58 \scriptsize{$\pm$ 0.3196}  & 92.206 \scriptsize{$\pm$ 0.2895} \\
\hline
\end{tabular}
\end{table*}

\begin{table*}[!t]
\centering
\caption{
Testing accuracy (\%) of various FL schemes on the unbalanced CIFAR-10 datasets (Non-i.i.d., \emph{K} = 10).
}\label{table:FL performance CIFAR10 unbalanced}
\begin{tabular}{c|c|cccc}
\hline
\multirow{2}{*}{\textbf{FL scheme}} & \textbf{Balanced} & \multicolumn{4}{c}{\textbf{Unbalanced}} \\
& $u = 0.5$ & $u = 0.6$ & $u = 0.7$ & $u = 0.8$ & $u = 0.9$ \\
\hline
\texttt{FedAvg}
& 71.27 \scriptsize{$\pm$ 3.0723} & 71.22 \scriptsize{$\pm$ 3.5824} & 69.92 \scriptsize{$\pm$ 3.4972} & 69.39 \scriptsize{$\pm$ 3.8465} & 67.86 \scriptsize{$\pm$ 4.7762} \\
\texttt{Power-of-Choice}
& 72.47 \scriptsize{$\pm$ 2.5712} & 70.37 \scriptsize{$\pm$ 4.6263} & 74.47 \scriptsize{$\pm$ 1.7044} & 73.84 \scriptsize{$\pm$ 2.3828} & \textbf{73.29 \scriptsize{$\pm$ 2.6068}} \\
\texttt{Newt}
& 67.81 \scriptsize{$\pm$ 5.8911} & 66.39 \scriptsize{$\pm$ 8.1328} & 67.98 \scriptsize{$\pm$ 6.9624} & 63.28 \scriptsize{$\pm$ 4.3386} & 59.21 \scriptsize{$\pm$ 10.3313} \\
\texttt{GS}
& 69.162 \scriptsize{$\pm$ 3.1439} & 68.68 \scriptsize{$\pm$ 2.3119} & 67.30 \scriptsize{$\pm$ 3.1538} & 68.76 \scriptsize{$\pm$ 2.4633} & 66.61 \scriptsize{$\pm$ 3.1235} \\
\texttt{TF-Aggregation}
& 10.00 \scriptsize{$\pm$ 0.00} & 10.00 \scriptsize{$\pm$ 0.00} & 10.00 \scriptsize{$\pm$ 0.00} & 10.00 \scriptsize{$\pm$ 0.00} & 10.00 \scriptsize{$\pm$ 0.00} \\
\texttt{FedNova}
& 70.87 \scriptsize{$\pm$ 3.2911} & 71.26 \scriptsize{$\pm$ 3.6441} & 69.89 \scriptsize{$\pm$ 3.7546} & 69.45 \scriptsize{$\pm$ 4.0064} & 67.51 \scriptsize{$\pm$ 4.8614} \\
\texttt{FedYOGI}
& 67.89 \scriptsize{$\pm$ 2.4015} & 68.31 \scriptsize{$\pm$ 1.1754} & 66.50 \scriptsize{$\pm$ 2.4422} & 66.39 \scriptsize{$\pm$ 2.8794} & 66.16 \scriptsize{$\pm$ 4.0516} \\
\texttt{FedProx}
& 70.79 \scriptsize{$\pm$ 3.2738} & 71.01 \scriptsize{$\pm$ 3.8370} & 69.89 \scriptsize{$\pm$ 3.3727} & 69.06 \scriptsize{$\pm$ 3.8049} & 67.57 \scriptsize{$\pm$ 5.4462} \\
\texttt{SCAFFOLD}
& 9.90 \scriptsize{$\pm$ 0.2394} & 10.25 \scriptsize{$\pm$ 0.7722} & 9.98 \scriptsize{$\pm$ 1.1011} & 9.72 \scriptsize{$\pm$ 0.2649} & 10 \scriptsize{$\pm$ 0.1113} \\
\textbf{\texttt{FedCote} (Ours)}
& \textbf{75.26 \scriptsize{$\pm$ 3.4509}} & \textbf{74.97 \scriptsize{$\pm$ 3.1347}} & \textbf{75.27 \scriptsize{$\pm$ 3.2723}} & \textbf{73.91 \scriptsize{$\pm$ 3.7634}} & 72.13 \scriptsize{$\pm$ 4.2337} \\
\hline
\texttt{Ideal}
& 77.04 \scriptsize{$\pm$ 2.0662} & 76.93 \scriptsize{$\pm$ 2.5807} & 76.79 \scriptsize{$\pm$ 2.8712} & 77.19 \scriptsize{$\pm$ 2.0573}  & 77.03 \scriptsize{$\pm$ 2.0914} \\
\hline
\end{tabular}
\end{table*}

\subsection{Performance of FedCote-II in Static Scenario}

\begin{figure}[!t]
\centering
\includegraphics[width= 1.9 in ]{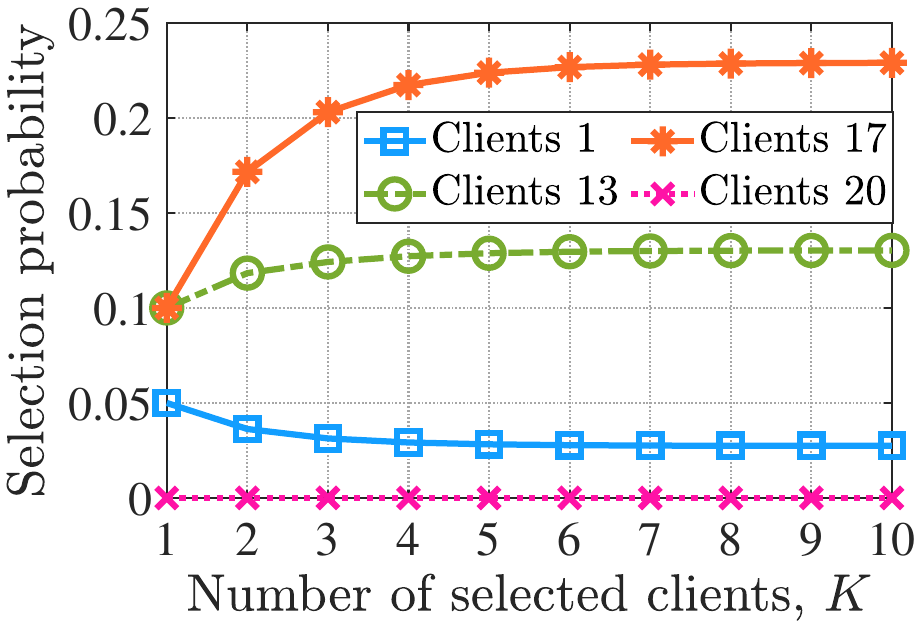}
\caption{Optimal selection probabilities of clients obtained from (\ref{eq:optimization problem}) under different \emph{K}.}
\label{fig:selection prob diff K}
\end{figure}

\begin{table}[!t]
\centering
\caption{\vspace{0.15cm} Ratio of \emph{C}$^{\emph{\text{N}}+{\emph{\text K}_{\emph{\text{apx}}}}-{\text{1}}}_{\emph{\text K}_{\emph{\text{apx}}}}$ to \emph{C}$^{\emph{\text{N}}+\emph{\text{K}}-{\text{1}}}_\emph{\text K}$ under \emph{N}=20 and \emph{K}=10.}\label{table:FedCoteII computation ratio}
\begin{tabular}{c|ccccc}
\hline
$K_{\rm apx}$ & 2 & 4 & 6 & 8 \\
\hline
Ratio (\%) & 0.0010 & 0.0442 & 0.8842 & 11.0837 \\
\hline
\end{tabular}
\label{table: ratio C}
\end{table}

In this subsection, we evaluate the effectiveness of the proposed \texttt{FedCote-II} in the static scenario, which can reduce computation complexity and fasten the optimization speed.

\subsubsection{Optimal Client Selection Probability under Different K}

Fig. \ref{fig:selection prob diff K} illustrates the optimal client selection probabilities $\{ s_i \}$, obtained from \eqref{eq:optimization problem}, under different values of $K$.
As observed, with increasing $K$, the value of $s_i$ gradually converges to a stationary solution.
Moreover, the gap between the optimal solutions of ${ s_i }$ under $K$ and $K+1$ diminishes as $K$ grows.
These findings align with the theoretical results presented in Proposition \ref{proposition:difference increasing K}.

\subsubsection{Performance under Different K$_{\text{apx}}$}

\begin{table*}[!t]
\centering
\caption{
Performance of FedCote-II under different values of \emph{K}$_{\text{apx}}$ on the non-i.i.d. MNIST dataset (\emph{K} = 10).
}\label{table:FedCote-II performance MNIST}
\begin{tabular}{c|cc|cc}
\hline
\multirow{2}{*}{$K_{\rm apx}$} & \multicolumn{2}{c|}{\textbf{Balanced}} & \multicolumn{2}{c}{\textbf{Unbalanced} ($u=0.1$)} \\
& \textbf{Training loss} & \textbf{Testing accuracy (\%)} & \textbf{Training loss} & \textbf{Testing accuracy (\%)} \\
\hline
2
& 0.2630 \scriptsize{$\pm$ 0.0423} & 91.59 \scriptsize{$\pm$ 1.5015} & 0.2439 \scriptsize{$\pm$ 0.0169} & 92.31 \scriptsize{$\pm$ 0.6076} \\
4
& 0.2379 \scriptsize{$\pm$ 0.0110} & 92.48 \scriptsize{$\pm$ 0.5614} & \textbf{0.2590 \scriptsize{$\pm$ 0.0143}} & \textbf{91.82 \scriptsize{$\pm$ 0.6028}} \\
6
& \textbf{0.2157 \scriptsize{$\pm$ 0.0127}} & \textbf{93.26 \scriptsize{$\pm$ 0.4433}} & 0.2424 \scriptsize{$\pm$ 0.0151} & 92.43 \scriptsize{$\pm$ 0.4798} \\
8
& 0.2194 \scriptsize{$\pm$ 0.0171} & 93.08 \scriptsize{$\pm$ 0.6843} & 0.2421 \scriptsize{$\pm$ 0.0178} & 92.34 \scriptsize{$\pm$ 0.5434} \\
\hline
\texttt{FedCote}
& 0.2174 \scriptsize{$\pm$ 0.0138} & 93.17 \scriptsize{$\pm$ 0.4425} & 0.2439 \scriptsize{$\pm$ 0.0169} & 92.31 \scriptsize{$\pm$ 0.6076 }\\
\hline
\end{tabular}
\end{table*}

\begin{table*}[!t]
\centering
\caption{
Performance of FedCote-II under different values of \emph{K}$_{\text{apx}}$ on the non-i.i.d. CIFAR-10 dataset (\emph{K} = 10).
}\label{table:FedCote-II performance CIFAR10}
\begin{tabular}{c|cc|cc}
\hline
\multirow{2}{*}{$K_{\rm apx}$} & \multicolumn{2}{c|}{\textbf{Balanced}} & \multicolumn{2}{c}{\textbf{Unbalanced} ($u=0.1$)} \\
& \textbf{Training loss} & \textbf{Testing accuracy (\%)} & \textbf{Training loss} & \textbf{Testing accuracy (\%)} \\
\hline
2
& 0.6199 \scriptsize{$\pm$ 0.2431} & 73.59 \scriptsize{$\pm$ 5.7986} & 0.8521 \scriptsize{$\pm$ 0.3080} & 68.01 \scriptsize{$\pm$ 7.2736} \\
4
& 0.5487 \scriptsize{$\pm$ 0.1020} & \textbf{75.28 \scriptsize{$\pm$ 3.1163}} & \textbf{0.6280 \scriptsize{$\pm$ 0.1215}} & \textbf{72.85 \scriptsize{$\pm$ 3.9682}} \\
6
& \textbf{0.5384 \scriptsize{$\pm$ 0.0974}} & 75.19 \scriptsize{$\pm$ 3.3896} & 0.6515 \scriptsize{$\pm$ 0.1145} & 71.62 \scriptsize{$\pm$ 4.4017} \\
8
& 0.5517 \scriptsize{$\pm$ 0.1101} & 74.57 \scriptsize{$\pm$ 3.7383} & 0.6388 \scriptsize{$\pm$ 0.0931} & 72.02 \scriptsize{$\pm$ 3.7200} \\
\hline
\texttt{FedCote}
& 0.5289 \scriptsize{$\pm$ 0.0965} & 75.26 \scriptsize{$\pm$ 3.4509} & 0.6384 \scriptsize{$\pm$ 0.1051} & 72.13 \scriptsize{$\pm$ 4.2337} \\
\hline
\end{tabular}
\end{table*}

According to the discussion in Section \ref{section:Enhanced FedCote}, the number of distinct combinations of $\mathcal{K}_r^z$, namely $C^{N+K-1}_K$, used for calculating the effective appearance probability ${\bar \beta}_i$ in \eqref{eq:bar beta i formulation}, increases with $K$, leading to a higher computational burden. This, in turn, increases the complexity of solving the client selection optimization problem in \eqref{eq:optimization problem}.
Following the experimental settings in Section \ref{section:FedCote Robustness}, we use a partial participation setting with $K=10$ for \texttt{FedCote}.
If we replace $K$ with a smaller value $K_{\rm apx}$, the number of distinct combinations of $\mathcal{K}_r^z$ becomes $C^{N+K_{\rm apx}-1}_{K_{\rm apx}}$.
Table \ref{table: ratio C} compares the ratio of $C^{N+K_{\rm apx}-1}_{K_{\rm apx}}$ to $C^{N+K-1}_K$ under different values of $K_{\rm apx}$.
As shown in Table \ref{table: ratio C}, the computational cost decreases exponentially as $K_{\rm apx}$ is reduced.

Tables \ref{table:FedCote-II performance MNIST} and \ref{table:FedCote-II performance CIFAR10} further compare the performance of \texttt{FedCote-II} under different values of $K_{\rm apx}$ with \texttt{FedCote} on non-i.i.d. MNIST and CIFAR-10 datasets.
As observed, when $K_{\rm apx}$ is small (e.g., $K_{\rm apx} = 2$), the variance in \texttt{FedCote-II} performance is slightly higher.
This is because, as shown in Fig. \ref{fig:selection prob diff K}, the optimal solutions for $s_i$ under $K_{\rm apx} = 2$ still exhibit a noticeable gap from the optimal solution for $K=10$.
Nevertheless, \texttt{FedCote-II} demonstrates robust performance across different values of $K_{\rm apx}$, achieving comparable training loss and testing accuracy to \texttt{FedCote} on both datasets.
These results not only validate the effectiveness of the client selection optimization problem in \eqref{eq:optimization problem}, but also confirm the feasibility of reducing computational complexity by replacing $K$ with $K_{\rm apx}$.

\subsubsection{Performance under K = 20}

\begin{table*}[!t]
\centering
\caption{
Testing accuracy (\%) of various FL schemes on balanced datasets with \emph{K} = 20.
}\label{table:FL performance MNIST CIFAR10 K20}
\begin{tabular}{c|cc|cc}
\hline
\multirow{2}{*}{\textbf{FL scheme}} & \multicolumn{2}{c|}{\textbf{MNIST}} & \multicolumn{2}{c}{\textbf{CIFAR-10}} \\
& \textbf{i.i.d.} & \textbf{Non-i.i.d.} & \textbf{i.i.d.} & \textbf{Non-i.i.d.} \\
\hline
\texttt{FedAvg}
& 96.11 \scriptsize{$\pm$ 0.1031} & 85.44 \scriptsize{$\pm$ 4.1275} & 84.73 \scriptsize{$\pm$ 0.3243} & 73.01 \scriptsize{$\pm$ 2.0789} \\
\texttt{Power-of-Choice}, \texttt{Newt}, \texttt{GS}
& 95.63 \scriptsize{$\pm$ 0.1244} & 86.13 \scriptsize{$\pm$ 1.8333} & 84.33 \scriptsize{$\pm$ 0.4856} & 71.18 \scriptsize{$\pm$ 1.7848} \\
\texttt{TF-Aggregation}
& 63.11 \scriptsize{$\pm$ 41.4053} & 88.32 \scriptsize{$\pm$ 2.5947} & 10.00 \scriptsize{$\pm$ 0.00}   & 10.00 \scriptsize{$\pm$ 0.00} \\
\texttt{FedNova}
& 96.11 \scriptsize{$\pm$ 0.1085} & 85.45 \scriptsize{$\pm$ 4.1240} & 84.82 \scriptsize{$\pm$ 0.4667}   & 73.12 \scriptsize{$\pm$ 2.5151} \\
\texttt{FedYOGI}
& 95.30 \scriptsize{$\pm$ 0.0857} & 82.81 \scriptsize{$\pm$ 5.4951} & 84.12 \scriptsize{$\pm$ 0.2850}   & 70.21 \scriptsize{$\pm$ 2.8150} \\
\texttt{FedProx}
& 96.08 \scriptsize{$\pm$ 0.1092} & 85.46 \scriptsize{$\pm$ 4.1367} & 84.95 \scriptsize{$\pm$ 0.4931}   & 72.52 \scriptsize{$\pm$ 2.7564} \\
\texttt{SCAFFOLD}
& 92.94 \scriptsize{$\pm$ 0.4439} & \textbf{92.24 \scriptsize{$\pm$ 0.7516}} & 10.47 \scriptsize{$\pm$ 1.0322}   & 9.72 \scriptsize{$\pm$ 0.4967} \\
\textbf{\texttt{FedCote-II} (Ours)}
& \textbf{96.11 \scriptsize{$\pm$ 0.1031}} & 91.22 \scriptsize{$\pm$ 1.6641} & \textbf{85.01 \scriptsize{$\pm$ 0.4758}} & \textbf{78.26 \scriptsize{$\pm$ 1.0859}} \\
\hline
\texttt{Ideal}
& 96.27 \scriptsize{$\pm$ 0.0723} & 92.02 \scriptsize{$\pm$ 1.1980} & 86.13 \scriptsize{$\pm$ 0.3015} & 78.26 \scriptsize{$\pm$ 2.4080} \\
\hline
\end{tabular}
\end{table*}

In this section, the number of selected clients per iteration, $K$, is increased from 10 to 20.
For \texttt{Power-of-Choice}, \texttt{Newt}, and \texttt{GS}, client selection without replacement results in all $N=20$ clients participating in training, aggregating the global model using \eqref{eq:weighted aggregation failure}.
This differs slightly from the settings of \texttt{FedAvg} described in Section \ref{section:Baselines}, where clients are selected with replacement, and the global model is aggregated using \eqref{global_model_failure}.
For \texttt{FedCote-II}, we set $K_{\rm apx} = 10$ and employ the optimal selection probabilities ${ s_i }$ obtained from \eqref{eq:optimization problem} with $K=10$ (depicted in Figs. \ref{fig:selection prob K10}).

\begin{figure}[!t]
\centering
\subfigure[MNIST, i.i.d.]{
\includegraphics[width= 1.69 in ]{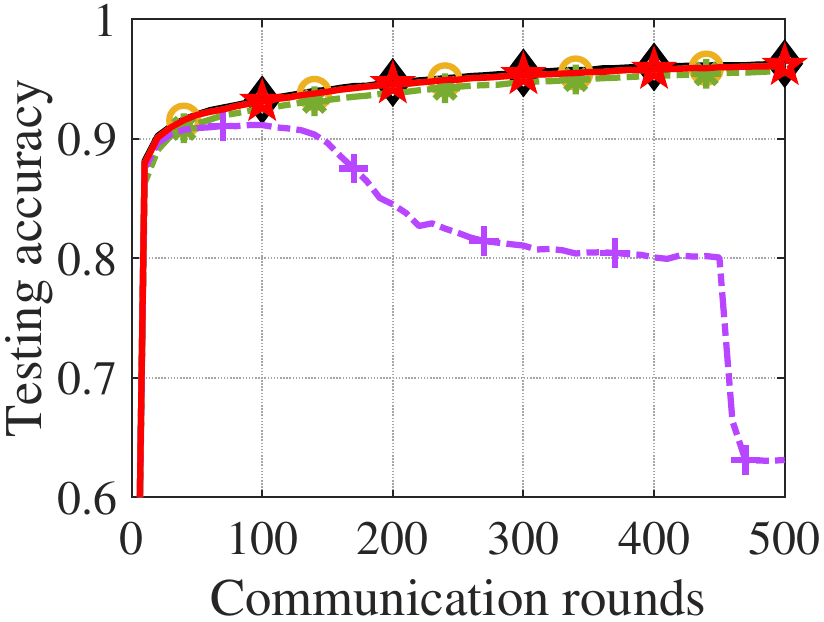}}
\subfigure[MNIST, non-i.i.d.]{
\includegraphics[width= 1.69 in ]{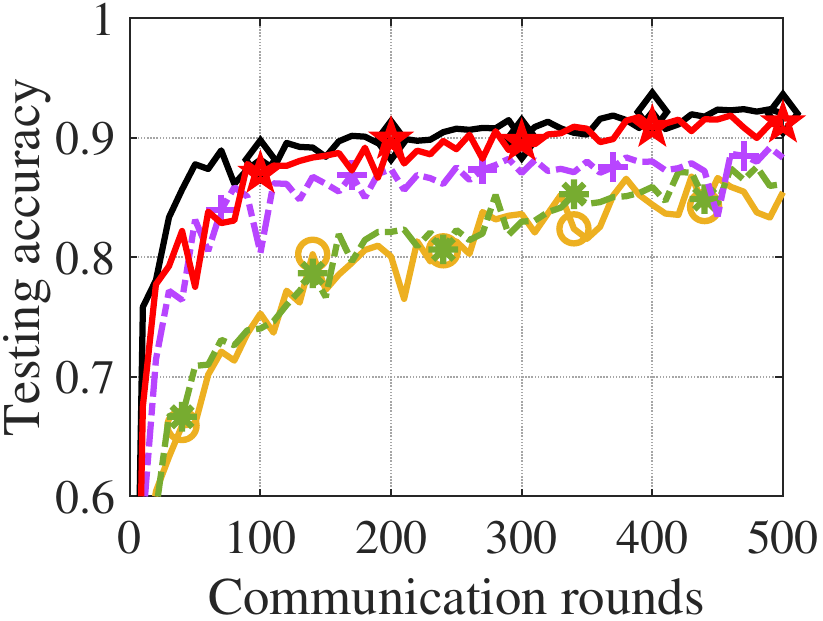}}
\subfigure[CIFAR-10, i.i.d.]{
\includegraphics[width= 1.69 in ]{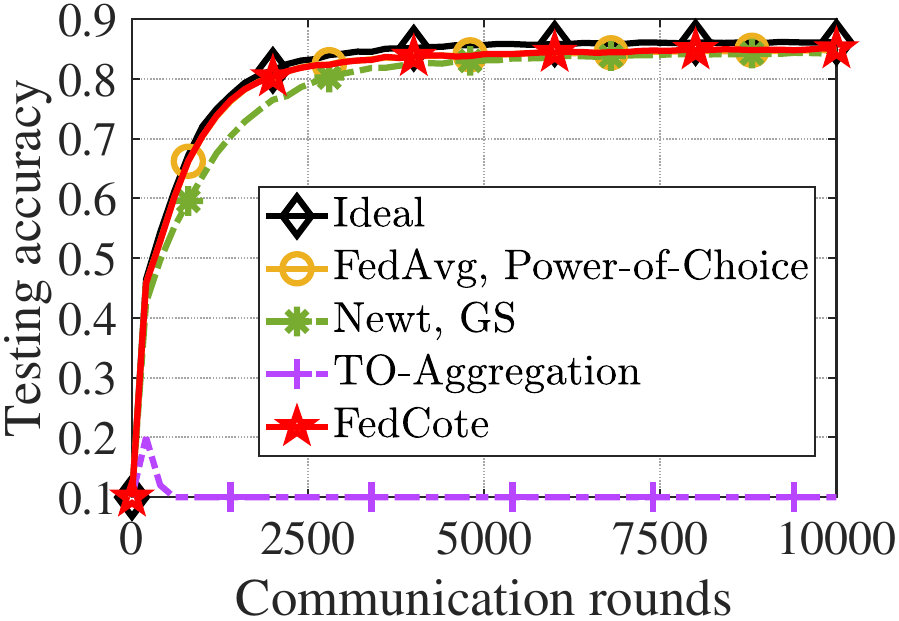}}
\subfigure[CIFAR-10, non-i.i.d.]{
\includegraphics[width= 1.69 in ]{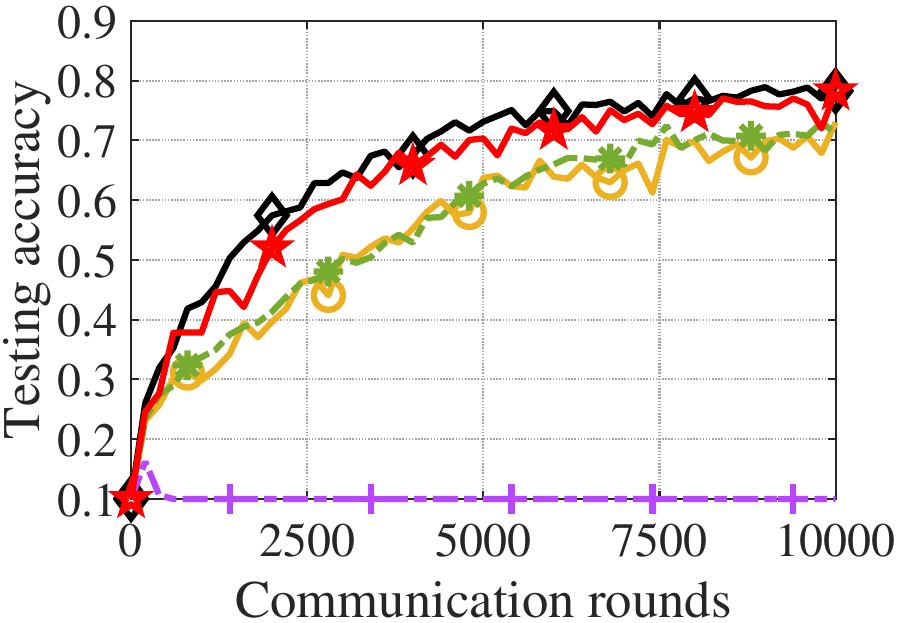}}
\caption{Convergence of various client selection schemes on balanced datasets (\emph{K} = 20).}
\label{fig:performance balanced K20}
\end{figure}

Based on the above settings, Table \ref{table:FL performance MNIST CIFAR10 K20} compares the testing accuracy of the global models across different FL schemes after completing training.
Additionally, Fig. \ref{fig:performance balanced K20} illustrates the convergence trends of various client selection schemes on the MNIST and CIFAR-10 datasets.
From Table \ref{table:FL performance MNIST CIFAR10 K20} and Fig. \ref{fig:performance balanced K20}, it is evident that \texttt{FedCote-II} achieves faster convergence and higher testing accuracy consistently compared to other benchmarks, for both i.i.d. and non-i.i.d. data cases.

Furthermore, based on Tables \ref{table:FL performance MNIST}, \ref{table:FL performance CIFAR10}, and \ref{table:FL performance MNIST CIFAR10 K20}, by comparing the testing accuracy of \texttt{FedCote} under $K=10$ with that of \texttt{FedCote-II} under $K=20$, we observe that increasing $K$ accelerates FL convergence and enhances testing accuracy, particularly for non-i.i.d. data cases.
Notably, on the more complex CIFAR-10 dataset, \texttt{FedCote-II} achieves a testing accuracy of 78.26\% with $K=20$, compared to 75.26\% with $K=10$, demonstrating the benefits of larger $K$.
This observation is consistent with the theoretical analysis in Corollary \ref{corollary:convergence FL failure}.

\subsection{Performance Comparison in Dynamic Scenario}\label{section:Performance Dynamic}

In more general dynamic scenarios, the transmission failure probabilities vary over time during the training process, which necessitates iteration-wise optimization of the selection probabilities $\{ s_i \}$.
Based on the results in Table \ref{table:FedCote-II performance MNIST} and Table \ref{table:FedCote-II performance CIFAR10}, we observe that $K_{\rm apx} = 4$ is sufficient to achieve stable and satisfactory performance.
Therefore, in the dynamic case, we adopt \texttt{FedCote-II} with $K_{\rm apx} = 4$, which enables efficient optimization of $\{ s_i \}$ at each iteration while maintaining low computational overhead.

Table \ref{table:FL performance dynamic} reports the testing accuracy of various FL schemes under dynamic transmission failures for both MNIST and CIFAR-10 with non-i.i.d. data distribution.
The results demonstrate that, similar to the static case, the proposed \texttt{FedCote-II} consistently outperforms the competing methods.
On MNIST and CIFAR-10, \texttt{FedCote-II} achieves accuracies of 92.06\% and 78.45\%, respectively, not only outperforming all baselines but also very close to the \texttt{Ideal} case.
This demonstrates that the proposed \texttt{FedCote} can adaptively adjust client selection probabilities in real time, effectively mitigating the impact of dynamic transmission failures.

\begin{table}[!t]
\centering
\caption{
Testing accuracy (\%) of various FL schemes in dynamic scenario $\qquad$
(\emph{K} = 10, \emph{K}$_{\text{apx}}$=4, non-i.i.d.).
}\label{table:FL performance dynamic}
\begin{tabular}{c|cc}
\hline
\textbf{FL scheme} & \textbf{MNIST} & \textbf{CIFAR-10} \\
\hline
\texttt{FedAvg} & 86.71 \scriptsize{$\pm$ 3.4133} & 74.39 \scriptsize{$\pm$ 2.3036} \\
\texttt{Power-of-Choice} & 75.93 \scriptsize{$\pm$ 0.4725} & 75.02 \scriptsize{$\pm$ 2.8424} \\
\texttt{Newt} & 83.31 \scriptsize{$\pm$ 5.6989} & 69.68 \scriptsize{$\pm$ 5.8326} \\
\texttt{GS} & 88.59 \scriptsize{$\pm$ 2.0003} & 76.62 \scriptsize{$\pm$ 1.2761} \\
\texttt{TF-Aggregation} & 9.80 \scriptsize{$\pm$ 0.00} & 10.00 \scriptsize{$\pm$ 0.00} \\
\texttt{FedNova} & 86.71 \scriptsize{$\pm$ 3.4096} & 74.20 \scriptsize{$\pm$ 1.9577} \\
\texttt{FedYOGI} & 88.90 \scriptsize{$\pm$ 0.8304} & 74.97 \scriptsize{$\pm$ 1.0329} \\
\texttt{FedProx} & 86.69 \scriptsize{$\pm$ 3.4340} & 74.21 \scriptsize{$\pm$ 2.3346} \\
\texttt{SCAFFOLD} & 46.03 \scriptsize{$\pm$ 13.7458} & 9.38 \scriptsize{$\pm$ 0.9060} \\
\textbf{\texttt{FedCote-II} (Ours)} & \textbf{92.06 \scriptsize{$\pm$ 1.7416}} & \textbf{78.45 \scriptsize{$\pm$ 2.1837}} \\
\hline
\texttt{Ideal} & 92.02 \scriptsize{$\pm$ 1.1980} & 78.26 \scriptsize{$\pm$ 2.4080} \\
\hline
\end{tabular}
\end{table}

\subsection{Convergence of Training Loss}\label{section:FedCote Justification}

\begin{figure}[!t]
\centering
\begin{minipage}[h]{1\linewidth}
\centering
\includegraphics[width= 3 in ]{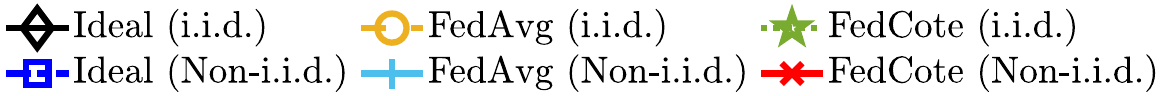}
\end{minipage}
\begin{minipage}[h]{1\linewidth}
\centering
\subfigure[MNIST.]{
\includegraphics[width= 1.69 in ]{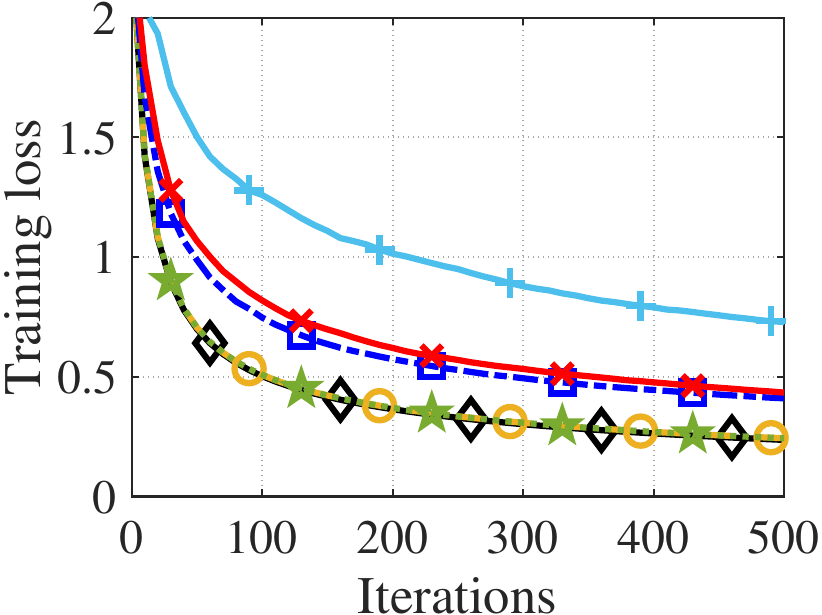}}
\subfigure[CIFAR-10.]{
\includegraphics[width= 1.69 in ]{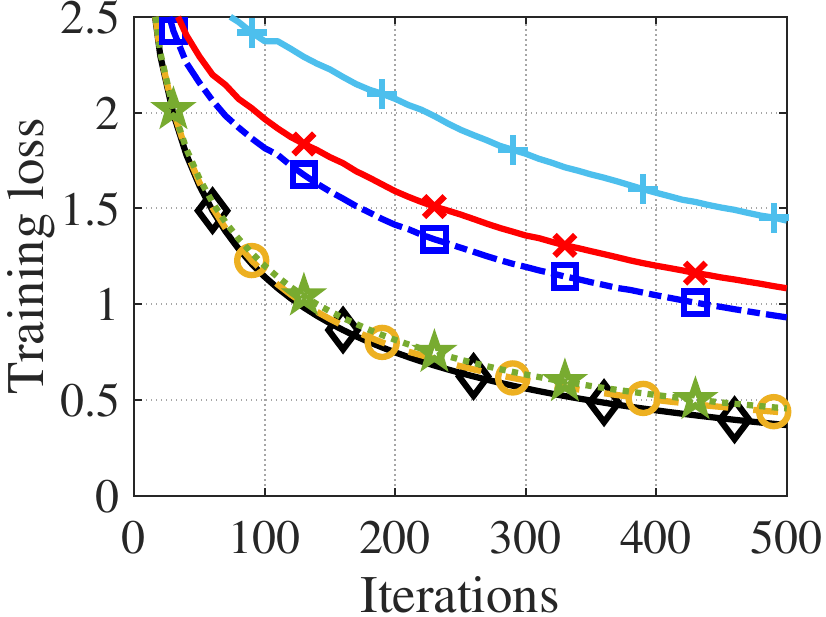}}
\caption{Average training loss of various FL schemes (Static, \emph{K} = 10).}
\label{fig:performance average training loss}
\end{minipage}
\end{figure}

We further provide numerical results to validate the convergence analysis in Theorem~\ref{theorem:convergence FL failure} and Corollary~\ref{corollary:convergence FL failure}.
Fig.~\ref{fig:performance average training loss} compares the average training loss, $\frac{1}{R} {\sum}_{r=1}^R F({\bar {\mathbf{w}}}_{r-1})$, of \texttt{Ideal}, \texttt{FedAvg}, and proposed \texttt{FedCote}, which clearly illustrates their convergence trends under different conditions.

Several key observations can be drawn.
First, as shown in Fig.~\ref{fig:performance average training loss}(a) and Fig.~\ref{fig:performance average training loss}(b), in an ideal setting without transmission failures, FL converges properly, while non-i.i.d. data slows the convergence rate.
Second, when local datasets are i.i.d., all FL schemes achieve proper convergence.
Third, under non-i.i.d. data, transmission failures aggravate data heterogeneity and cause \texttt{FedAvg} to converge to a biased solution.
Finally, unlike \texttt{FedAvg}, the proposed \texttt{FedCote} converges reliably (close to the \texttt{Ideal} scheme) even in the non-i.i.d. case, by mitigating the divergence between the effective and actual label distributions, i.e., enforcing $\chi^2_{\bm{\bar \alpha} \| \bm{\alpha}_g} = 0$, through client selection.
These results are consistent with the theoretical insights in Theorem~\ref{theorem:convergence FL failure} and Corollary~\ref{corollary:convergence FL failure}.

\section{Conclusion}\label{section:Conclusion}

In this paper, we have investigated FL in wireless edge networks, particularly focusing on the joint impacts of unreliable network conditions and data heterogeneity.
Through a novel theoretical analysis, we have demonstrated that in non-i.i.d. scenarios with label distribution skew, transmission failures distort the effective label distributions of local samples, deviating from the global dataset's actual label distribution, which negatively affects FL performance (Theorem \ref{theorem:convergence FL failure} and Observation \ref{observation:G larger than Vic}).
To address this issue, we have proposed \texttt{FedCote}, a client selection approach that mitigates this divergence without relying on wireless resource scheduling.
Experimental results have shown that \texttt{FedCote} significantly improves robustness against transmission failures, particularly in non-i.i.d. data settings.
However, the computational complexity of solving the client selection optimization problem increases with the number of selected clients.
To address this, we have introduced \texttt{FedCote-II}, which reduces computational overhead while maintaining satisfactory learning performance.

There remain several promising directions for future research.
First, although this work focuses on the standard FL setting where all clients communicate directly with a central server, it would be valuable to investigate transmission failures under different network topologies.
Extending the proposed client selection optimization method to more general distributed networks would further enhance its generality and practical applicability.
Second, while we focus on supervised learning, future work could extend the client selection approach to semi-supervised and unsupervised learning scenarios.
Third, although this work optimizes client selection probabilities alone, jointly optimizing both client selection probabilities and aggregation weights would be efficient and can further mitigate the impact of transmission failures on FL convergence without altering the existing network infrastructure.
Last but not the least, while this work adopts digital communication systems between the server and clients, an interesting direction is to extend the proposed client selection optimization method to advanced frameworks such as over-the-air FL (AirComp-FL).

\bibliographystyle{IEEEtran}
\bibliography{refs_journal}

\vfill

\clearpage

\onecolumn

\begin{appendices}

\section{Proof of Proposition \ref{proposition:data heterogeneity}}\label{appendix:proof ineq nabla Fi - F}

Based on \eqref{eq:nabla Fi F alpha c}, the discrepancy between local and global gradients is bounded by
\begin{small}
\begin{align}\label{eq:nabla Fi - F}
\| \nabla F_i({\mathbf{w}}) - \nabla F ({\mathbf{w}}) \|^2
= &
\Big\|
\nabla F_i({\mathbf{w}}) - {\sum}_{c=1}^C \alpha_{i,c} \nabla F_{g,c}
+ {\sum}_{c=1}^C \alpha_{i,c} \nabla F_{g,c} - \nabla F ({\mathbf{w}})
\Big\|^2
\notag \\
\overset{(a)}{\leq} &
2 \Big\|
{\sum}_{c=1}^C \alpha_{i,c}  \left( \nabla F_{i,c} ({\mathbf{w}}) - \nabla F_{g,c} ({\mathbf{w}}) \right)
\Big\|^2
+
2 \Big\| {\sum}_{c=1}^C \left( \alpha_{i,c} - \alpha_{g,c} \right) \nabla F_{g,c} ({\mathbf{w}})
\Big\|^2
\notag \\
= &
2 \Big\| {\sum}_{c=1}^C \alpha_{i,c}  \left( \nabla F_{i,c} ({\mathbf{w}}) - \nabla F_{g,c} ({\mathbf{w}}) \right) \Big\|^2
+ 2 \Big\| {\sum}_{c=1}^C \frac{\left( \alpha_{i,c} - \alpha_{g,c} \right)}{\sqrt{\alpha_{g,c}}} \sqrt{\alpha_{g,c}} \nabla F_{g,c} ({\mathbf{w}}) \Big\|^2
\notag \\
\overset{(b)}{\leq} &
2 {\sum}_{c=1}^C \alpha_{i,c}  \| \nabla F_{i,c} ({\mathbf{w}}) - \nabla F_{g,c} ({\mathbf{w}}) \|^2
+ 2 {\sum}_{c=1}^C \frac{\left( \alpha_{i,c} - \alpha_{g,c} \right)^2}{\alpha_{g,c}}  {\sum}_{c=1}^C \alpha_{g,c} \| \nabla F_{g,c} ({\mathbf{w}}) \|^2
,
\end{align}
\end{small}
\vspace{-0.3cm}

\noindent
where inequality (a) follows from the relation $\| x_1 + x_2 \|^2 \leq 2 \| x_1\|^2 + 2 \| x_2\|^2$,
and inequality (b) is due to Jensen's Inequality and the Cauchy-Schwarz Inequality.
\hfill $\blacksquare$

\section{Proof of Lemma \ref{lemma:beta}}\label{sec:Proof_lemma_beta}

Based on the selection scheme described in Section \ref{sec:FL procedure}, at each iteration, $K$ clients are selected independently and with replacement according to the selection probabilities $\{ s_i \}_{i=1}^N$, where $\sum_{i=1}^N s_i = 1$.
This results in $N^K$ possible permutations of the selection set $\mathcal{K}_r$, denoted by $\mathcal{K}_r^g$ for $g \in [N^K]$, with the appearance probability of each permutation given by
\begin{small}
\begin{equation}
\Pr({\mathcal{K}}_{r} = {\mathcal{K}}_r^g) = {\prod}_{{i} \in {\mathcal{K}}_r^g} s_{i}.
\end{equation}
\end{small}
\vspace{-0.3cm}

\noindent
Since transmission failures occur independently across clients, we have
\begin{small}
\begin{equation}
\Pr \bigg( {\sum}_{{i} \in {\mathcal{K}}_{r}} \mathds{1}^{r}_{i} \neq 0 \bigg) = 1 - {\prod}_{{i} \in {\mathcal{K}}_{r}} \epsilon_{i}.
\end{equation}
\end{small}
\vspace{-0.3cm}

\noindent
Thus, we have
\begin{small}
\begin{align}\label{beta_proof}
&
\mathbb{E}_{{\mathcal{K}}_{r}, \mathds{1}^{r}_{i}}
\Bigg[
\frac{ {\sum}_{{i} \in {\mathcal{K}}_{r}} \mathds{1}^{r}_{i}
\mathbf{w}^{r,E}_{i} }{ {\sum}_{{i} \in {\mathcal{K}}_{r}} \mathds{1}^{r}_{i}}
\Bigg|
\sum_{{i} \in {\mathcal{K}}_{r}} \mathds{1}^{r}_{i} \neq 0
\Bigg]
= {\mathbb{E}}_{\mathcal{K}_{r}}
\Bigg[
{\mathbb{E}}_{\mathds{1}^{r}_{i}}
\bigg[
\frac{ {\sum}_{{i} \in {\mathcal{K}}_{r}} \mathds{1}^{r}_{i} \mathbf{w}^{r,E}_{i} }{ {\sum}_{{i} \in {\mathcal{K}}_{r}} \mathds{1}^{r}_{i}}
\bigg|
\sum_{{i} \in {\mathcal{K}}_{r}} \mathds{1}^{r}_{i} \neq 0
\bigg]
\Bigg]
\notag \\
\overset{(a)}{=} &
{\mathbb{E}}_{\mathcal{K}_{r}}
\Bigg[
\sum_{k=1}^K
\sum_{\mathcal{S}_r \bigcup {\bar{\mathcal{S}}}_r = \mathcal{K}_{r}, \atop |\mathcal{S}_r|=k, |{\bar{\mathcal{S}}}_r|=K-k}
\Pr
\bigg(
\mathds{1}^{r}_{i_1}  =  1 \, \forall i_1  \in  \mathcal{S}_r, \mathds{1}^{r}_{i_2}  =  0 \, \forall i_2  \in  {\bar{\mathcal{S}}}_r
\bigg|
\sum_{{i} \in {\mathcal{K}}_{r}} \mathds{1}^{r}_{i}  \neq  0
\bigg)
\cdot  \frac{ {\sum}_{i_1 \in \mathcal{S}_r}  \mathbf{w}^{r,E}_{i} }{k}
\Bigg]
\notag \\
\overset{(b)}{=} &
\sum_{g=1}^{N^K}
\Bigg(
\prod_{{i} \in {\mathcal{K}}_r^g}  s_{i}
\Bigg)
\cdot
\Bigg(
\sum_{k=1}^K
\sum_{\mathcal{S}_r^g \bigcup {\bar{\mathcal{S}}}_r^g = {\mathcal{K}}_r^g, \atop |\mathcal{S}_r^g|=k, |{\bar{\mathcal{S}}}_r^g|=K-k}
\frac{ {\prod}_{i_1 \in \mathcal{S}_r^g}  (1  -  \epsilon_{{i_1}})
 {\prod}_{i_2 \in {\bar{\mathcal{S}}}_r^g}  \epsilon_{{i_2}}}{1- {\prod}_{{i} \in {\mathcal{K}}_r^g} \epsilon_{i}}  \cdot  \frac{{\sum}_{i_1 \in \mathcal{S}_r^g} \mathbf{w}^{r,E}_{i} }{k}
\Bigg)
\notag \\
\triangleq &
\sum_{i=1}^N {\bar \beta}_i \mathbf{w}^{r,E}_{i}
,
\end{align}
\end{small}
\vspace{-0.3cm}

\noindent
where in equality (a), $\mathcal{S}_r$ represents the subset of clients in $\mathcal{K}_r$ with successful transmission of local models to the server, and ${\bar{\mathcal{S}}}_r$ denotes the clients with transmission failure due to failures.
In equality (b), $\prod_{i_1 \in \mathcal{S}_r^g} (1-\epsilon_{{i_1}}) \prod_{i_2 \in {\bar{\mathcal{S}}}_r^g} \epsilon_{{i_2}} $ is the probability that clients in $\mathcal{S}_r^g$ have successful transmissions, while those in ${\bar{\mathcal{S}}}_r^g$ experience transmission failures.
Using the derivations in \eqref{beta_proof}, we obtain \eqref{average_beta} for some non-negative $\bar \beta_i$, $i \in [N]$.
\hfill $\blacksquare$

\section{Proof of Theorem \ref{theorem:convergence FL failure}}\label{appendix:proof theorem convergence FL failure}

Our analysis considers only the ``successful'' communication rounds, where the server correctly receives updated local models from at least one client in $\mathcal{K}_r$.
Accordingly, all derivations are based on the condition that $\sum_{{i} \in \mathcal{K}_{r}} \mathds{1}^{r}_{i} \neq 0$, $\forall r \in [R]$.
For simplicity, in the subsequent proof, we denote the conditional expectation $\mathbb{E}[ \, \cdot \,  | \sum_{{i} \in \mathcal{K}_{r}} \mathds{1}^{r}_{i} \neq 0]$ as $\mathbb{E}[ \cdot ]$.

\subsection{Proof of convergence rate}

With Assumption \ref{assumption:L continuous}, we have
\begin{small}
\begin{align}\label{ineq:F_w_r_bound}
\mathbb{E}[F({\bar {\mathbf{w}}}_{r})]
\leq &
\mathbb{E}[F({\bar {\mathbf{w}}}_{r-1})]
+ \mathbb{E}\left[\langle\nabla  F({\bar {\mathbf{w}}}_{r-1}), {\bar {\mathbf{w}}}_{r} - {\bar {\mathbf{w}}}_{r-1}\rangle\right]
+ \frac{L}{2} \mathbb{E}\left[ \| {\bar {\mathbf{w}}}_{r} - {\bar {\mathbf{w}}}_{r-1}  \|^2 \right]
.
\end{align}
\end{small}
\vspace{-0.3cm}

We need the following three key lemmas which are proved in subsequent subsections.

\begin{lemma}\label{lemma:bound nabla F wr wr-1}
Under Assumptions \ref{assumption:class heterogeneity}, \ref{assumption:garident norm} and \ref{assumption:L continuous}, it holds that
\begin{small}
\begin{align}\label{ineq:bound nabla F wr wr-1}
& \mathbb{E}\left[ \left\langle\nabla  F({\bar {\mathbf{w}}}_{r-1}), {\bar {\mathbf{w}}}_{r} - {\bar {\mathbf{w}}}_{r-1} \right\rangle \right]
\notag \\
\leq
& -\frac{\gamma E}{2} \left\| \nabla  F({\bar {\mathbf{w}}}_{r-1}) \right\|^2
+ \gamma L^2 {\sum}_{i = 1}^N  {\bar \beta}_i {\sum}_{t = 2}^{E} \left\| \mathbf{w}^{r,t-1}_{i} - \mathbf{\bar w}_{r-1} \right\|^2
+ 2\gamma E
\Big( \chi^2_{\bm{\bar \beta}\|\mathbf{p}} {\sum}_{c=1}^C {\sum}_{i = 1}^N p_i \alpha_{i,c} V_{i,c}^2
+ \chi^2_{\bm{\bar \alpha} \| \bm{\alpha}_g} G^2 \Big)
,
\end{align}
\end{small}
\vspace{-0.3cm}

\noindent
where $\chi^2_{\bm{\bar \beta}\|\mathbf{p}}$ and $\chi^2_{\bm{\bar \alpha} \| \bm{\alpha}_g}$ are chi-square divergences defined in \eqref{eq:convergence bound of FL failure}.
\end{lemma}

\begin{lemma}\label{lemma:Proof_Thm1_lemma_2}
With \eqref{eq:local updating} and \eqref{global_model_failure}, the difference between the global models of two consecutive iterations is given by
\begin{small}
\begin{equation}\label{eq:bar wr - wr-1}
{\bar {\mathbf{w}}}_{r} - {\bar {\mathbf{w}}}_{r-1}
=
- \gamma \frac{\sum_{{i} \in \mathcal{K}_{r}} {\mathds{1}}^{r}_{i} \sum_{t = 1}^{E} \nabla F_{i} (\mathbf{w}^{r,t-1}_{i}) }{\sum_{{i} \in \mathcal{K}_{r}} {\mathds{1}}^{r}_{i}}
,
\end{equation}
\end{small}
\vspace{-0.3cm}

\noindent
which results in
\begin{small}
\begin{align}\label{ineq:Proof_Thm1_lemma_2_formulation}
&
\mathbb{E} \left[  \| {\bar {\mathbf{w}}}_{r} - {\bar {\mathbf{w}}}_{r-1}  \|^2 \right]
\notag \\
\leq &
4 \gamma^2 E^2 \left \| \nabla F ({\bar {\mathbf{w}}}_{r-1}) \right\|^2
+ 2 \gamma^2 E L^2 {\sum}_{i = 1}^{N} {\bar \beta}_i {\sum}_{t = 2}^{E} \| \mathbf{w}^{r,t-1}_{i} - {\bar {\mathbf{w}}}_{r-1} \|^2
+ 8 \gamma^2 E^2 {\sum}_{i = 1}^{N} {\bar \beta}_i
{\sum}_{c=1}^C
\big( \alpha_{i,c} V_{i,c}^2 + \chi^2_{\bm{\alpha}_i \| \bm{\alpha}_g} G^2 \big)
.
\end{align}
\end{small}
\end{lemma}

\begin{lemma}\label{Proof_Thm1_lemma_3}
The difference between the local model at each round $r$ and the global model from the previous round is bounded by
\begin{small}
\begin{align}\label{Proof_Thm1_lemma_3_formulation}
{\sum}_{t = 2}^{E} \left\| \mathbf{w}^{r,t-1}_{i} - {\bar {\mathbf{w}}}_{r-1} \right\|^2
\leq
\frac{\gamma^2 E^3}{1 - \frac{3}{2} \gamma^2 E^2 L^2} \left \| \nabla F(\mathbf{\bar w}_{r-1}) \right\|^2
+ \frac{2 \gamma^2 E^3}{1 - \frac{3}{2} \gamma^2 E^2 L^2}
{\sum}_{c=1}^C
\big( \alpha_{i,c} V_{i,c}^2 + \chi^2_{\bm{\alpha}_i \| \bm{\alpha}_g} G^2 \big)
.
\end{align}
\end{small}
\end{lemma}

By substituting \eqref{ineq:bound nabla F wr wr-1} into the second term on the RHS of \eqref{ineq:F_w_r_bound}, \eqref{ineq:Proof_Thm1_lemma_2_formulation} into the third term, along with the result from \eqref{Proof_Thm1_lemma_3_formulation}, we have
\begin{small}
\begin{align}
\mathbb{E}[F({\bar {\mathbf{w}}}_{r})]
\leq &
\mathbb{E}[F({\bar {\mathbf{w}}}_{r-1})]
- \Big( \frac{\gamma E}{2} - 2 \gamma^2 E^2 L - \frac{\gamma^3 E^3 L^2 + \gamma^4 E^4 L^3}{1 - \frac{3}{2} \gamma^2 E^2 L^2} \Big)
\left\| \nabla  F({\bar {\mathbf{w}}}_{r-1}) \right\|^2
\notag \\
& + \Big( 4 \gamma^2 E^2 L + \frac{2\gamma^3 E^3 L^2 + 2\gamma^4 E^4 L^3}{1 - \frac{3}{2} \gamma^2 E^2 L^2} \Big)
{\sum}_{i = 1}^{N} {\bar \beta}_i
{\sum}_{c=1}^C
\big( \alpha_{i,c} V_{i,c}^2 + \chi^2_{\bm{\alpha}_i \| \bm{\alpha}_g} G^2 \big)
\notag \\
&
+ 2 \gamma E \Big( \chi^2_{\bm{\bar \beta}\|\mathbf{p}}
{\sum}_{c=1}^C {\sum}_{i = 1}^N p_i \alpha_{i,c} V_{i,c}^2
+ \chi^2_{\bm{\bar \alpha} \| \bm{\alpha}_g} G^2 \Big)
.
\end{align}
\end{small}
\vspace{-0.3cm}

Next, summing the above items from ${r} = 1$ to $R$ and dividing both sides by the product of the learning rate and the total number of local gradient descent steps, $\gamma T ( = \gamma RE)$, yields
\begin{small}
\begin{align}\label{Proof_Thm1_lemma_4_formulation}
&
\Big(
\underbrace{
\frac{1}{2} - 2 \gamma E L - \frac{\gamma^2 E^2 L^2 + \gamma^3 E^3 L^3}{1 - \frac{3}{2} \gamma^2 E^2 L^2}
}_{\rm \triangleq (\ref{Proof_Thm1_lemma_4_formulation}a)}
\Big)
\frac{\sum_{r=1}^R \mathbb{E}[ \left\| \nabla  F({\bar {\mathbf{w}}}_{r-1}) \right\|^2 ]}{R}
\notag \\
\leq &
\underbrace{ \frac{1}{\gamma T} }_{\rm \triangleq (\ref{Proof_Thm1_lemma_4_formulation}b)}
\left( \mathbb{E}[F({\bar {\mathbf{w}}}_{0})]
- \mathbb{E}[F({\bar {\mathbf{w}}}_{R})] \right)
+ \Big( \underbrace{4 \gamma E L}_{\rm \triangleq (\ref{Proof_Thm1_lemma_4_formulation}c)} + \underbrace{\frac{2\gamma^2 E^2 L^2 + 2\gamma^3 E^3 L^3}{1 - \frac{3}{2} \gamma^2 E^2 L^2}}_{\rm \triangleq (\ref{Proof_Thm1_lemma_4_formulation}d)} \Big)
{\sum}_{i = 1}^{N} {\bar \beta}_i
{\sum}_{c=1}^C
\big( \alpha_{i,c} V_{i,c}^2 + \chi^2_{\bm{\alpha}_i \| \bm{\alpha}_g} G^2 \big)
\notag \\
&
+ 2 \chi^2_{\bm{\bar \beta}\|\mathbf{p}} {\sum}_{c=1}^C {\sum}_{i = 1}^N p_i \alpha_{i,c} V_{i,c}^2
+ 2 \chi^2_{\bm{\bar \alpha} \| \bm{\alpha}_g} G^2
.
\end{align}
\end{small}
\vspace{-0.3cm}

Let the learning rate $\gamma = K^{\frac{1}{2}} / (6 L{T}^{\frac{1}{2}}) $ and the number of local updating steps $E \leq T^{\frac{1}{4}}/K^{\frac{3}{4}}$, where $T \geq K^{3}$ to ensure $E \geq 1$.
Consequently, we have ${\rm (\ref{Proof_Thm1_lemma_4_formulation}b)} = 6L(TK)^{-\frac{1}{2}}$.
Since $\gamma E L \leq (T {\bar K})^{-\frac{1}{4}}/6$, we have ${\rm (\ref{Proof_Thm1_lemma_4_formulation}c)} \leq \frac{2}{3} (TK)^{-\frac{1}{4}}$ and

\vspace{-0.3cm}
\begin{small}
\begin{align}
{\rm (\ref{Proof_Thm1_lemma_4_formulation}d)}
\leq
\frac{\frac{1}{18}(TK)^{-\frac{1}{2}}
+ \frac{1}{108}(TK)^{-\frac{3}{4}} }{1 - \frac{1}{24} (TK)^{-\frac{1}{2}}}
\overset{(a)}{\leq}
\frac{\frac{1}{18}(TK)^{-\frac{1}{2}}
+ \frac{1}{108}(TK)^{-\frac{3}{4}} }{1 - \frac{1}{24}}
=
\frac{4}{69(TK)^{\frac{1}{2}}}
+ \frac{2}{207(TK)^{\frac{3}{4}}}
,
\end{align}
\end{small}
\vspace{-0.1cm}

\noindent
where inequality (a) is due to $TK \geq 1$.
Then,
\begin{small}
\begin{align}
{\rm (\ref{Proof_Thm1_lemma_4_formulation}a)}
= \frac{1}{2} - \frac{\rm (\ref{Proof_Thm1_lemma_4_formulation}c)}{2} - \frac{\rm (\ref{Proof_Thm1_lemma_4_formulation}d)}{2}
\geq
\frac{1}{2}
\Big(
1 -  \frac{2}{3 ( T {\bar K})^{\frac{1}{4}}}
 -  \frac{4}{69 ( T {\bar K})^{\frac{1}{2}}}
 -  \frac{2}{207 ( T {\bar K})^{\frac{3}{4}}}
\Big)
\geq
\frac{1}{2}
\Big(
1 -  \frac{2}{3}
 -  \frac{4}{69}
 -  \frac{2}{207}
\Big)
=
\frac{55}{414}
.
\end{align}
\end{small}
\vspace{-0.3cm}

Finally, by substituting above coefficients and $\mathbb{E}[F({\bar {\mathbf{w}}}_{R})] \geq \underline{F}$ from Assumption \ref{assumption:Fi lower bound} into \eqref{Proof_Thm1_lemma_4_formulation}, Theorem \ref{theorem:convergence FL failure} is proved.
\hfill $\blacksquare$

\subsection{Proof of Lemma \ref{lemma:bound nabla F wr wr-1}}

Based on \eqref{eq:bar wr - wr-1}, we have
\begin{small}
\begin{align}\label{eq:bound nabla F wr wr-1}
& \mathbb{E}\left[ \left\langle \nabla F({\bar {\mathbf{w}}}_{r-1}), {\bar {\mathbf{w}}}_{r} - {\bar {\mathbf{w}}}_{r-1} \right\rangle \right]
\notag \\
=
& \mathbb{E}\Big[ \Big\langle \nabla  F({\bar {\mathbf{w}}}_{r-1}),- \gamma \frac{\sum_{{i} \in \mathcal{K}_{r}} {\mathds{1}}^{r}_{i} \sum_{t = 1}^{E} \nabla F_{i} (\mathbf{w}^{r,t-1}_{i}) }{\sum_{{i} \in \mathcal{K}_{r}} {\mathds{1}}^{r}_{i}} \Big \rangle\Big]
\notag \\
\overset{(a)}{=}
&
- \gamma {\sum}_{t = 1}^{E} \Big\langle\nabla  F({\bar {\mathbf{w}}}_{r-1}),
{\sum}_{i = 1}^N {\bar \beta}_i \nabla F_{i} (\mathbf{w}^{r,t-1}_{i}) \Big\rangle
\notag \\
\overset{(b)}{=}
&
- \frac{\gamma}{2}  {\sum}_{t = 1}^{E} \| \nabla  F({\bar {\mathbf{w}}}_{r-1}) \|^2
 -  \frac{\gamma}{2}  {\sum}_{t = 1}^{E}  \left\|  {\sum}_{i = 1}^N {\bar \beta}_i \nabla  F_{i} (\mathbf{w}^{r,t-1}_{i}) \right\|^2
+ \frac{\gamma}{2} {\sum}_{t = 1}^{E} \left\| \nabla  F({\bar {\mathbf{w}}}_{r-1}) -  {\sum}_{i = 1}^N {\bar \beta}_i \nabla F_{i} (\mathbf{w}^{r,t-1}_{i}) \right\|^2
\notag \\
\leq
&
-\frac{\gamma E}{2} \left\| \nabla  F({\bar {\mathbf{w}}}_{r-1}) \right\|^2
+ \frac{\gamma}{2} {\sum}_{t = 1}^{E} \left\| \nabla  F({\bar {\mathbf{w}}}_{r-1}) -  {\sum}_{i = 1}^N {\bar \beta}_i \nabla F_{i} (\mathbf{w}^{r,t-1}_{i}) \right\|^2
\notag \\
\overset{(c)}{\leq}
&
-\frac{\gamma E}{2} \left\| \nabla  F({\bar {\mathbf{w}}}_{r-1}) \right\|^2
+ \gamma E
\underbrace{
\left\| \nabla  F({\bar {\mathbf{w}}}_{r-1}) - {\sum}_{i = 1}^N  {\bar \beta}_i \nabla F_i({\bar {\mathbf{w}}}_{r-1}) \right\|^2
}_{\rm \triangleq (\ref{eq:bound nabla F wr wr-1}d)}
+ \gamma \underbrace{{\sum}_{t = 1}^{E} \left\| {\sum}_{i = 1}^N {\bar \beta}_i (\nabla F_i({\bar {\mathbf{w}}}_{r-1}) -  \nabla F_i(\mathbf{w}^{r,t-1}_{i}) ) \right\|^2
}_{\rm \triangleq (\ref{eq:bound nabla F wr wr-1}e)}
,
\end{align}
\end{small}
\vspace{-0.3cm}

\noindent
where equality (a) is obtained using Lemma \ref{lemma:beta},
equality (b) follows from the identity $\langle {\mathbf{x}}_1,{\mathbf{x}}_2 \rangle = \frac{1}{2}( \| {\mathbf{x}}_1 \|^2 + \| {\mathbf{x}}_2 \|^2 - \| {\mathbf{x}}_1 - {\mathbf{x}}_2 \|^2 )$,
and inequality (c) arises from the property $\|x_1+x_2\|^2 \leq 2\|x_1\|^2 + 2\|x_2\|^2$.

In \eqref{eq:bound nabla F wr wr-1}, the term (\ref{eq:bound nabla F wr wr-1}d) can be further bounded as
\begin{small}
\begin{align}\label{eq:bound B1}
{\rm (\ref{eq:bound nabla F wr wr-1}d)}
\overset{(a)}{=} &
\left\| {\sum}_{c=1}^C {\sum}_{i = 1}^N ( p_i \alpha_{i,c} - {\bar \beta}_i \alpha_{i,c} ) \nabla F_{i,c} ({\bar {\mathbf{w}}}_{r-1}) \right\|^2
\notag \\
=
& \left\| {\sum}_{c=1}^C {\sum}_{i = 1}^N ( p_i \alpha_{i,c} - {\bar \beta}_i \alpha_{i,c} ) \nabla F_{i,c} ({\bar {\mathbf{w}}}_{r-1}) \right.
- {\sum}_{c=1}^C {\sum}_{i = 1}^N ( p_i \alpha_{i,c} - {\bar \beta}_i \alpha_{i,c} ) \nabla F_{c} ({\bar {\mathbf{w}}}_{r-1})
\notag \\
& \;\;
\left.
+ {\sum}_{c=1}^C {\sum}_{i = 1}^N ( p_i \alpha_{i,c} - {\bar \beta}_i \alpha_{i,c} ) \nabla F_{c} ({\bar {\mathbf{w}}}_{r-1}) \right\|^2
\notag \\
\overset{(b)}{\leq} &
2 \, \bigg\| \sum_{c=1}^C \sum_{i = 1}^N ( p_i \alpha_{i,c}  -  {\bar \beta}_i \alpha_{i,c} ) \left( \nabla F_{i,c} ({\bar {\mathbf{w}}}_{r-1})  -  \nabla F_{c} ({\bar {\mathbf{w}}}_{r-1}) \right) \bigg\|^2
+ 2 \left\| {\sum}_{c=1}^C \Big( \alpha_{g,c} - {\sum}_{i = 1}^N {\bar \beta}_i \alpha_{i,c} \Big) \nabla F_{c} ({\bar {\mathbf{w}}}_{r-1})
\right\|^2
\notag \\
= &
2 \Big\| {\sum}_{c=1}^C {\sum}_{i = 1}^N \frac{p_i \alpha_{i,c} - {\bar \beta}_i \alpha_{i,c}}{\sqrt{p_i \alpha_{i,c}}} \sqrt{p_i \alpha_{i,c}}
\left( \nabla F_{i,c} ({\bar {\mathbf{w}}}_{r-1}) - \nabla F_{c} ({\bar {\mathbf{w}}}_{r-1}) \right) \Big\|^2
+ 2
\Big\| {\sum}_{c=1}^C \frac{\alpha_{g,c} - \sum_{i = 1}^N {\bar \beta}_i \alpha_{i,c}}{\sqrt{\alpha_{g,c}}} \sqrt{\alpha_{g,c}} \nabla F_{c} ({\bar {\mathbf{w}}}_{r-1})
\Big\|^2
\notag \\
\overset{(c)}{\leq}
& 2 \underbrace{{\sum}_{c=1}^C {\sum}_{i = 1}^N \frac{(p_i \alpha_{i,c} - {\bar \beta}_i \alpha_{i,c})^2}{p_i \alpha_{i,c}}}_{\rm \triangleq (\ref{eq:bound B1}e)}
{\sum}_{c=1}^C {\sum}_{i = 1}^N p_i \alpha_{i,c} \left\| \nabla F_{i,c} ({\bar {\mathbf{w}}}_{r-1}) - \nabla F_{c} ({\bar {\mathbf{w}}}_{r-1}) \right\|^2
\notag \\
&
 +  2 {\sum}_{c=1}^C  \frac{(\alpha_{g,c}  -  \sum_{i = 1}^N  {\bar \beta}_i \alpha_{i,c})^2}{\alpha_{g,c}}
{\sum}_{c=1}^C  \alpha_{g,c}  \left\| \nabla F_{c} ({\bar {\mathbf{w}}}_{r-1}) \right\|^2
\notag \\
\overset{(d)}{\leq}
& 2 \chi^2_{\bm{\bar \beta}\|\mathbf{p}} {\sum}_{c=1}^C {\sum}_{i = 1}^N p_i \alpha_{i,c} V_{i,c}^2
+ 2 \chi^2_{\bm{\bar \alpha} \| \bm{\alpha}_g} G^2
,
\end{align}
\end{small}
\vspace{-0.3cm}

\noindent
where equality (a) follows from \eqref{eq:nabla Fi alpha c},
inequality (b) is a consequence of the global label distribution $\alpha_{g,c} = {\sum}_{i = 1}^N p_i \alpha_{i,c}$ and the relation $\|x_1+x_2\|^2 \leq 2\|x_1\|^2 + 2\|x_2\|^2$,
and equality (c) is established using the Cauchy-Schwarz Inequality.
In inequality (c), the term (\ref{eq:bound B1}e) is bounded by
\begin{small}
\begin{align}\label{ref:lemma proof term chi beta p}
{\rm (\ref{eq:bound B1}e)}
=
{\sum}_{c=1}^C {\sum}_{i = 1}^N \frac{(p_i - {\bar \beta}_i)^2 \alpha_{i,c}}{p_i}
=
 \underbrace{{\sum}_{i = 1}^N \frac{(p_i - {\bar \beta}_i)^2}{p_i}}_{\triangleq \chi^2_{\bm{\bar \beta}\|\mathbf{p}}}
\underbrace{{\sum}_{c=1}^C \alpha_{i,c}}_{= 1}
=
\chi^2_{\bm{\bar \beta}\|\mathbf{p}}
.
\end{align}
\end{small}
\vspace{-0.3cm}

\noindent
Combining this with Assumptions \ref{assumption:class heterogeneity} and \ref{assumption:garident norm}, along with the definition of the chi-square divergence $\chi^2_{\bm{\bar \alpha} \| \bm{\alpha}_g}$ in \eqref{eq:convergence bound of FL failure}, leads to inequality (d) in \eqref{eq:bound B1}.

Meanwhile, the term (\ref{eq:bound nabla F wr wr-1}e) is bounded by
\begin{small}
\begin{align}\label{eq:bound B2}
{\rm (\ref{eq:bound nabla F wr wr-1}e)}
\overset{(a)}{\leq}
&
{\sum}_{i = 1}^N  {\bar \beta}_i {\sum}_{t = 1}^{E} \big\| \nabla F_i({\bar {\mathbf{w}}}_{r-1}) -  \nabla F_i(\mathbf{w}^{r,t-1}_{i}) \big\|^2
\notag \\
\overset{(b)}{\leq}
&
L^2 {\sum}_{i = 1}^N  {\bar \beta}_i {\sum}_{t = 1}^{E} \big\| \mathbf{w}^{r,t-1}_{i} - \mathbf{\bar w}_{r-1} \big\|^2
\overset{(c)}{=}
L^2 {\sum}_{i = 1}^N  {\bar \beta}_i {\sum}_{t = 2}^{E} \big\| \mathbf{w}^{r,t-1}_{i} - \mathbf{\bar w}_{r-1} \big\|^2
,
\end{align}
\end{small}
\vspace{-0.3cm}

\noindent
where inequality (a) is derived from Jensen's Inequality, inequality (b) is due to Assumption \ref{assumption:L continuous}, and equality (c) follows from \eqref{eq:local SGD_a}.

Finally, by substituting \eqref{eq:bound B1} and \eqref{eq:bound B2} into \eqref{eq:bound nabla F wr wr-1}, we can directly obtain Lemma \ref{lemma:bound nabla F wr wr-1}.
\hfill $\blacksquare$

\subsection{Proof of Lemma \ref{lemma:Proof_Thm1_lemma_2}}

According to \eqref{eq:bar wr - wr-1}, we have
\begin{small}
\begin{align}\label{Proof_Thm1_lemma_2_step1}
\mathbb{E}\left[ \| {\bar {\mathbf{w}}}_{r} - {\bar {\mathbf{w}}}_{r-1} \|^2 \right]
= &
\gamma^2 \mathbb{E} \bigg[ \bigg\| \frac{ {\sum}_{{i} \in \mathcal{K}_{r}} {\mathds{1}}^{r}_{i} {\sum}_{t = 1}^{E} \nabla F_{i} (\mathbf{w}^{r,t-1}_{i}) }{{\sum}_{{i} \in \mathcal{K}_{r}} {\mathds{1}}^{r}_{i}} \bigg\|^2 \bigg]
\notag \\
\overset{(a)}{\leq} &
2 \gamma^2 \underbrace{ \mathbb{E} \bigg[ \bigg\|
\frac{ {\sum}_{{i} \in \mathcal{K}_{r}} {\mathds{1}}^{r}_{i}  {\sum}_{t = 1}^{E} ( \nabla F_{i} (\mathbf{w}^{r,t-1}_{i}) - \nabla F_{i} ({\bar {\mathbf{w}}}_{r-1}) ) }{{\sum}_{{i} \in \mathcal{K}_{r}} {\mathds{1}}^{r}_{i}}
\bigg\|^2 \bigg]  }_{\rm \triangleq (\ref{Proof_Thm1_lemma_2_step1}b)}
+ 2 \gamma^2 \underbrace{ \mathbb{E} \bigg[ \bigg\| \frac{ {\sum}_{{i} \in \mathcal{K}_{r}} {\mathds{1}}^{r}_{i} {\sum}_{t = 1}^{E} \nabla F_{i} ({\bar {\mathbf{w}}}_{r-1}) }{\sum_{{i} \in \mathcal{K}_{r}} {\mathds{1}}^{r}_{i}} \bigg\|^2 \bigg]
}_{\rm \triangleq (\ref{Proof_Thm1_lemma_2_step1}c)}
,
\end{align}
\end{small}
\vspace{-0.3cm}

\noindent
where inequality (a) is derived from the relation $\|x_1+x_2\|^2 \leq 2\|x_1\|^2 + 2\|x_2\|^2$.

The term (\ref{Proof_Thm1_lemma_2_step1}b) is bounded by
\begin{small}
\begin{align}\label{bound:B_3}
{\rm (\ref{Proof_Thm1_lemma_2_step1}b)}
\leq
& E \cdot \mathbb{E}  \bigg[
\frac{ {\sum}_{{i} \in \mathcal{K}_{r}} {\mathds{1}}^{r}_{i} {\sum}_{t = 1}^{E}  \| \nabla F_{i} (\mathbf{w}^{r,t-1}_{i})  - \nabla F_{i} ({\bar {\mathbf{w}}}_{r-1}) \|^2 }{{\sum}_{{i} \in \mathcal{K}_{r}} {\mathds{1}}^{r}_{i}}  \bigg]
\notag \\
\overset{(a)}{=} &
E {\sum}_{i = 1}^{N} {\bar \beta}_i {\sum}_{t = 1}^{E} \left\| \nabla F_{i} (\mathbf{w}^{r,t-1}_{i}) - \nabla F_{i} ({\bar {\mathbf{w}}}_{r-1}) \right\|^2
\overset{(b)}{\leq} 
E L^2 {\sum}_{i = 1}^{N} {\bar \beta}_i {\sum}_{t = 2}^{E} \| \mathbf{w}^{r,t-1}_{i} - {\bar {\mathbf{w}}}_{r-1} \|^2
,
\end{align}
\end{small}
\vspace{-0.3cm}

\noindent
where equality (a) follows from Lemma \ref{lemma:beta}, and inequality (b) is based on Assumption \ref{assumption:L continuous} and \eqref{eq:local SGD_a}.
Meanwhile, term (\ref{Proof_Thm1_lemma_2_step1}c) is bounded by
\begin{small}
\begin{align}\label{bound:B_4}
{\rm (\ref{Proof_Thm1_lemma_2_step1}c)}
=
& E^2 \mathbb{E} \Bigg[ \Bigg \| \frac{ {\sum}_{{i} \in \mathcal{K}_{r}} {\mathds{1}}^{r}_{i} \nabla F_{i} ({\bar {\mathbf{w}}}_{r-1}) }{\sum_{{i} \in \mathcal{K}_{r}} {\mathds{1}}^{r}_{i}} \Bigg\|^2 \Bigg]
\notag \\
\overset{(a)}{\leq} &
2E^2 \mathbb{E} \bigg[ \bigg \| \frac{ {\sum}_{{i} \in \mathcal{K}_{r}} {\mathds{1}}^{r}_{i} \left( \nabla F_{i} ({\bar {\mathbf{w}}}_{r-1}) - \nabla F ({\bar {\mathbf{w}}}_{r-1}) \right) }{\sum_{{i} \in \mathcal{K}_{r}} {\mathds{1}}^{r}_{i}} \bigg\|^2 \bigg]
+ 2E^2 \mathbb{E} \left[ \left\| \nabla F ({\bar {\mathbf{w}}}_{r-1}) \right\|^2 \right]
\notag \\
\overset{(b)}{\leq} &
2E^2 \mathbb{E} \bigg[ \frac{ {\sum}_{{i} \in \mathcal{K}_{r}} {\mathds{1}}^{r}_{i} \left \| \nabla F_{i} ({\bar {\mathbf{w}}}_{r-1}) - \nabla F ({\bar {\mathbf{w}}}_{r-1}) \right\|^2 }{\sum_{{i} \in \mathcal{K}_{r}} {\mathds{1}}^{r}_{i}} \bigg]
+ 2E^2 \mathbb{E} \left[ \left\| \nabla F ({\bar {\mathbf{w}}}_{r-1}) \right\|^2 \right]
\notag \\
\overset{(c)}{=} &
2E^2 {\sum}_{i = 1}^{N} {\bar \beta}_i \left \| \nabla F_{i} ({\bar {\mathbf{w}}}_{r-1}) - \nabla F ({\bar {\mathbf{w}}}_{r-1}) \right\|^2
+ 2E^2 \mathbb{E} \left[ \left\| \nabla F ({\bar {\mathbf{w}}}_{r-1}) \right\|^2 \right]
\notag \\
\overset{(d)}{\leq} &
4E^2 {\sum}_{i = 1}^{N} {\bar \beta}_i {\sum}_{c=1}^C \left( \alpha_{i,c} V_{i,c}^2 + \chi^2_{\bm{\alpha}_i \| \bm{\alpha}_g} G^2 \right)
+ 2E^2 \mathbb{E} \left[ \left\| \nabla F ({\bar {\mathbf{w}}}_{r-1}) \right\|^2 \right]
,
\end{align}
\end{small}
\vspace{-0.3cm}

\noindent
where
inequality (a) also relies on the relation $\|x_1+x_2\|^2 \leq 2\|x_1\|^2 + 2\|x_2\|^2$,
inequality (b) is a consequence of Jensen's Inequality,
equality (c) is derived from Lemma \ref{lemma:beta},
and inequality (d) follows from \eqref{ineq:nabla Fi - F 2}.

Finally, by substituting \eqref{bound:B_3} and \eqref{bound:B_4} into \eqref{Proof_Thm1_lemma_2_step1}, we obtain Lemma \ref{lemma:Proof_Thm1_lemma_2}.

\hfill $\blacksquare$

\subsection{Proof of Lemma \ref{Proof_Thm1_lemma_3}}

Based on \eqref{eq:local updating}, the local model for each client at the $r$-th iteration is updated by
\begin{small}
\begin{align}
\mathbf{w}_{i}^{r,t-1}
= \mathbf{\bar w}_{r-1} - \gamma {\sum}_{e = 1}^{t-1} \nabla F_i (\mathbf{w}_{i}^{r,e-1})
.
\end{align}
\end{small}
\vspace{-0.3cm}

\noindent
Consequently, the difference between the local and global models is bounded by
\begin{small}
\begin{align}\label{w_i-w_bar}
& {\sum}_{t = 2}^{E} \left\| \mathbf{w}^{r,t-1}_{i} - \mathbf{\bar w}_{r-1} \right\|^2
=
{\sum}_{t = 2}^{E} \left\| \gamma {\sum}_{e = 1}^{t-1} \nabla F_i (\mathbf{w}_{i}^{r,e-1}) \right\|^2
\notag  \\
\leq
& \gamma^2 {\sum}_{t = 2}^{E} (t - 1) {\sum}_{e = 1}^{t-1} \left \| \nabla F_i (\mathbf{w}_{i}^{r,e-1}) \right\|^2
\notag  \\
\leq
& 3 \gamma^2 {\sum}_{t = 2}^{E} (t - 1) {\sum}_{e = 1}^{t-1}
\Big( \left\| \nabla F_i (\mathbf{w}_{i}^{r,e-1})  -  \nabla F_i (\mathbf{\bar w}_{r-1}) \right\|^2
+ \left \| \nabla F_i (\mathbf{\bar w}_{r-1})  -  \nabla F(\mathbf{\bar w}_{r-1}) \right\|^2
+ \left \| \nabla F (\mathbf{\bar w}_{r-1}) \right\|^2  \Big) \notag \\
\overset{(a)}{\leq}
&
3 \gamma^2 L^2 {\sum}_{t = 2}^{E} (t - 1) {\sum}_{e = 1}^{t-1} \left \| \mathbf{w}_{i}^{r,e-1} - \mathbf{\bar w}_{r-1} \right\|^2
+ 6 \gamma^2 {\sum}_{t = 2}^{E} (t - 1) {\sum}_{e = 1}^{t-1} {\sum}_{c=1}^C \big( \alpha_{i,c} V_{i,c}^2 + \chi^2_{\bm{\alpha}_i \| \bm{\alpha}_g} G^2 \big)
\notag  \\
&
+ 3 \gamma^2 {\sum}_{t = 2}^{E} (t - 1) {\sum}_{e = 1}^{t-1} \left \| \nabla F(\mathbf{\bar w}_{r-1}) \right\|^2
\notag \\
=
&
3 \gamma^2 L^2 \underbrace{
{\sum}_{t = 2}^{E} (t - 1) {\sum}_{e = 1}^{t-1} \left \| \mathbf{w}_{i}^{r,e-1} - \mathbf{\bar w}_{r-1} \right\|^2
}_{\rm \triangleq (\ref{w_i-w_bar}b)}
+ 6 \gamma^2 \underbrace{
{\sum}_{t = 2}^{E} (t - 1)^2
}_{\rm \triangleq (\ref{w_i-w_bar}c)}
{\sum}_{c=1}^C \big( \alpha_{i,c} V_{i,c}^2 + \chi^2_{\bm{\alpha}_i \| \bm{\alpha}_g} G^2 \big)
\notag \\
&
+ 3 \gamma^2
\underbrace{
{\sum}_{t = 2}^{E} (t - 1)^2
}_{\rm \triangleq (\ref{w_i-w_bar}c)}
\left \| \nabla F(\mathbf{\bar w}_{r-1}) \right\|^2
,
\end{align}
\end{small}
\vspace{-0.3cm}

\noindent
where inequality (a) follows from Assumption \ref{assumption:L continuous} and \eqref{ineq:nabla Fi - F 2}.

The term (\ref{w_i-w_bar}b) is bounded by
\begin{small}
\begin{align}\label{Proof_Thm1_lemma_3_step2}
{\rm (\ref{w_i-w_bar}b)}
\overset{(a)}{=} &
{\sum}_{t = 2}^{E} (t - 1) {\sum}_{e = 2}^{t-1} \left\| \mathbf{w}_{i}^{r,e-1} - \mathbf{\bar w}_{r-1} \right\|^2
\notag \\
\overset{(b)}{=} &
{\sum}_{t = 2}^{E-1} {\sum}_{e = t+1}^{E} (e - 1) \left\| \mathbf{w}_{i}^{r,t-1} - \mathbf{\bar w}_{r-1} \right\|^2
=
{\sum}_{t = 2}^{E-1} \left( {\sum}_{e = t}^{E-1} e \cdot \left\| \mathbf{w}_{i}^{r,t-1} - \mathbf{\bar w}_{r-1} \right\|^2 \right)
\notag \\
\overset{(c)}{\leq} &
\frac{E^2}{2} {\sum}_{t = 2}^{E-1} \left\| \mathbf{w}_{i}^{r,t-1} - \mathbf{\bar w}_{r-1} \right\|^2
\leq 
\frac{E^2}{2} {\sum}_{t = 2}^{E} \left\| \mathbf{w}_{i}^{r,t-1} - \mathbf{\bar w}_{r-1} \right\|^2
.
\end{align}
\end{small}
\vspace{-0.3cm}

\noindent
where equality (a) is derived from \eqref{eq:local SGD_a},
equality (b) results from interchanging the indices of the two summations in equality (a),
and inequality (c) follows from the fact that
$\sum_{e = t}^{E-1} e = \frac{(E-1+t)(E-1-t+1)}{2} < \frac{(E+t)(E-t)}{2} = \frac{E^2-t^2}{2} < \frac{E^2}{2}$.
Meanwhile, the term (\ref{w_i-w_bar}c) is bounded by
\begin{small}
\begin{align}\label{Proof_Thm1_lemma_3_step31}
{\rm (\ref{w_i-w_bar}c)}
=
{\sum}_{t = 2}^{E} (t - 1)^2
=
{\sum}_{t = 1}^{E-1} t^2
\overset{(a)}{=}
\frac{(E-1)E(2E-1)}{6}
\leq
\frac{E^3}{3}
,
\end{align}
\end{small}
\vspace{-0.3cm}

\noindent
where equality (a) is calculated using $\sum_{x=1}^m x^2 = m(m+1)(2m+1)/6$.

Substituting \eqref{Proof_Thm1_lemma_3_step2} and \eqref{Proof_Thm1_lemma_3_step31} into \eqref{w_i-w_bar} yields
\begin{small}
\begin{align}\label{wi-wbar2}
\sum_{t = 2}^{E} \left\| \mathbf{w}^{r,t-1}_{i} - {\bar {\mathbf{w}}}_{r-1} \right\|^2
\leq
\frac{3}{2} \gamma^2 E^2 L^2 \sum_{t = 2}^{E} \left\| \mathbf{w}_{i}^{r,t-1} - \mathbf{\bar w}_{r-1} \right\|^2
+  2 \gamma^2 E^3 \sum_{c=1}^C
\left( \alpha_{i,c} V_{i,c}^2 + \chi^2_{\bm{\alpha}_i \| \bm{\alpha}_g} G^2 \right)
 + \gamma^2 E^3 \left \| \nabla F(\mathbf{\bar w}_{r-1}) \right\|^2
.
\end{align}
\end{small}
\vspace{-0.3cm}

Finally, rearranging terms in \eqref{wi-wbar2} leads to Lemma \ref{Proof_Thm1_lemma_3}.
\hfill $\blacksquare$

\section{Label and weight distributions for experiments in Figures \ref{fig:value Vic G MNIST CIFAR-10}, \ref{fig:G Vic ratio MNIST CIFAR-10}, and \ref{fig:Vic value versus pi alpha}}\label{sec:label_distribution_experiments_Vic_G_value}

For the experiments presented in Figures \ref{fig:value Vic G MNIST CIFAR-10}, \ref{fig:G Vic ratio MNIST CIFAR-10}, and \ref{fig:Vic value versus pi alpha}, each client is assigned samples from a single class out of the total 10 classes.
The detailed class distribution for each client is provided in Table \ref{table:Class and weight distributions}.
\begin{table}[!h]
\small
\centering
\caption{Class and weight distributions.}
\begin{tabular}{c|c|c|c|c|c|c|c}
\hline
\multirow{2}{*}{\textbf{Client} ($i$)} & \multirow{2}{*}{\textbf{Class} ($c$)} & \multicolumn{2}{c|}{\textbf{Weight} ($p_i$)} & \multirow{2}{*}{\textbf{Client} ($i$)} & \multirow{2}{*}{\textbf{Class} ($c$)} & \multicolumn{2}{c}{\textbf{Weight} ($p_i$)} \\ \cline{3-4} \cline{7-8}
& & \textbf{Balanced} & \textbf{Unbalanced} & & & \textbf{Balanced} & \textbf{Unbalanced} \\
\hline
1 & 1 & \multirow{10}{*}{$\frac{1}{20}$} & \multirow{10}{*}{$\frac{1}{100}$} & 2 & 1 & \multirow{10}{*}{$\frac{1}{20}$} & \multirow{10}{*}{$\frac{9}{100}$} \\
3 & 2 & & & 4 & 2 & & \\
5 & 3 & & & 6 & 3 & & \\
7 & 4 & & & 8 & 4 & & \\
9 & 5 & & & 10 & 5 & & \\
11 & 6 & & & 12 & 6 & & \\
13 & 7 & & & 14 & 7 & & \\
15 & 8 & & & 16 & 8 & & \\
17 & 9 & & & 18 & 9 & & \\
19 & 10 & & & 20 & 10 & & \\
\hline
\end{tabular}
\label{table:Class and weight distributions}
\end{table}

From Table \ref{table:Class and weight distributions}, the label distribution $\{ \alpha_{i,c} \}$ for each client $i \in [N]$ is given by
\begin{align}
\alpha_{i,c}
=
\begin{cases}
1, \; \text{if class-$c$ samples is assigned to client $i$};
\\
0, \; \text{otherwise}.
\end{cases}
\end{align}

Additionally, as shown in Table \ref{table:Class and weight distributions}, in the balanced scenario, where each client holds an equal number of samples, the weight for each client is $p_i = \frac{1}{20}$.
In the unbalanced scenario, where half of the clients holding 90\% of the total training samples, half of the clients have $p_i = \frac{1}{100}$, while the remaining clients  have $p_i = \frac{9}{100}$.

\section{Convergence of General Case} \label{proof: theorem unfixed failure prob}

For the general case, we consider unfixed transmission failure probabilities $\epsilon^{r}_{i}$, which may vary across iterations during the training process.
Analogous to Lemma \ref{lemma:beta}, we establish the following property for this general setting, as presented in Lemma \ref{lemma:beta general}.

\begin{lemma}\label{lemma:beta general}
In the FL procedure described in Algorithm \ref{algorithm:FL under transmission failure}, the global aggregation in \eqref{global_model_failure} satisfies
\begin{align}\label{average_beta_r}
\mathbb{E}_{{\mathcal{K}}_{r},\mathds{1}^{r}_{i}}
\bigg[
\frac{ \sum_{{i} \in {\mathcal{K}}_{r}} \mathds{1}^{r}_{i}
\mathbf{w}^{r,E}_{i}
 }{ \sum_{{i} \in {\mathcal{K}}_{r}} \mathds{1}^{r}_{i}} \bigg|
\sum_{{i} \in {\mathcal{K}}_{r}} \mathds{1}^{r}_{i} \neq 0
\bigg]
=
\sum_{i=1}^N {\bar \beta}^{r}_{i}
\mathbf{w}^{r,E}_{i}
,
\end{align}
where ${\bar \beta}^{r}_{i} \in [0,1]$ with $\sum_{i=1}^N {\bar \beta}^{r}_{i} = 1$, and $\mathbb{E}[\cdot]$ denotes the expectation taken over ${\mathcal{K}}_{r}$ and $\{ \mathds{1}^{r}_{i} \}$.
If the transmission failure probability $\epsilon^{r}_{i} = 0$ $\forall i \in [N]$ and $r \in [R]$, then $\mathds{1}^{r}_{i}=1$ and ${\bar \beta}^{r}_{i} = s_i$.
\end{lemma}

From \eqref{average_beta_r}, one can see that ${\bar \beta}^{r}_{i}$ represents the effective appearance probability of client $i$ in the global aggregation at iteration $r$.
Its explicit formulation can be obtained from \eqref{eq:bar beta i formulation} by substituting $\epsilon_{i}$, $\epsilon_{j}$, and $\epsilon_{i'}$ with $\epsilon^{r}_{i}$, $\epsilon^{r}_{j}$, and $\epsilon^{r}_{i'}$, respectively.
The main convergence result is as follows.

\begin{theorem}\label{theorem:convergence FL failure unfixed}
(General case)
Under the same conditions as Theorem \ref{theorem:convergence FL failure}, the convergence of FL Algorithm 1 with unfixed transmission failure probabilities is upper bounded by

\vspace{-0.1cm}
\begin{small}
\begin{align}\label{eq:convergence bound of FL failure unfixed}
\frac{1}{R} {\sum}_{r=1}^R
\mathbb{E}[ \left\| \nabla  F({\bar {\mathbf{w}}}_{r-1}) \right\|^2 ]
\leq &
\frac{2484L}{55(TK)^{\frac{1}{2}}}
\left( \mathbb{E}[F({\bar {\mathbf{w}}}_{0})]
- \underline{F} \right)
\notag \\
& +
\underbrace{
\bigg( \frac{276}{55(TK)^{\frac{1}{4}}} + \frac{24}{55(TK)^{\frac{1}{2}}} + \frac{4}{55(TK)^{\frac{3}{4}}} \bigg)
\frac{1}{R}{\sum}_{r = 1}^{R}
{\sum}_{i = 1}^{N} {\bar \beta}_{r,i} {\sum}_{c=1}^C
\left( \alpha_{i,c} V_{i,c}^2
+ \chi^2_{\bm{\alpha}_i \| \bm{\alpha}_g} G^2 \right)
}_{\text{\small (\ref{eq:convergence bound of FL failure unfixed}a) caused by non-i.i.d. data}}
\notag \\
&
+
\underbrace{
\frac{828}{55}
\frac{1}{R}{\sum}_{r = 1}^{R}
\left(
\chi^2_{\bm{\bar \beta}_r \|\mathbf{p}} {\sum}_{c=1}^C {\sum}_{i = 1}^N p_i \alpha_{i,c} V_{i,c}^2
+
\chi^2_{\bm{\bar \alpha}_r \| \bm{\alpha}_g} G^2
\right)
}_{\text{\small (\ref{eq:convergence bound of FL failure unfixed}b) caused by transmission failure and non-i.i.d. data}}
,
\end{align}
\end{small}
\vspace{-0.1cm}

\noindent
where in term (\ref{eq:convergence bound of FL failure}b), $\chi^2_{\bm{\bar \beta}_r\|\mathbf{p}} \triangleq \sum_{i = 1}^N \frac{({\bar \beta}^{r}_{i} - p_i)^2}{p_i}$ represents the chi-square divergence between the effective appearance probabilities $\{ \bar{\beta}_{i,r} \}$ and the weights $\{ p_i \}$.
Meanwhile, $\chi^2_{\bm{\bar \alpha}_r \| \bm{\alpha}_g} \triangleq \sum_{c=1}^C \frac{( \sum_{i = 1}^N (p_i - {\bar \beta}^{r}_{i}) \alpha_{i,c})^2}{\alpha_{g,c}} = \sum_{c=1}^C \frac{( \alpha_{g,c} - \sum_{i = 1}^N {\bar \beta}^{r}_{i} \alpha_{i,c})^2}{\alpha_{g,c}}$ quantifies the divergence between the actual global label distribution $\{ \alpha_{g,c} \}$ and the effective global label distribution $\{ {\bar \alpha}_{c,r} \}$ at iteration $r$, where ${\bar \alpha}_{c,r} \triangleq \sum_{i = 1}^N {\bar \beta}^{r}_{i} \alpha_{i,c}$.
\end{theorem}

\emph{Proof:}
The proof of Theorem \ref{theorem:convergence FL failure unfixed} follows similarly to that of Theorem \ref{theorem:convergence FL failure} (in Appendix \ref{appendix:proof theorem convergence FL failure}), with the fixed transmission failure probabilities ${\bar \beta}_{i}$ replaced by the iteration-dependent probabilities ${\bar \beta}^{r}_{i}$ for each iteration $r$.
\hfill $\blacksquare$

The upper bound in \eqref{eq:convergence bound of FL failure unfixed} provides insights similar to those in Theorem \ref{theorem:convergence FL failure}.
Furthermore, in conjunction with Observation \ref{observation:G larger than Vic}, the label-related heterogeneity component in (\ref{eq:convergence bound of FL failure unfixed}b), namely $\chi^2_{\bm{\bar \alpha}_r \| \bm{\alpha}_g} G^2$, emerges as the dominant factor influencing FL convergence.
Then, analogous to Corollary \ref{corollary:convergence FL failure}, we obtain Corollary \ref{corollary:convergence FL failure unfixed}.

\begin{corollary}\label{corollary:convergence FL failure unfixed}
Under the same conditions as Theorem \ref{theorem:convergence FL failure unfixed} and given Observation \ref{observation:G larger than Vic}, if $\chi^2_{\bm{\bar \alpha}_r \| \bm{\alpha}_g}=0$ for each iteration $r$, then we approximately have

\vspace{-0.1cm}
\begin{small}
\begin{align}\label{eq:convergence bound of FL failure corollary unfixed}
\frac{1}{R} {\sum}_{r=1}^R
\mathbb{E}[ \left\| \nabla  F({\bar {\mathbf{w}}}_{r-1}) \right\|^2 ]
\leq &
\frac{2484L}{55(TK)^{\frac{1}{2}}}
\left( \mathbb{E}[F({\bar {\mathbf{w}}}_{0})]
- \underline{F} \right)
\notag \\
& +
\bigg( \frac{276}{55(TK)^{\frac{1}{4}}} + \frac{24}{55(TK)^{\frac{1}{2}}} + \frac{4}{55(TK)^{\frac{3}{4}}} \bigg)
\frac{1}{R}{\sum}_{r = 1}^{R}
{\sum}_{i = 1}^{N} {\bar \beta}_{r,i} {\sum}_{c=1}^C
\left( \alpha_{i,c} V_{i,c}^2
+ \chi^2_{\bm{\alpha}_i \| \bm{\alpha}_g} G^2 \right)
.
\end{align}
\end{small}
\end{corollary}

\section{Proof of Proposition \ref{proposition:bar beta i formulation}}\label{sec:Appearance Probability}

\subsection{Derivation of $\beta_{z, i}$}

To prove Proposition \ref{proposition:bar beta i formulation}, the key is to derive the value of $\beta_{z,i}$ in \eqref{eq:bar beta i formulation}.
Given selection set $\mathcal{K}_r^z$, the appearance probability of each selected client $i \in \mathcal{K}_r^z$ in global aggregation is
\begin{small}
\begin{align}\label{eq:beta_zi}
\beta_{z, i}
=
\frac{1}{1 - {\prod}_{j \in \mathcal{K}_r^z} \epsilon_j}
{\sum}_{k=1}^K \frac{1}{k}
{\sum}_{\mathcal{K}' \subseteq \mathcal{K}_r^z \backslash i, \atop |\mathcal{K}'| = k-1}
\bigg(
\underbrace{
(1 - \epsilon_i)
{\prod}_{i' \in \mathcal{K}'} (1-\epsilon_{i'})
}_{\rm \triangleq (\ref{eq:beta_zi}a)}
\underbrace{
{\prod}_{j' \in \mathcal{K}_r^z\backslash i \backslash \mathcal{K}'} \epsilon_{j'}
}_{\rm \triangleq (\ref{eq:beta_zi}b)}
\bigg)
,
\end{align}
\end{small}
\vspace{-0.2cm}

\noindent
where $k$ is the number of clients with successful transmissions, and $\mathcal{K}_r^z \backslash i$ denotes the set obtained by removing one occurrence of $i$ from $\mathcal{K}_r^z$
(if $i$ appears multiple times in $\mathcal{K}_r^z$, only one occurrence is removed).
The term $(1 - {\prod}_{j \in \mathcal{K}_r^z} \epsilon_j)$ represents the probability of $\sum_{{i} \in \mathcal{K}_{r}} \mathds{1}^{r}_{i} \neq 0$.
In the summation over $\mathcal{K}'$, client $i$ and the other $(k-1)$ clients in $\mathcal{K}'$ constitute the $k$ successful transmissions, corresponding to the probability in (\ref{eq:beta_zi}a), while the remaining $(K-k)$ clients experiences transmission failures, corresponding to the probability in (\ref{eq:beta_zi}b).

After performing the mathematical derivation, we can simplify the formulation in \eqref{eq:beta_zi} as follows:
\begin{small}
\begin{align}\label{eq:beta_zi_5}
\beta_{z, i}
=
\frac{1 - \epsilon_i}{(1 - \prod_{j \in \mathcal{K}_r^z} \epsilon_j)K}
\Bigg(
1
+
{\sum}_{k = 1}^{K-1}
\bigg(
\frac{1}{C^{K-1}_{k}}
{\sum}_{\mathcal{K}' \subseteq \mathcal{K}_r^z\backslash i, \atop |\mathcal{K}'| = k}
{\prod}_{i' \in \mathcal{K}'} \epsilon_{i'}
\bigg)
\Bigg)
.
\end{align}
\end{small}
\vspace{-0.2cm}

\noindent
where the detailed derivation can be found in the subsequent Appendix \ref{section:Proof Formulation beta zi}.

As described in Section \ref{sec:FL procedure}, $K$ clients are selected with replacement at each iteration.
Consequently, clients in $\mathcal{K}_r^z$ may appear multiple times.
For each repeated client $i \in \mathcal{K}_r^z$, the value of $\beta_{z, i}$ remains the same.
Thus, with $n_{z,i}$ denoting the number of times each client $i \in [N]$ appears in $\mathcal{K}_r^z$, the value of $\beta_{z,i}$ in \eqref{eq:bar beta i formulation} is obtained.

\subsection{Proof of Formulation (\ref{eq:beta_zi_5})}\label{section:Proof Formulation beta zi}

For simplifying the expression in \eqref{eq:beta_zi}, we introduce the following Lemma \ref{lemma:product1-ek} and Lemma \ref{lemma:sum_equal_1K_divide_C}, which are provided in the subsequent subsections.
\begin{lemma}\label{lemma:product1-ek}
The product of the successful probability of all $K$ selected clients in a selection set $\mathcal{K}$ satisfies
\begin{small}
\begin{align}\label{eq:product1-ek}
{\prod}_{i \in \mathcal{K}} (1-\epsilon_i)
=
1
+
{\sum}_{k=1}^K (-1)^k
{\sum}_{\mathcal{K}' \subseteq \mathcal{K}, \, |\mathcal{K}'| = k} {\prod}_{i' \in \mathcal{K}'} \epsilon_{i'}
,
\end{align}
\end{small}
\vspace{-0.3cm}

\noindent
where $\mathcal{K}'$ represents all possible subsets of $\mathcal{K}$.
\end{lemma}

\begin{lemma}\label{lemma:sum_equal_1K_divide_C}
For $k \in [K-1]$,  the following holds:
\begin{small}
\begin{align}\label{eq:sum_equal_1K_divide_C}
{\sum}_{k'=0}^{k}
\frac{(-1)^{k-k'} C^{k}_{k'}}{K-k'}
=
\frac{1}{K C^{K-1}_{k}}
.
\end{align}
\end{small}
\vspace{-0.3cm}
\end{lemma}

Using Lemma \ref{lemma:product1-ek}, we can combine it with \eqref{eq:beta_zi}, leading to
\begin{small}
\begin{align}\label{eq:beta_zi_3}
\beta_{z, i}
= &
\frac{1 - \epsilon_i}{1 - \prod_{j \in \mathcal{K}_r^z} \epsilon_j}
\sum_{k=0}^{K-1} \frac{1}{K-k}
\sum_{\mathcal{K}' \subseteq \mathcal{K}_r^z \backslash i, \atop |\mathcal{K}'| = k}
\bigg(
\prod_{i' \in \mathcal{K}'} \epsilon_{i'}
\prod_{j' \in \mathcal{K}_r^z\backslash i \backslash \mathcal{K}'} (1-\epsilon_{j'})
\bigg)
\notag \\
= &
\frac{1 - \epsilon_i}{1 - \prod_{j \in \mathcal{K}_r^z} \epsilon_j}
\sum_{k=0}^{K-1} \frac{1}{K-k}
\sum_{\mathcal{K}' \subseteq \mathcal{K}_r^z\backslash i, \atop |\mathcal{K}'| = k}
\Bigg(
\prod_{i' \in \mathcal{K}'} \epsilon_{i'}
\bigg(
\underbrace{
1 +
\sum_{k'=1}^{K-k-1}
(-1)^{k'}
\sum_{\mathcal{\tilde S}' \subseteq \mathcal{K}_r^z\backslash i \backslash \mathcal{K}', \atop |\mathcal{\tilde S}'|=k'} \prod_{j' \in \mathcal{\tilde S}'} \epsilon_{j'}
}_{\rm \triangleq (\ref{eq:beta_zi_3}b)}
\bigg)
\Bigg)
\notag \\
\overset{(a)}{=} &
\frac{1 - \epsilon_i}{1 - \prod_{j \in \mathcal{K}_r^z} \epsilon_j}
\Bigg(
\underbrace{
\sum_{k=0}^{K-1} \frac{1}{K-k}
\bigg(
\sum_{\mathcal{K}' \subseteq \mathcal{K}_r^z \backslash i, \atop |\mathcal{K}'| = k}
\prod_{i' \in \mathcal{K}'} \epsilon_{i'}
\bigg)
}_{\rm \triangleq (\ref{eq:beta_zi_3}c)}
+
\sum_{k=0}^{K-2} \frac{1}{K-k}
\bigg(
\sum_{\mathcal{K}' \subset \mathcal{K}_r^z \backslash i, \atop |\mathcal{K}'| = k}
\prod_{i' \in \mathcal{K}'} \epsilon_{i'}
\underbrace{
\sum_{k'=1}^{K-k-1}
(-1)^{k'}
\sum_{\mathcal{\tilde S}' \subseteq \mathcal{K}_r^z \backslash i \backslash \mathcal{K}', \atop |\mathcal{\tilde S}'|=k'} \prod_{j' \in \mathcal{\tilde S}'} \epsilon_{j'}
}_{\rm \triangleq (\ref{eq:beta_zi_3}d)}
\bigg)
\Bigg)
,
\end{align}
\end{small}
\vspace{-0.3cm}

\noindent
where the term (\ref{eq:beta_zi_3}b) follows from Lemma \ref{lemma:product1-ek},
and equality (a) holds because (\ref{eq:beta_zi_3}b) equals 1 when $k = K-1$.

For term (\ref{eq:beta_zi_3}c), we obtain
\begin{small}
\begin{align}\label{eq:A22}
{\rm (\ref{eq:beta_zi_3}c)}
=
&
\frac{1}{K}
+
{\sum}_{k=1}^{K-1} \frac{1}{K-k}
\bigg(
\sum_{\mathcal{K}' \subseteq \mathcal{K}_r^z \backslash i, \atop |\mathcal{K}'| = k}
\prod_{i' \in \mathcal{K}'} \epsilon_{i'}
\bigg)
\overset{(a)}{=}
\frac{1}{K}
+
{\sum}_{k = 1}^{K-1}
\frac{(-1)^{k-k} C^{k}_{k}}{K-k}
\bigg(
\sum_{\mathcal{K}' \subseteq \mathcal{K}_r^z \backslash i, \atop |\mathcal{K}'| = k}
\prod_{i' \in \mathcal{K}'} \epsilon_{i'}
\bigg)
,
\end{align}
\end{small}
\vspace{-0.3cm}

\noindent
where equality (a) follows from the fact that $(-1)^{k-k}=1$ and $C^{k}_{k}=1$.

For term (\ref{eq:beta_zi_3}d), by setting $k_{\text{sum}} = k + k'$, we obtain
\begin{small}
\begin{align}\label{eq:A23}
{\rm (\ref{eq:beta_zi_3}d)}
= &
\sum_{k_{\text{sum}} = 1}^{K-1}
\sum_{k=0}^{k_{\text{sum}}-1}
\bigg(
\frac{(-1)^{k_{\text{sum}}-k}}{K-k}
\sum_{\mathcal{K}' \subset \mathcal{K}_r^z \backslash i, \atop |\mathcal{K}'| = k}
\sum_{\mathcal{\tilde S}' \subseteq \mathcal{K}_r^z \backslash i \backslash \mathcal{K}', \atop |\mathcal{\tilde S}'|=k_{\text{sum}}-k}
\Big(
\prod_{i' \in \mathcal{K}'} \epsilon_{i'}
 \prod_{j' \in \mathcal{\tilde S}'} \epsilon_{j'}
\Big)
\bigg)
\notag \\
\overset{(a)}{=} &
\sum_{k_{\text{sum}} = 1}^{K-1}
\sum_{k=0}^{k_{\text{sum}}-1}
\bigg(
\frac{(-1)^{k_{\text{sum}}-k}}{K-k}
\sum_{\mathcal{K}' \subseteq \mathcal{K}_r^z \backslash i, \atop |\mathcal{K}'| = k_{\text{sum}}}
\Big(
C^{k_{\text{sum}}}_k
\prod_{i' \in \mathcal{K}'} \epsilon_{i'}
\Big)
\bigg)
\overset{(b)}{=} 
\sum_{k = 1}^{K-1}
\bigg(
\sum_{k'=0}^{k-1}
\frac{(-1)^{k-k'}  C^{k}_{k'}}{K-k'}
\bigg)
\bigg(
\sum_{\mathcal{K}' \subseteq \mathcal{K}_r^z \backslash i, \atop |\mathcal{K}'| = k}
\prod_{i' \in \mathcal{K}'} \epsilon_{i'}
\bigg)
,
\end{align}
\end{small}
\vspace{-0.3cm}

\noindent
where equality (a) results from counting the number of different combinations,
and equality (b) comes from substituting $k_{\text{sum}}$ and $k$ with $k$ and $k'$, respectively.

Next, by combining \eqref{eq:beta_zi_3} with \eqref{eq:A22} and \eqref{eq:A23}, the effective appearance probability of each selected client $i \in \mathcal{K}_r^z$ is
\begin{small}
\begin{align}\label{eq:beta_zi_4}
\beta_{z, i}
=
\frac{1 - \epsilon_i}{1 - \prod_{j \in \mathcal{K}_r^z} \epsilon_j}
\Bigg(
\frac{1}{K}
+
\sum_{k = 1}^{K-1}
\bigg(
\sum_{k'=0}^{k}
\frac{(-1)^{k-k'} C^{k}_{k'}}{K-k'}
\bigg)
\bigg(
\sum_{\mathcal{K}' \subseteq \mathcal{K}_r^z\backslash i, \atop |\mathcal{K}'| = k}
\prod_{i' \in \mathcal{K}'} \epsilon_{i'}
\bigg)
\Bigg)
.
\end{align}
\end{small}
\vspace{-0.3cm}

Finally, substituting \eqref{eq:sum_equal_1K_divide_C} into \eqref{eq:beta_zi_4}, the effective appearance probability for each selected client $i \in \mathcal{K}_r^z$ becomes \eqref{eq:beta_zi_5}.

\hfill $\blacksquare$

\subsection{Proof of Lemma \ref{lemma:product1-ek}}\label{sec:lemma proof product}

We employ the mathematical induction method to prove Lemma \ref{lemma:product1-ek}.

\begin{enumerate}[(i)]
\item
When $K=1$, there is only one element $i$ in $\mathcal{K}$, both sides of \eqref{eq:product1-ek} equals $1 - \epsilon_i$, thereby satisfying \eqref{eq:product1-ek}.

\item
Assume that \eqref{eq:product1-ek} holds.
Then, if $K$ increases to $K+1$ and a new element $j$ is added to the original set $\mathcal{K}$, making $\mathcal{K}$ become $\mathcal{K} \cup j$, we have
\begin{small}
\begin{align}\label{eq:product1-ek cup j}
\prod_{i \in \mathcal{K} \cup j} (1 - \epsilon_i)
= &
\Big(\prod_{i \in \mathcal{K}} (1 - \epsilon_i)\Big)(1 - \epsilon_j)
\overset{(a)}{=} 
\bigg( 1 + \sum_{k=1}^K (-1)^k \sum_{\mathcal{K}' \subseteq \mathcal{K}, \atop |\mathcal{K}'| = k} \prod_{i \in \mathcal{K}'} \epsilon_i \bigg)
(1 - \epsilon_j)
\notag \\
= &
1
+ \underbrace{(- \epsilon_j)}_{\rm \triangleq (\ref{eq:product1-ek cup j}a)}
+ \underbrace{\sum_{k=1}^K (-1)^k \sum_{\mathcal{K}' \subseteq \mathcal{K}, \atop |\mathcal{K}'| = k} \prod_{i \in \mathcal{K}'} \epsilon_i}_{\rm \triangleq (\ref{eq:product1-ek cup j}b)}
+ \underbrace{
\bigg( \sum_{k=1}^{K-1} (-1)^k \sum_{\mathcal{K}' \subseteq \mathcal{K}, \atop |\mathcal{K}'| = k} \prod_{i \in \mathcal{K}'} \epsilon_i \bigg)
(- \epsilon_j)
}_{\rm \triangleq (\ref{eq:product1-ek cup j}c)}
+ \underbrace{
(-1)^K \Big(\prod_{i \in \mathcal{K}} \epsilon_i\Big) (- \epsilon_j)
}_{\rm \triangleq (\ref{eq:product1-ek cup j}d)}
,
\end{align}
\end{small}
\vspace{-0.3cm}

\noindent
where equality (a) comes from \eqref{eq:product1-ek}, and
\begin{small}
\begin{align}\label{eq:A132}
{\rm (\ref{eq:product1-ek cup j}d)}
=
(-1)^{K+1} \prod_{i \in \mathcal{K}\cup j} \epsilon_i
.
\end{align}
\end{small}
\vspace{-0.3cm}

\noindent
Besides, since
\begin{small}
\begin{align}\label{eq:A12 A131}
{\rm (\ref{eq:product1-ek cup j}a)} + {\rm (\ref{eq:product1-ek cup j}c)}
=
\bigg( \sum_{k=0}^{K-1} (-1)^k \sum_{\mathcal{K}' \subseteq \mathcal{K}, \atop |\mathcal{K}'| = k} \prod_{i \in \mathcal{K}'} \epsilon_i \bigg)
(- \epsilon_j)
=
\sum_{k=1}^{K} (-1)^k \sum_{\mathcal{K}' \subseteq \mathcal{K}, \atop |\mathcal{K}'| = k-1} \Big( \prod_{i \in \mathcal{K}'} \epsilon_i \Big)
\epsilon_j
,
\end{align}
\end{small}
\vspace{-0.3cm}

\noindent
we have
\begin{small}
\begin{align}\label{eq:A11 A12 A131}
{\rm (\ref{eq:product1-ek cup j}a)} + {\rm (\ref{eq:product1-ek cup j}b)} + {\rm (\ref{eq:product1-ek cup j}c)}
=
\sum_{k=1}^{K} (-1)^k \sum_{\mathcal{K}' \subseteq \mathcal{K}\cup j, \, |\mathcal{K}'| = {k}} \prod_{i \in \mathcal{K}'} \epsilon_i
.
\end{align}
\end{small}
\vspace{-0.3cm}

Finally, combining \eqref{eq:product1-ek cup j} with \eqref{eq:A132} and \eqref{eq:A11 A12 A131}, we have
\begin{small}
\begin{align}
\prod_{i \in \mathcal{K} \cup j} (1 - \epsilon_i)
=
1
+
\sum_{k=1}^{K+1} (-1)^k \sum_{\mathcal{K}' \subseteq \mathcal{K} \cup j, \atop |\mathcal{K}'| = {k}} \prod_{i \in \mathcal{K}'} \epsilon_i
.
\end{align}
\end{small}
\vspace{-0.3cm}
\end{enumerate}

Based on the above induction process, we have proved Lemma \ref{lemma:product1-ek}.
\hfill $\blacksquare$

\subsection{Proof of Lemma \ref{lemma:sum_equal_1K_divide_C}}\label{sec:lemma sum_equal_1K_divide_C}

We adopt the mathematical induction method to prove Lemma \ref{lemma:sum_equal_1K_divide_C}.
\begin{enumerate}[(i)]
\item
When $k=1$, we have
\begin{small}
\begin{equation}
\sum_{k'=0}^{1}
\frac{(-1)^{1-k'} C^{1}_{k'}}{K-k'}
=
\frac{1}{K-1} -\frac{1}{K}
=
\frac{1}{K(K-1)}
=
\frac{1}{K C^{K-1}_{1}}
,
\end{equation}
\end{small}
\vspace{-0.3cm}

\noindent
which satisfies \eqref{eq:sum_equal_1K_divide_C}.

\item
Assume that \eqref{eq:sum_equal_1K_divide_C} holds.
Then, if $k$ increases to $k+1$, the left-hand side of \eqref{eq:sum_equal_1K_divide_C} becomes
\begin{small}
\begin{align}
\sum_{k'=0}^{k+1}
\frac{(-1)^{k+1-k'} C^{k+1}_{k'}}{K-k'}
\overset{(a)}{=} &
\sum_{k'=0}^{k}
\frac{(-1)^{k+1-k'} C^{k}_{k'}}{K-k'}
+
\sum_{k'=1}^{k+1}
\frac{(-1)^{k+1-k'} C^{k}_{k'-1}}{K-k'}
= 
- \sum_{k'=0}^{k}
\frac{(-1)^{k-k'} C^{k}_{k'}}{K-k'}
+
\sum_{k'=0}^{k}
\frac{(-1)^{k-k'} C^{k}_{k'}}{K-1-k'}
\notag \\
\overset{(b)}{=} &
-
\frac{1}{K C^{K-1}_{k}}
+
\frac{1}{(K-1) C^{K-2}_{k}}
=
\frac{1}{KC^{K-1}_{k+1}}
,
\end{align}
\end{small}
\vspace{-0.3cm}

\noindent
where equality (a) is due to
\begin{small}
\begin{align}
C^{k+1}_{k'}
= &
\frac{(k+1)!}{(k+1-k')! k'!}
=
\frac{k+1}{k+1-k'}
\frac{k!}{(k-k')! k'!}
=
\left(1 + \frac{k'}{k+1-k'} \right)
\frac{k!}{(k-k')! k'!}
\notag \\
= &
\frac{k!}{(k-k')! k'!}
+
\frac{k!}{(k-(k'-1))! (k'-1)!}
\notag \\
= &
\begin{cases}
C^{k}_{k'}, & \text{if } k' = 0;\\
C^{k}_{k'} + C^{k}_{k'-1}, & \text{if } k' = 1,2,\cdots,k; \\
C^{k}_{k'-1}, & \text{if } k' = k+1,
\end{cases}
\end{align}
\end{small}
\vspace{-0.3cm}

\noindent
and equality (b) is due to \eqref{eq:sum_equal_1K_divide_C}.

\end{enumerate}
Based on the above induction process, we have proved Lemma \ref{lemma:sum_equal_1K_divide_C}.
\hfill $\blacksquare$

\section{Proof of Proposition \ref{proposition:beta_zi_failure_diff}}\label{sec:proposition beta_zi_failure_diff}

\subsection{Derivation of (\ref{eq:beta_zi_failure_prob_diff})}

Before presenting the detailed derivation procedure, we introduce the following Lemma \ref{lemma:bar_beta_i_failure_prob}, which is proved in the subsequent Appendix \ref{sec:proof lemma:bar_beta_i_failure_prob}.
\begin{lemma}\label{lemma:bar_beta_i_failure_prob}
The effective appearance probability ${\bar \beta}_i$ in \eqref{eq:bar beta i formulation} is equivalent to
\begin{small}
\begin{align}\label{eq:bar_beta_i_1_epsilon}
{\bar \beta}_i
= &
s_i(1 - \epsilon_i)
\sum_{j_0,j_1,\cdots,j_{K-2}=1}^N
\prod_{k=0}^{K-2} s_{j_k}
\frac{1 + \sum_{k=0}^{K-2} \prod_{k'=0}^{k} \epsilon_{j_{k'}}}{1 - \epsilon_{i}\prod_{k=0}^{K-2} \epsilon_{j_k}}
.
\end{align}
\end{small}
\end{lemma}

Next, \eqref{eq:bar_beta_i_1_epsilon} can be rewritten as
\begin{small}
\begin{align}\label{eq:bar_beta_i_1_epsilon 2}
{\bar \beta}_i
= &
s_i
\sum_{j_0,j_1,\cdots,j_{K-2}=1}^N
\prod_{k=0}^{K-2} s_{j_k}
\frac{1 - \epsilon_i + \sum_{k=0}^{K-2} \prod_{k'=0}^{k} \epsilon_{j_{k'}} - \epsilon_i \sum_{k=0}^{K-2} \prod_{k'=0}^{k} \epsilon_{j_{k'}}}{1 - \epsilon_{i}\prod_{k=0}^{K-2} \epsilon_{j_k}}
\notag \\
= &
s_i
\sum_{j_0,j_1,\cdots,j_{K-2}=1}^N
\prod_{k=0}^{K-2} s_{j_k}
\frac{1 + \epsilon_{j_0} + \epsilon_{j_0} \sum_{k=1}^{K-2} \prod_{k'=1}^{k} \epsilon_{j_{k'}} - \epsilon_i - \epsilon_i \sum_{k=0}^{K-3} \prod_{k'=0}^{k} \epsilon_{j_{k'}}  - \epsilon_i \prod_{k=0}^{K-2} \epsilon_{j_{k}}}{1 - \epsilon_{i}\prod_{k=0}^{K-2} \epsilon_{j_k}}
\notag \\
\overset{(a)}{=} &
s_i
\sum_{j_0,j_1,\cdots,j_{K-2}=1}^N
\prod_{k=0}^{K-2} s_{j_k}
\frac{1 + \epsilon_{j_0} + \epsilon_{j_0} \sum_{k=1}^{K-2} \prod_{k'=1}^{k} \epsilon_{j_{k'}} - \epsilon_i - \epsilon_i \sum_{k=1}^{K-2} \prod_{k'=1}^{k} \epsilon_{j_{k'}}  - \epsilon_i \prod_{k=0}^{K-2} \epsilon_{j_{k}}}{1 - \epsilon_{i}\prod_{k=0}^{K-2} \epsilon_{j_k}}
\notag \\
= &
s_i
\sum_{j_0,j_1,\cdots,j_{K-2}=1}^N
\prod_{k=0}^{K-2} s_{j_k}
\frac{1 - \epsilon_i \prod_{k=0}^{K-2} \epsilon_{j_{k}}
+ ( \epsilon_{j_0} - \epsilon_i ) ( 1 + \sum_{k=1}^{K-2} \prod_{k'=1}^{k} \epsilon_{j_{k'}} )}{1 - \epsilon_{i}\prod_{k=0}^{K-2} \epsilon_{j_k}}
\notag \\
= &
s_i
\Bigg(
1 +
\sum_{j_0=1}^N s_{j_0} (\epsilon_{j_0} - \epsilon_i)
\sum_{j_1,j_2,\cdots, j_{K-2}=1}^N
\prod_{k=1}^{K-2} s_{j_k}
\frac{1 + \sum_{k=1}^{K-2} \prod_{k'=1}^{k} \epsilon_{j_{k'}}}{1 - \epsilon_{i}\epsilon_{i'} \prod_{k=1}^{K-2} \epsilon_{j_k}}
\Bigg)
,
\end{align}
\end{small}
\vspace{-0.3cm}

\noindent
where equality (a) holds because, for all indices $j_0,j_1,\cdots,j_{K-2}$ ranging from 1 to $N$,
\begin{small}
\begin{align}
\sum_{j_0,j_1,\cdots,j_{K-2}=1}^N \prod_{k=0}^{K-2} s_{j_k}
\frac{\sum_{k=0}^{K-3} \prod_{k'=0}^{k} \epsilon_{j_{k'}}}{1 - \epsilon_{i}\prod_{k=0}^{K-2} \epsilon_{j_k}}
=
\sum_{j_0,j_1,\cdots,j_{K-2}=1}^N \prod_{k=0}^{K-2} s_{j_k}
\frac{\sum_{k=1}^{K-2} \prod_{k'=1}^{k} \epsilon_{j_{k'}}}{1 - \epsilon_{i}\prod_{k=0}^{K-2} \epsilon_{j_k}}
.
\end{align}
\end{small}
\vspace{-0.3cm}

\noindent
Finally, by replacing the index $j_0$ in \eqref{eq:bar_beta_i_1_epsilon 2} with index $i'$ yields equation \eqref{eq:beta_zi_failure_prob_diff} as stated in Proposition \ref{proposition:beta_zi_failure_diff}.
\hfill $\blacksquare$

\subsection{Proof of Lemma \ref{lemma:bar_beta_i_failure_prob}}\label{sec:proof lemma:bar_beta_i_failure_prob}

Based on \eqref{eq:bar beta i formulation}, we have
\begin{small}
\begin{align}\label{eq:bar beta i formulation 2}
{\bar \beta}_i
=
&
\sum_{z=1}^{C^{N+K-1}_K}
\bigg( \prod_{i \in \mathcal{K}_r^z} s_i \bigg)
\frac{K!}{\prod_{i \in [N]} n_{z,i}!}
\frac{n_{z,i}}{K}
\frac{1 - \epsilon_i}{1 - \prod_{j \in \mathcal{K}_r^z} \epsilon_j}
\sum_{k = 0}^{K-1}
\bigg(
\frac{1}{C^{K-1}_{k}}
\sum_{\mathcal{K}' \subseteq \mathcal{K}_r^z \backslash i, \, |\mathcal{K}'| = k}
\prod_{j' \in \mathcal{K}'} \epsilon_{j'}
\bigg)
\notag \\
\overset{(a)}{=}
&
s_i
\sum_{z=1, \, {i \in \mathcal{K}_r^z}}^{C^{N+K-1}_{K}}
\bigg( \prod_{j \in \mathcal{K}_r^z \backslash i} s_j \bigg)
\frac{(K-1)!}{(n_{z,i}-1)! \prod_{j \in [N] \backslash i} n_{z,j}!}
\Bigg(
\frac{1 - \epsilon_i}{1 - \prod_{j \in \mathcal{K}_r^z} \epsilon_j}
\sum_{k = 0}^{K-1}
\bigg(
\frac{1}{C^{K-1}_{k}}
\sum_{\mathcal{K}' \subseteq \mathcal{K}_r^z \backslash i, \, |\mathcal{K}'| = k}
\prod_{j' \in \mathcal{K}'} \epsilon_{j'}
\bigg)
\Bigg)
,
\end{align}
\end{small}
\vspace{-0.3cm}

\noindent
where $\mathcal{K}_r^z \backslash i$ denotes the set obtained by removing one occurrence of $i$ from $\mathcal{K}_r^z$, and $\mathcal{K}'$ represents any subset of the set $\mathcal{K}_r^z \backslash i$.
Equality (a) follows from $n_{z,i} = 0$ if $i \notin \mathcal{K}_r^z$.

If one of the $K$ selected clients in the selection set $\mathcal{K}_r$ is fixed as client $i$, the remaining $(K-1)$ clients in $\mathcal{K}_r$ have $C^{N+K-2}_{K-1}$ possible combinations.
Based on this, let ${\mathcal{\tilde K}}_r^z$ denote the selection set formed by randomly selecting $(K-1)$ clients with replacement from a total $N$ clients, where $|{ \mathcal{\tilde K}}_r^z| = K-1$.
Consequently, \eqref{eq:bar beta i formulation 4} can be reformulated as
\begin{small}
\begin{align}\label{eq:bar beta i formulation 4}
{\bar \beta}_i
=
&
s_i(1 - \epsilon_i)
\sum_{z=1}^{C^{N+K-2}_{K-1}}
\underbrace{
\bigg( \prod_{j \in \mathcal{\tilde K}_r^z} s_j \bigg)
\frac{(K-1)!}{\prod_{j \in [N]} {\tilde n}_{z,j}!}
}_{\triangleq \Pr(\mathcal{\tilde K}_r^z)}
\Bigg(
\frac{1}{1 - \epsilon_i \prod_{j \in \mathcal{\tilde K}_r^z} \epsilon_j}
\sum_{k = 0}^{K-1}
\bigg(
\frac{1}{C^{K-1}_{k}}
\sum_{\mathcal{K}' \subseteq \mathcal{\tilde K}_r^z, \, |\mathcal{K}'| = k}
\prod_{j' \in \mathcal{K}'} \epsilon_{j'}
\bigg)
\Bigg)
\notag \\
=
&
s_i(1 - \epsilon_i)
\sum_{z=1}^{C^{N+K-2}_{K-1}}
\bigg( \prod_{j \in \mathcal{\tilde K}_r^z} s_j \bigg)
\frac{1}{1 - \epsilon_i \prod_{j \in \mathcal{\tilde K}_r^z} \epsilon_j}
\Bigg(
\underbrace{
\sum_{k = 0}^{K-1}
\frac{(K-1)!}{\prod_{j \in [N]} {\tilde n}_{z,j}!}
\frac{1}{C^{K-1}_{k}}
\sum_{\mathcal{K}' \subseteq \mathcal{\tilde K}_r^z, \, |\mathcal{K}'| = k}
\prod_{j' \in \mathcal{K}'} \epsilon_{j'}
}_{(\ref{eq:bar beta i formulation 4}a)}
\Bigg)
.
\end{align}
\end{small}
\vspace{-0.3cm}

\noindent
Here, $\Pr(\mathcal{\tilde K}_r^z)$ is the selection probability of each set $\mathcal{\tilde K}_r^z$, ${\tilde n}_{z,i}$ represents the number of times each client $j \in [N]$ appears in $\mathcal{\tilde K}_r^z$, and $\frac{(K-1)!}{\prod_{j \in [N]} {\tilde n}_{z,j}!}$ gives the number of distinct permutations of selected clients in $\mathcal{\tilde K}_r^z$.

Define $\mathcal{\hat K}_r^{z,k}$ as the subset obtained by randomly selecting $k$ clients without replacement from $\mathcal{\tilde K}_r^z$.
Let ${\hat n}_{z,j}^k$ represent the number of times each client $j \in [N]$ appears in $\mathcal{\hat K}_r^{z,k}$.
The number of distinct permutations of selected clients in $\mathcal{\hat K}_r^{z,k}$ is given by
\begin{small}
\begin{align}\label{eq:permutations number hat Krzk}
\prod_{j \in [N]} C^{{\tilde n}_{z,j}}_{{\hat n}_{z,j}^k}
=
\prod_{j \in [N]} \frac{{\tilde n}_{z,j}!}{({\tilde n}_{z,j} - {\hat n}_{z,j}^k)! {\hat n}_{z,j}^k!}
.
\end{align}
\end{small}
\vspace{-0.3cm}

\noindent
By combining \eqref{eq:bar beta i formulation 4} with \eqref{eq:permutations number hat Krzk}, we derive
\begin{small}
\begin{align}\label{eq:bar beta i formulation permutation num}
(\ref{eq:bar beta i formulation 4}a)
= &
\sum_{k = 0}^{K-1}
\frac{(K-1)!}{\prod_{j \in [N]} {\tilde n}_{z,j}!}
\frac{1}{C^{K-1}_{k}}
\sum_{\mathcal{\hat K}_r^{z,k} \subseteq \mathcal{\tilde K}_r^z}
\bigg(
\prod_{j \in [N]} C^{{\tilde n}_{z,j}}_{{\hat n}_{z,j}^k}
\prod_{{\hat j} \in \mathcal{\hat K}_r^{z,k}} \epsilon_{\hat j}
\bigg)
= 
\sum_{k = 0}^{K-1}
\sum_{\mathcal{\hat K}_r^{z,k} \subseteq \mathcal{\tilde K}_r^z}
\bigg(
\frac{(K-1)!}{\prod_{j \in [N]} {\tilde n}_{z,j}!}
\frac{1}{C^{K-1}_{k}}
\prod_{j \in [N]} C^{{\tilde n}_{z,j}}_{{\hat n}_{z,j}^k}
\bigg)
\prod_{{\hat j} \in \mathcal{\hat K}_r^{z,k}} \epsilon_{\hat j}
\notag \\
= &
\sum_{k = 0}^{K-1}
\sum_{\mathcal{\hat K}_r^{z,k} \subseteq \mathcal{\tilde K}_r^z}
\bigg(
\underbrace{
\frac{k!}{\prod_{j \in [N]} {\hat n}_{z,j}^k!}
}_{(\ref{eq:bar beta i formulation permutation num}a)}
\underbrace{
\frac{(K-1-k)!}{\prod_{j \in [N]} ({\tilde n}_{z,j} - {\hat n}_{z,j}^k)!}
}_{(\ref{eq:bar beta i formulation permutation num}b)}
\bigg)
\prod_{{\hat j} \in \mathcal{\hat K}_r^{z,k}} \epsilon_{\hat j}
\notag \\
= &
\sum_{{\hat j}_0 \in \mathcal{\tilde K}_r^z}
\sum_{{\hat j}_1 \in \mathcal{\tilde K}_r^z \backslash {\hat j}_0}
\cdots
\sum_{{\hat j}_{K-2} \in \mathcal{\tilde K}_r^z \backslash \{ \bigcup_{i=0}^{K-3} {\hat j}_0 \}}
\bigg(
1
+ \epsilon_{{\hat j}_0} + \epsilon_{{\hat j}_0}\epsilon_{{\hat j}_1}
+ \cdots
+ \prod_{k'=0}^{K-2} \epsilon_{{\hat j}_{k'}}
\bigg)
,
\end{align}
\end{small}
\vspace{-0.3cm}

\noindent
where (\ref{eq:bar beta i formulation permutation num}a) represents the number of different permutations for the selected clients in $\mathcal{\hat K}_r^{z,k}$,
while (\ref{eq:bar beta i formulation permutation num}b) corresponds to the permutations for the clients in $\mathcal{\tilde K}_r^z \backslash \mathcal{\hat K}_r^{z,k}$.

Substituting \eqref{eq:bar beta i formulation permutation num} into \eqref{eq:bar beta i formulation 2}, we obtain
\begin{small}
\begin{align}\label{eq:bar beta i formulation 3}
{\bar \beta}_i
\overset{(a)}{=}
&
s_i(1 - \epsilon_i)
\sum_{z=1}^{C^{N+K-2}_{K-1}}
\bigg(
\prod_{j \in \mathcal{\tilde K}_r^z} s_j
\bigg)
\bigg(
\sum_{{\hat j}_0 \in \mathcal{\tilde K}_r^z}
\sum_{{\hat j}_1 \in \mathcal{\tilde K}_r^z \backslash {\hat j}_0}
\cdots
\sum_{{\hat j}_{K-2} \in \mathcal{\tilde K}_r^z \backslash \{ \bigcup_{i=0}^{K-3} {\hat j}_0 \}}
\frac{
1
+ \epsilon_{{\hat j}_0} + \epsilon_{{\hat j}_0}\epsilon_{{\hat j}_1}
+ \cdots
+ \prod_{k'=0}^{K-2} \epsilon_{{\hat j}_{k'}}
}{1 - \epsilon_{i} \prod_{j \in \mathcal{\tilde K}_r^z} \epsilon_j}
\bigg)
\notag \\
=
&
s_i(1 - \epsilon_i)
\sum_{z=1}^{C^{N+K-2}_{K-1}}
\bigg(
\sum_{{\hat j}_0 \in \mathcal{\tilde K}_r^z} s_{{\hat j}_0}
\sum_{{\hat j}_1 \in \mathcal{\tilde K}_r^z \backslash {\hat j}_0} s_{{\hat j}_1}
\cdots
\sum_{{\hat j}_{K-2} \in \mathcal{\tilde K}_r^z \backslash \{ \bigcup_{i=0}^{K-3} {\hat j}_0 \}} s_{{\hat j}_{K-2}}
\frac{
1
+ \epsilon_{{\hat j}_0} + \epsilon_{{\hat j}_0}\epsilon_{{\hat j}_1}
+ \cdots
+ \prod_{k'=0}^{K-2} \epsilon_{{\hat j}_{k'}}
}{1 - \epsilon_{i} \prod_{j \in \mathcal{\tilde K}_r^z} \epsilon_j}
\bigg)
\notag \\
\overset{(b)}{=}
&
s_i(1 - \epsilon_i)
\sum_{j_0=1}^N s_{j_0}
\sum_{j_1=1}^N s_{j_1}
\cdots
\sum_{j_{K-2}=1}^N s_{j_{K-2}}
\frac{1 + \epsilon_{j_0} + \epsilon_{j_0}\epsilon_{j_1} + \cdots + \prod_{k'=0}^{K-2} \epsilon_{{j}_{k'}} }{1 - \epsilon_{i} \prod_{k=0}^{K-2} \epsilon_{{j}_{k}}}
\notag \\
= &
s_i(1 - \epsilon_i)
\sum_{j_0,j_1,\cdots,j_{K-2}=1}^N
\prod_{k=0}^{K-2} s_{j_k}
\frac{1 + \sum_{k=0}^{K-2} \prod_{k'=0}^{k} \epsilon_{j_{k'}}}{1 - \epsilon_{i}\prod_{k=0}^{K-2} \epsilon_{j_k}}
,
\end{align}
\end{small}
\vspace{-0.3cm}

\noindent
where in equality (a), the summation
$
\sum_{{\hat j}_0 \in \mathcal{\tilde K}_r^z}
\sum_{{\hat j}_1 \in \mathcal{\tilde K}_r^z \backslash {\hat j}_0}
\cdots
\sum_{{\hat j}_{K-2} \in \mathcal{\tilde K}_r^z \backslash \{ \bigcup_{i=0}^{K-3} {\hat j}_0 \}}
$
enumerates all permutations of the clients in $\mathcal{\tilde K}_r^z$.
Summing over all possible combinations of $\mathcal{\tilde K}_r^z$ yields all permutations of selecting $K$ clients from the $N$ clients, as shown in equality (b).
\hfill $\blacksquare$

\section{Gradient of Objective Function in (\ref{eq:gradient descent})}\label{sec:Objective Function Gradient}

According to \eqref{eq:definition chi square alpha} and \eqref{eq:objective function}, the gradient of objective function is given by
\begin{small}
\begin{align}
\frac{\partial {\rm(\ref{eq:objective function})}}{\partial s_i}
=
- 2
{\sum}_{c=1}^C
\bigg(
\frac{\alpha_c - \sum_{j=1}^N {\bar \beta}_j \alpha_{j,c}}{\alpha_c}
{\sum}_{j=1}^N
\alpha_{j,c}
\frac{ \partial {\bar \beta}_j  }{\partial s_i}
\bigg)
,
\end{align}
\end{small}
\vspace{-0.3cm}

\noindent
where based on \eqref{eq:bar beta i formulation}, the gradient ${ \partial {\bar \beta}_j  }/{\partial s_i}$ is computed by
\begin{small}
\begin{align}
\frac{
\partial
{\bar \beta}_j  }{\partial s_i}
= &
{\sum}_{z=1, \, n_{z,j} > 0}^{C^{N+K-1}_K}
\frac{K!}{\prod_{i' \in [N]} n_{z,i'}!}
\beta_{z,j}
\frac{
\partial
\big(
\prod_{j' \in \mathcal{K}_r^z} s_{j'}
\big)
}{\partial s_i}
\notag \\
= &
{\sum}_{z=1, \, n_{z,j} > 0, \, n_{z,i} > 0}^{C^{N+K-1}_K}
\frac{K!}{{\prod}_{i' \in [N]} n_{z,i'}!}
\beta_{z,j}
n_{z,i} s_i^{n_{z,i}-1}
{\prod}_{j' \in \mathcal{K}_r^z \backslash \{ i \}}
s_{j'}
.
\end{align}
\end{small}
\vspace{-0.3cm}

\noindent
Here, $n_{z,j}$ and $n_{z,i}$ denote the number of times clients $j$ and $i$ appear in the selection set $\mathcal{K}_r^z$, respectively.
Additionally, the term $\mathcal{K}_r^z \backslash \{ i \}$ refers to the set obtained by removing all occurrences of client $i$ from the set $\mathcal{K}_r^z$.
\hfill $\blacksquare$

\section{Proof of Proposition \ref{proposition:difference increasing K}}\label{sec:proposition difference increasing K}

Since \eqref{eq:bar beta i formulation} and \eqref{eq:beta_zi_failure_prob_diff} are equivalent, the effective appearance probability ${\bar \beta}_i$ when selecting $K$ clients per iteration, denoted as $[ {\bar \beta}_i ]_{K}$, is computed by
\begin{small}
\begin{align}\label{eq:beta_zi_failure_prob_diff_K1}
[ {\bar \beta}_i ]_{K}
=
s_i
\Bigg(
1 +
\sum_{i'=1, i' \neq i}^N s_{i'} (\epsilon_{i'} - \epsilon_i)
\underbrace{
\sum_{j_1,j_2,\cdots, j_{K-2}=1}^N
\prod_{k=1}^{K-2} s_{j_k}
\frac{1 + \sum_{k=1}^{K-2} \prod_{k'=1}^{k} \epsilon_{j_{k'}}}{1 - \epsilon_{i}\epsilon_{i'} \prod_{k=1}^{K-2} \epsilon_{j_k}}
}_{(\ref{eq:beta_zi_failure_prob_diff_K1}a)}
\Bigg)
.
\end{align}
\end{small}
\vspace{-0.3cm}

Similarly, when selecting $(K+1)$ clients, we have
\begin{small}
\begin{align}\label{eq:beta_zi_failure_prob_diff_K2}
[ {\bar \beta}_i ]_{K+1}
=
s_i
\Bigg(
1 +
\sum_{i'=1, i' \neq i}^N s_{i'} (\epsilon_{i'} - \epsilon_i)
\underbrace{
\sum_{j_1,j_2,\cdots, j_{K-1}=1}^N
\prod_{k=1}^{K-1} s_{j_k}
\frac{1 + \sum_{k=1}^{K-1} \prod_{k'=1}^{k} \epsilon_{j_{k'}}}{1 - \epsilon_{i}\epsilon_{i'} \prod_{k=1}^{K-1} \epsilon_{j_k}}
}_{(\ref{eq:beta_zi_failure_prob_diff_K2}a)}
\Bigg)
.
\end{align}
\end{small}
\vspace{-0.2cm}

Based on \eqref{eq:beta_zi_failure_prob_diff_K1} and \eqref{eq:beta_zi_failure_prob_diff_K2}, the difference between $[ {\bar \beta}_i ]_{K}$ and $[ {\bar \beta}_i ]_{K+1}$ is computed as
\begin{small}
\begin{align}\label{eq:difference beta_zi_failure_prob_diff_K1_K2}
&
[ {\bar \beta}_i ]_{K}
-
[ {\bar \beta}_i ]_{K+1}
\notag \\
= &
s_i
\sum_{i'=1, i' \neq i}^N s_{i'} (\epsilon_{i'} - \epsilon_i)
\Big(
{\text (\ref{eq:beta_zi_failure_prob_diff_K1}a)}
-
{\text (\ref{eq:beta_zi_failure_prob_diff_K2}a)}
\Big)
\notag \\
= &
s_i
\sum_{i'=1, i' \neq i}^N s_{i'} (\epsilon_{i'} - \epsilon_i)
\Bigg(
\sum_{j_1,j_2,\cdots, j_{K-2}=1}^N
\prod_{k=1}^{K-2} s_{j_k}
\underbrace{\sum_{j_{K-1}=1}^N s_{j_{K-1}}}_{=1}
\frac{1 + \sum_{k=1}^{K-2} \prod_{k'=1}^{k} \epsilon_{j_{k'}}}{1 - \epsilon_{i}\epsilon_{i'} \prod_{k=1}^{K-2} \epsilon_{j_k}}
\notag \\
&
\qquad\qquad\qquad\qquad\qquad\quad
-
\sum_{j_1,j_2,\cdots, j_{K-2}=1}^N
\prod_{k=1}^{K-2} s_{j_k}
\sum_{j_{K-1}=1}^N s_{j_{K-1}}
\frac{1 + \sum_{k=1}^{K-2} \prod_{k'=1}^{k} \epsilon_{j_{k'}} + \prod_{k'=1}^{K-1} \epsilon_{j_{k'}}}{1 - \epsilon_{i}\epsilon_{i'} \prod_{k=1}^{K-1} \epsilon_{j_k}}
\Bigg)
\notag \\
= &
s_i
\sum_{i'=1, i' \neq i}^N s_{i'} (\epsilon_{i'} - \epsilon_i)
\sum_{j_1,j_2,\cdots, j_{K-2}=1}^N
\prod_{k=1}^{K-2} s_{j_k}
\sum_{j_{K-1}=1}^N s_{j_{K-1}}
\Bigg(
\frac{1 + \sum_{k=1}^{K-2} \prod_{k'=1}^{k} \epsilon_{j_{k'}}}{1 - \epsilon_{i}\epsilon_{i'} \prod_{k=1}^{K-2} \epsilon_{j_k}}
-
\frac{1 + \sum_{k=1}^{K-2} \prod_{k'=1}^{k} \epsilon_{j_{k'}} + \prod_{k'=1}^{K-1} \epsilon_{j_{k'}}}{1 - \epsilon_{i}\epsilon_{i'} \prod_{k=1}^{K-1} \epsilon_{j_k}}
\Bigg)
\notag \\
= &
s_i
\sum_{i'=1, i' \neq i}^N s_{i'} (\epsilon_{i'} - \epsilon_i)
\sum_{j_1,j_2,\cdots, j_{K-2}=1}^N
\prod_{k=1}^{K-2} s_{j_k}
\sum_{j_{K-1}=1}^N s_{j_{K-1}}
\Bigg(
\frac{\big(1 + \sum_{k=1}^{K-2} \prod_{k'=1}^{k} \epsilon_{j_{k'}} \big) \big(1 - \epsilon_{j_{K-1}} \big)\epsilon_{i}\epsilon_{i'} \prod_{k=1}^{K-2} \epsilon_{j_k}}
{\big(1 - \epsilon_{i}\epsilon_{i'} \prod_{k=1}^{K-2} \epsilon_{j_k} \big) \big(1 - \epsilon_{i}\epsilon_{i'} \prod_{k=1}^{K-1} \epsilon_{j_k} \big)}
-
\frac{\prod_{k'=1}^{K-1} \epsilon_{j_{k'}}}{1 - \epsilon_{i}\epsilon_{i'} \prod_{k=1}^{K-1} \epsilon_{j_k}}
\Bigg)
.
\end{align}
\end{small}
\vspace{-0.2cm}

According to \eqref{eq:si0 ei0}, $s_i = 0$ when $\epsilon_i = 1$.
Therefore, the value of \eqref{eq:difference beta_zi_failure_prob_diff_K1_K2} depends only on the clients with $\epsilon_i \in [0,1)$.
As $K$ increases, both $\epsilon_{i}\epsilon_{i'} \prod_{k=1}^{K-2} \epsilon_{j_k}$ and $\prod_{k=1}^{K-1} \epsilon_{j_k}$ approach zero.
Consequently, the value of \eqref{eq:difference beta_zi_failure_prob_diff_K1_K2} also approaches zero, leading to the diminishing difference between $\left[ {\bar \beta}_i \right]_{K}$ and $\left[ {\bar \beta}_i \right]_{K+1}$.
\hfill $\blacksquare$

\section{Stability of Different Aggregation Schemes}\label{sec:convergence TF-Aggregation}

\subsection{Influence of Transmission Failure Probability on the Stability of Baseline TF-Aggregation \cite{salehi2021federated}}

According to \cite[Lemma 2]{salehi2021federated}, the aggregation rule in \eqref{eq:optimization problem TO} yields an unbiased estimation of the global model under full participation, i.e.,
\begin{align}\label{eq:TO unbiased estimation}
\mathbb{E}_{\mathcal{K}_{r}, \mathds{1}^{r}_{i}} [{\bar {\mathbf{w}}}_{r}]
\overset{(a)}{=} & \frac{1}{K} \mathbb{E}_{\mathcal{K}_{r}}
\bigg[ {\sum}_{{i} \in {\mathcal{K}}_{r}} \frac{p_i}{s_i \left(1- \epsilon^{r}_{i} \right)} \mathbf{w}^{r,E}_{i} \cdot \left(1- \epsilon^{r}_{i} \right) \bigg]
= \frac{1}{K} {\sum}_{{i} \in {\mathcal{K}}_{r}}
\mathbb{E}_{\mathcal{K}_{r}} \bigg[ \frac{p_i}{s_i} \mathbf{w}^{r,E}_{i} \bigg] \notag \\
= & \frac{1}{K} {\sum}_{{i} \in {\mathcal{K}}_{r}} \bigg( {\sum}_{i=1}^{N} \frac{p_i}{s_i} \mathbf{w}^{r,E}_{i} \cdot s_i \bigg)
= {\sum}_{i=1}^{N} p_i \mathbf{w}^{r,E}_{i},
\end{align}
where equality (a) follows from $\Pr(\mathds{1}^{r}_{i}=1)=1-\epsilon^{r}_{i}$ and $\Pr(\mathds{1}^{r}_{i}=0)= \epsilon^{r}_{i}$.
The unbiasedness property underpins the convergence guarantee of \texttt{TF-Aggregation} \cite[Appendix B]{salehi2021federated}.
Then, from the convergence analysis \cite[Corollary 1]{salehi2021federated}, the summation of all clients' aggregation weights
\begin{equation}\label{eq:summation weights}
\sum_{i=1}^N \frac{p_i}{s_i (1- \epsilon^{r}_{i} )}
\end{equation}
directly affects convergence.
An increase in the transmission failure probability $\epsilon_i$ enlarges the summation in \eqref{eq:summation weights}, thereby degrading the convergence rate.
Accordingly, \cite{salehi2021federated} formulates the client selection problem in \eqref{eq:optimization problem FedAvg TO} to minimize \eqref{eq:summation weights}, which allocate larger selection probabilities $s_i$ to clients with higher $\epsilon_i$.

However, the unbiasedness in \eqref{eq:TO unbiased estimation} holds only in expectation.
At each communication round, the summation of the selected clients' aggregation weights
\begin{equation}
\frac{1}{K} \sum_{{i} \in {\mathcal{K}}_{r}}
\mathds{1}^{r}_{i} \frac{p_i}{s_i \left(1- \epsilon^{r}_{i} \right)}
\end{equation}
is not guaranteed to be equal to one.
When $\epsilon^{r}_{i}$ approaches one, the denominator becomes excessively large, undermining the stability of global aggregation in \eqref{eq:optimization problem TO}.

\subsection{Stability of Proposed FedCote under Transmission Failures}

In contrast, the aggregation scheme \eqref{global_model_failure} adopted in our proposed \texttt{FedCote} guarantees that the summation of aggregation weights always satisfies
\begin{align}\label{global_model_failure_summation}
\frac{\sum_{{i} \in {\mathcal{K}}_{r}} \mathds{1}^{r}_{i} \mathbf{w}^{r,E}_{i} }{\sum_{{i} \in {\mathcal{K}}_{r}} \mathds{1}^{r}_{i}}
=1,
\;\;
\forall r \in [R],
\end{align}
where ${\mathds{1}^{r}_{i}}=1$ indicates successful receipt of client ${i}$'s local model, and ${\mathds{1}^{r}_{i}}=0$ otherwise.

As shown in Lemma \ref{lemma:beta} and Theorem \ref{theorem:convergence FL failure}, transmission failures introduce divergence between the actual and effective global label distributions, thereby affecting both convergence rate and direction.
Nevertheless, the weight normalization property in \eqref{global_model_failure_summation}
guarantees that the aggregated global model converges stably.

\end{appendices}

\end{document}